%% file: ms.tex
\documentclass[final,5p,times,twocolumn]{elsarticle}
\usepackage{graphicx}
\usepackage{amssymb}
\usepackage{amsmath}

\usepackage{import}
\usepackage{color}
\usepackage[colorlinks,urlcolor=blue,bookmarks=false]{hyperref}
\usepackage{booktabs}
\usepackage{symbol_short_cuts}
\usepackage{xfrac}
\usepackage[british]{babel}
\usepackage{array}

\usepackage[nodots]{numcompress}

\journal{Renewable Energy}

\let\oldhat\hat
\renewcommand{\vec}[1]{\mathbf{#1}}
\renewcommand{\hat}[1]{\oldhat{\mathbf{#1}}}
\newcommand{\regtm}{\textsuperscript{\textregistered}{}}

\begin{document}

\begin{frontmatter}



\title{Quasi-Steady Model of a Pumping Kite Power System}


\author[a]{Rolf van der Vlugt}
\author[a,b]{Anna Bley}
\author[a]{Michael Noom}
\author[a]{Roland Schmehl\corref{cor1}}
\ead{r.schmehl@tudelft.nl}
\address[a]{Delft University of Technology, Faculty  of Aerospace Engineering, Kluyverweg 1, 2629 HS Delft, Netherlands}
\address[b]{Kitepower B.V., Kluyverweg 1, 2629 HS Delft, Netherlands}
	
\cortext[cor1]{Corresponding author. Tel.: +31 15 278 5318.}

\begin{abstract}
The traction force of a kite can be used to drive a cyclic motion for extracting wind energy from the atmosphere.
This paper presents a novel quasi-steady modelling framework for predicting the power generated over a full pumping cycle.
The cycle is divided into traction, retraction and transition phases, each described by an individual set of analytic equations.
The effect of gravity on the airborne system components is included in the framework.
A trade-off is made between modelling accuracy and computation speed such that the model is specifically useful for system optimisation and scaling in economic feasibility studies. 
Simulation results are compared to experimental measurements of a 20 kW kite power system operated up to a tether length of 720 m.
Simulation and experiment agree reasonably well, both for moderate and for strong wind conditions, indicating that the effect of gravity has to be taken into account for a predictive performance simulation.
\end{abstract}


\end{frontmatter}

%\linenumbers

\section{Introduction}
\label{sec:intro}
%%%%%%%%%%%%%%%%%%%%%%%%%%%%%%%%%%%%%%%%%%%%%%%%%%%%%%%%%%%%%%%%%%%%%%%%%%%%%%%%%

The pumping kite concept provides a simple yet effective solution for wind energy conversion at a potentially low cost \cite{Cherubini2015}.
Important aspects of the technology are the performance characteristics of implemented concepts and how these depend on the operational and environmental parameters.
Various modelling frameworks have been proposed to predict the traction force and power generated by a tethered wing, both for the production of electricity \cite{Loyd1980, Argatov2009, Fagiano2010, Costello2015, Fechner2016b, Argatov2015} and for the propulsion of ships \cite{Wellicome1984, Wellicome1985, Naaijen2007, Dadd2010, Dadd2011, Fagiano2012c}. The analysis presented in \cite{Wellicome1984, Wellicome1985} has been validated experimentally, yet not assessed for its potential to predict the power generated over a full cycle of a pumping system. Dynamic models have been proposed by \cite{Williams2008a, Houska2007b, Fagiano2014, Zgraggen2014, Fechner2015, DeLellis2016} to address challenges in the field of control or by \cite{Schmidt2017} for state estimation.
Recent studies have used measurement data from full-scale demonstrator systems to analyse the turning dynamics of kites and to assess flight control algorithms \cite{Erhard2013b,Jehle2014}.

The current challenge is to formulate a model that does not require advanced control algorithms, while accurately predicting the power generated over a pumping cycle. 
For this purpose it is important to critically revise commonly used simplifying assumptions, for example, regarding the wind velocity gradient, the tether shape, the mass of tether and kite and the aerodynamic properties of the wing. 
%Valid arguments are given for applying these assumptions, yet the experimental evidence for the validity of these assumptions is not provided. 
%
%The current research focuses on the situation where the mechanical power supplied by a tethered wing is converted by a ground generator. 
The model is intended for optimisation of pumping cycle kite power systems and for predicting the achievable cost of energy.
Section~\ref{sec:model} first describes the analytical framework assuming a massless system, which is then extended to account for the effect of gravity on all airborne system components. An experimental setup, consisting of a fully operational pumping kite power system is presented in Sect.~\ref{sec:experimentalsetup}. To validate the described model, measured and computed results are compared in Sect.~\ref{sec:results}.
The preliminary results of this study had been presented at the Airborne Wind Energy Conference 2015 in Delft \cite{Schmehl2015a}.

%%%%%%%%%%%%%%%%%%%%%%%%%%%%%%%%%%%%%%%%%%%%%%%%%%%%%%%%%%%%%%%%%%%%%%%%%%%%%%%%%
\section{Computational approach}
\label{sec:model}
\begin{figure}[h]
\centering
\scriptsize
\def\svgwidth{250pt}
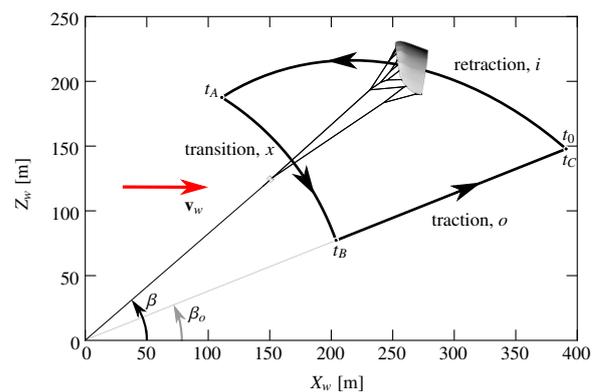
\caption{Idealised flight trajectory of a traction kite of a pumping cycle. The trajectory segment in the traction phase does not resolve the crosswind flight manoeuvres. Adapted from \cite{Fechner2013a}.}
\label{fig:trajectory}
\end{figure}

%%%%%%%%%%%%%%%%%%%%%%%%%%%%%%%%%%%%%%%%%%%%%%%%%%%%%%%%%%%%%%%%%%%%%%%%%%%%%%%%%
For the theoretical analysis the pumping cycle is divided into the three characteristic phases illustrated in Fig.~\ref{fig:trajectory}: the retraction phase, from $t_0$ until $t_A$, the transition phase, from $t_A$ until $t_B$, and the traction phase, from $t_B$ until $t_C$, closing the cycle.
The depicted side view of the idealised flight trajectory in the wind reference frame includes the  wind velocity $\vec{v}_w$ in direction of the $X_w$-axis and the elevation angle as $\beta$. A detailed presentation of the forces governing the flight operation of a kite including the gravitational and inertial effects is provided in \cite{Schmehl2013,Noom2013}.
In the following we discuss several assumptions that reduce the complexity of the computational approach to achieve a substantial speed-up of the simulations.

Firstly, the study is limited to kites with relatively large surface-to-mass ratio. For such kites the timescale of dynamic processes is generally very short compared to the timescales of typical flight manoeuvres or complete pumping cycles. As consequence the flight operation is dominated by the balance of aerodynamic, tether and gravitational forces and can be approximated as a transition through quasi-steady flight states.
The analysis is further limited to typical tether lengths during pumping operation which are much larger than the geometrical dimensions of the kite. 
At very short tether length, as occurring during launching and landing, inertial forces such as centrifugal forces, can contribute substantially.

Secondly, the tether is assumed to be inelastic. It is represented by a straight line although the effect of sagging due to distributed gravitational loading is taken into account.
Thirdly, the aerodynamic properties of the kite are assumed to be constant throughout each phase. Lastly, the atmospheric properties are assumed to be constant over time but varying with altitude.
This is taken into account by assuming altitude profiles for both the wind velocity and the air density.

\subsection{Atmospheric Wind Model}
\label{sec:atmospheric}
Conventional tower-based wind turbines have a constant hub height and operate within a limited atmospheric layer close to the ground. Pumping kite power systems on the other hand can harvest energy from a much larger and variable altitude range. Because the wind velocity $v_w$ increases substantially between the minimum and maximum altitude of the kite it is important to include the wind velocity profile in the simulation.
In the atmospheric boundary layer up to 500 m altitude the functional dependency can be estimated by the logarithmic wind law \cite{Stull2000}
\begin{equation}\label{eq:windshear}
v_w = v_{w,ref}~ \frac{\ln (z/z_0)}{\ln (z_{ref}/z_0)} ,
\end{equation}
where $v_{w,ref}$ is a known reference wind speed at a reference altitude $z_{ref}$ and $z_0$ is the aerodynamic roughness length. The logarithmic profile suits best to model a neutral boundary layer, which typically develops in overcast or windy conditions.  

The decrease of air density $\rho$ with increasing altitude can be approximated by the barometric altitude formula for constant temperature \cite{Stull2000}
\begin{equation}\label{eq:airdensity}
\rho = \rho_0~ \exp \left(-\frac{z}{H_{\rho}}\right) ,
\end{equation} 
where $\rho_0 = 1.225$ kg/m$^3$  is the standard atmospheric density at sea level at the standard temperature of $T^0 = 15^\circ$C and $H_{\rho} = 8.55$ km is the scale height for density.  

\subsection{Basic Modelling Framework}
\label{sec:basic_model}

\begin{figure}[h]
\centering
\def\svgwidth{250pt}
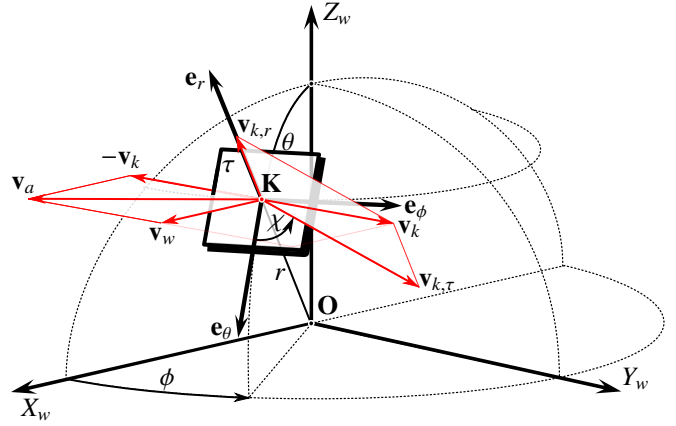
\caption{Decomposition of kite velocity $\vec{v}_k$ into radial component $\vec{v}_{k,r}$ and tangential component $\vec{v}_{k,\tau}$, definition of apparent wind velocity $\vec{v}_a=\vec{v}_w-\vec{v}_k$. Course angle $\chi$ is measured in the tangential plane $\tau$, spherical coordinates $(r, \theta, \phi)$ defined in the wind reference frame $X_w, Y_w, Z_w$, where $X_w$ represents the wind direction \cite{Schmehl2013}.}
\label{fig:coordinatesystem}
\end{figure}

Starting point for the analysis is the wind reference frame which has its origin $\vec{O}$ coinciding with the tether exit point from the ground station and has its $Z_w$-axis pointing vertically upwards and its $X_w$-axis aligned with the wind direction.
The kite is represented by a geometrical point.
To describe its position $\vec{K}$ and velocity $\vec{v}_k$ we follow the approaches in \cite{Argatov2009, Schmehl2013, Noom2013} and use a spherical coordinate system $(r, \theta, \phi)$.
As depicted in Fig.~\ref{fig:coordinatesystem}, the position is described by the radial distance $r$, the polar angle $\theta$ and the azimuth angle $\phi$. The direction of flight in the local tangential plane $\tau$ is described by the course angle $\chi$.

The apparent wind velocity describes the flow velocity relative to the kite 
\begin{equation}\label{eq:Vapp_definition}
\vec{v}_{a} = \vec{v}_{w} - \vec{v}_{k} .
\end{equation}
This vector can be described in spherical coordinates as follows
\begin{equation}\label{eq:Vapp_definition2}
\vec{v}_{a} = 
     \begin{bmatrix}
       \sin\theta\cos\phi   \\[0.3em]
       \cos\theta\cos\phi    \\[0.3em]
       -\sin\phi       
     \end{bmatrix} v_w - 
   \begin{bmatrix}
          1 \\[0.3em]
          0  \\[0.3em]
          0      
        \end{bmatrix} v_{k,r} -
\begin{bmatrix}
       0 \\[0.3em] 
       \cos \chi \\[0.3em]
        \sin \chi 
     \end{bmatrix} v_{k,\tau} ,
\end{equation}
where $v_{k,r}$ and $v_{k,\tau}$ represent the radial and tangential contributions to the kite velocity, respectively.

The straight tether implies that the radial kite velocity is identical with the reeling velocity
\begin{equation}\label{eq:v_a_r_v_t}
   v_{k,r} = v_t .
\end{equation}
Introducing the reeling factor 
\begin{equation}\label{eq:reeloutfactor}
f=\frac{v_{k,r}}{v_w}
\end{equation}
and the tangential velocity factor
\begin{equation}\label{eq:lambda_definition}
\lambda=\frac{v_{k,\tau}}{v_w} ,
\end{equation}
Eq.~(\ref{eq:Vapp_definition2}) can be formulated as
\begin{equation}\label{eq:v_a3}
\vec{v}_{a} 
=
\begin{bmatrix}
       \sin\theta\cos\phi  - f \\[0.3em]
       \cos\theta\cos\phi -\lambda\cos\chi   \\[0.3em]
       -\sin\phi  -\lambda\sin\chi     
     \end{bmatrix} v_w .
\end{equation}
%
%The apparent wind velocity can also be decomposed into radial and tangential components
%%
%\begin{equation}\label{eq:v_a_r_and_t}
%\vec{v}_{a} =  \vec{v}_{a,r} +  \vec{v}_{a,\tau} .
%\end{equation}
%%
The meaning of the velocity variables in this expression can be summarised as follows. The reeling factor $f$ is controlled by the ground station, the course angle $\chi$ is controlled by the steering system and the tangential velocity factor $\lambda$ is a dependent variable, which is determined by the force equilibrium.

The integral aerodynamic force acting on the airborne system components can be decomposed into lift and drag vectors
\begin{equation}
   \vec{F}_a = \vec{L} + \vec{D}.
\end{equation}
The lift and drag forces contributed solely by the wing are calculated as
\begin{equation}\label{eq:lift}
L  = \frac{1}{2}\,\rho\, C_L\, v_{a}^2\, S , 
\end{equation}
and 
\begin{equation}\label{eq:drag}
D_k  = \frac{1}{2}\,\rho\, C_{D,k} \,v_{a}^2\, S ,
\end{equation}
where $C_L$ and $C_{D,k}$ are the aerodynamic lift and drag coefficients, respectively, and $S$ the projected surface area of the wing.

The aerodynamic drag of the tether is taken into account by adding one fourth of the tether drag area to the kite drag area as proposed in \cite{Argatov2009} and numerically validated in \cite{Argatov2011}. The total aerodynamic drag $D$ of the airborne system is then estimated as
\begin{equation}\label{eq:totaldrag}
D  = D_k + D_t, 
\end{equation}
where
\begin{equation}\label{eq:tetherdrag}
D_t  = \frac{1}{8}\,\rho\, d_t\, r\, C_{D,c}\, v_{a}^2,
\end{equation}
with $d_t$ being the tether diameter, $r$ the tether length, $C_{D,c}$ the drag coefficient of a cylinder in cross flow and $v_{a}$ the apparent wind velocity at the kite. With the tether being subjected to a relative velocity $v_{a,t}$ between $10$ and $30$  m/s and a kinematic viscosity of $\nu = 1.47 \times 10^{-5}$, the Reynolds number Re = $v_{a,t}~d_t/\nu$  is estimated to be between $2.7 \times 10^3$ and $8.2 \times 10^3$. In this range $C_{D,c}$ has a constant value of $1.1$ \cite{Hoerner1975}. 
As consequence a total aerodynamic drag coefficient of the airborne system components can be defined as
\begin{equation}\label{eq:totaldragcoeff}
C_D = C_{D,k} + \frac{1}{4} \frac{d_t\, r}{S}  C_{D,c}.
\end{equation}
\subsection{Analytic Model for Negligible Effect of Mass}
\label{sec:analytic_model}

\begin{figure}[h]
\centering
\def\svgwidth{250pt}
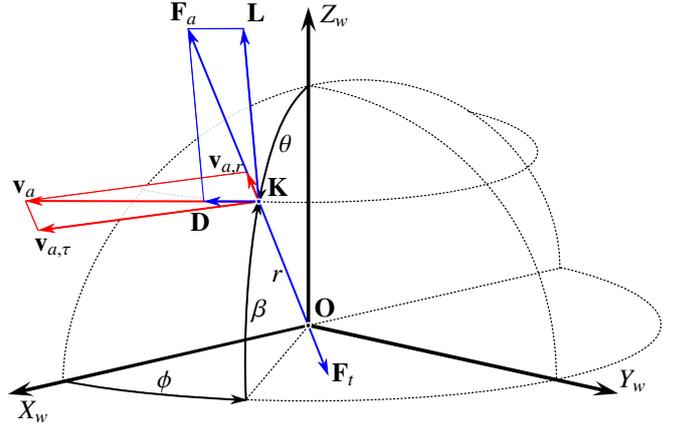
\caption{Geometrical similarity of the force and velocity diagrams. $\vec{v}_a$ and $\vec{F}_a$ are decomposed in the plane spanned by the two vectors. $\vec{D}$ is aligned with $\vec{v}_a$ by definition, whereas $\vec{v}_{a,r}$ is aligned with $\vec{F}_a$ when assuming a straight tether and a negligible effect of mass. Adapted from \cite{Schmehl2013}.}\label{fig:trianglesimilarity}
\end{figure}

For the massless case, the radial and tangential components of the apparent wind velocity and the lift and drag components of the aerodynamic force are related as follows
\begin{equation}\label{eq:triangle_similarity}
\kappa = \frac{v_{a,\tau}}{v_{a,r}} = \frac{L}{D}.
\end{equation}
The ratio of the relative velocity components is denoted as kinematic ratio and represented by the symbol $\kappa$.
Equation~(\ref{eq:triangle_similarity}) can be derived from the geometrical similarity of the force and velocity diagrams illustrated in Figs.~\ref{fig:trianglesimilarity} and \ref{fig:velocity_force_diagram}.
%
%This similarity can be explained as follows: vectors $\vec{v}_a$ and $\vec{F}_a$ span the plane in which both vectors are decomposed.
%The aerodynamic drag component $\vec{D}$ is aligned with $\vec{v}_a$ by definition.
%The radial component $\vec{v}_{a,r}$ is aligned with $\vec{F}_a$ as consequence of the straight tether assumption in combination with the quasi steady state assumption. Quasi steady state implies that on each moment in time the system has reached a situation where all forces acting on the kite have reached an equilibrium.
%The two alignments are the reason for the geometric similarity.
%
%Equation (\ref{eq:triangle_similarity}) corresponds to \cite[Eq. (19)]{Argatov2009} and to \cite[Eq. (11)]{Loyd1980} for the special case of $\phi=0$. 
%The radial component of the apparent wind velocity follows from Eq. (\ref{eq:v_a3}) as
%%
%\begin{equation}\label{eq:v_a_r}
%v_{a,r} = (\sin\theta\cos\phi  - f)v_w .
%\end{equation}
%
%From the vector sum in Eq. (\ref{eq:v_a_r_and_t}) follows
%
%\begin{equation}\label{eq:v_a_magnitude}
%v_{a} = \sqrt{(v_{a,r})^2 + (v_{a,\tau})^2}
%\end{equation}
%{\color{red} Commented out the magnitude equation. I think we can skip this step.}
%
Starting from the decomposition $\vec{v}_a = \vec{v}_{a,r} + \vec{v}_{a,\tau}$ and using the radial component of Eq.~(\ref{eq:v_a3}) in conjunction with Eq.~(\ref{eq:triangle_similarity}) to eliminate the tangential component results in the following expression for the nondimensional apparent wind velocity
\begin{equation}\label{eq:Vapp_final}
\frac{v_{a}}{v_w} = (\sin\theta\cos\phi  - f) \sqrt{1 + \left( \frac{L}{D}\right) ^2} .
\end{equation}
%
%By definition, the magnitude of the apparent wind velocity cannot be negative which constrains the reel-out factor as follows
%%
%\begin{equation}
%f < \sin\theta\cos\phi .
%\end{equation}
%%
%This corresponds to the fundamental flight requirement of a massless system that the component of the wind velocity along the tether needs to be higher than the tether reeling velocity.
On the other hand, inserting the radial and tangential velocity components of Eq.~(\ref{eq:v_a3}) into Eq.~(\ref{eq:triangle_similarity}) and solving for the tangential velocity factor $\lambda$ results in

\begin{equation}\label{eq:tangentialkitevelocityfactor}
\lambda = a + \sqrt{ a^2 + b^2 - 1 + \left(\frac{L}{D}\right)^2 \left( b -f \right)^2} ,
\end{equation}
with trigonometric coefficients
\begin{align}
\label{eq:a_trigcoeff}
a & = \cos\theta \cos\phi \cos\chi - \sin\phi \sin\chi , \\
\label{eq:b_trigcoeff}
b & = \sin\, \theta \cos\phi .
\end{align}
%

%An equation for the tangential kite velocity was also derived in \cite[Eq. (3)]{Argatov2010}. By definition, the tangential velocity factor $\lambda$ cannot be negative. Analyzing Eq. (\ref{eq:tangentialkitevelocityfactor}) for this condition results in the following constraint
%%
%\begin{equation}\label{eq:lambda_limit}
%\sin\theta \cos\phi <  \frac{\sqrt{1 + \left( \frac{L}{D}\right)^2\left( 1- f^2\right)} + f\left(\frac{L}{D}\right)^2}{1 + \left( \frac{L}{D}\right)^2} ,
%%\frac{\sqrt{\left( \frac{L}{D}\right)^2\left( 1- f^2\right)  + 1} + f\left( \frac{L}{D}\right)^2}{\left( \frac{L}{D}\right)^2 + 1}
%\end{equation}
%%
%which indicates that there is a maximum azimuth angle $\phi_{max}$ and elevation angle $\beta_{max}$ for physically possible flight conditions. 
%
The quasi-steady motion of a massless kite is governed by the equilibrium of the tether force and the resultant aerodynamic force 
\begin{equation}\label{eq:T_Fa}
\vec{F}_t + \vec{F}_a = 0.
\end{equation}
Inserting Eqs.~(\ref{eq:lift}) and (\ref{eq:drag}) into Eq.~(\ref{eq:T_Fa}) results in
\begin{equation}\label{eq:tetherforce_a}
F_t = \frac{1}{2} \rho C_R v_{a}^2 S ,
\end{equation}
with the resultant aerodynamic force coefficient
\begin{equation}\label{eq:aerocoeff}
C_R = \sqrt{C_D^2 + C_L^2} .
\end{equation}
Using Eq.~(\ref{eq:Vapp_final}) to substitute the apparent wind velocity in Eq.~(\ref{eq:tetherforce_a}) gives the following equation for the normalised tether force \cite[Eq.~(48)]{Argatov2009}
%
%\begin{equation}\label{eq:tetherforce}
%F_t =  (\cos\beta \cos\phi - f)^2 K q
%\end{equation}
%
%where $K$ is the kite constant
%
%\begin{equation}\label{eq:kiteconstant}
%K = C_D A\left[1 + \left( \frac{L}{D}\right)^2\right]^{\frac{3}{2}}
%\end{equation}
%
%{\color{red}(I don't think we should use kite constant, because a part in here is later defined as effective lift-to-drag $b/c$ which also depends on operational conditions. I propose the following equation:)}
%
\begin{equation}\label{eq:tetherforce}
\frac{F_t}{q S} = C_R \left[ 1 + \left( \frac{L}{D}\right) ^2\right](\sin\theta\cos\phi  - f)^2 ,
\end{equation}
with the dynamic wind pressure at the altitude of the kite calculated as
\begin{equation}\label{eq:dynamicpressure}
q =\frac{1}{2} \rho v_w^2 ,
\end{equation}
with the air density and wind velocity described by Eqs.~(\ref{eq:airdensity}) and (\ref{eq:windshear}), respectively.
%The air density and wind velocity are dependent on the kite altitude as explained in Sec. \ref{sec:atmospheric}.
%
%Using Eq. (\ref{eq:aerocoeff}) the aerodynamic coefficient term can be formulated as
%%
%\begin{equation}\label{eq:aerocoeffterm}
%C_R \left[ 1 + \left( \frac{L}{D}\right) ^2\right] = C_R \left( \frac{C_R}{C_D} \right)^2 .
%\end{equation}
%%
%Equation (\ref{eq:tetherforce}) has been published previously by \cite[Eq. (48)]{Argatov2009}.

The generated traction power is determined as the product of tether force and reeling velocity
\begin{equation}
P =  F_t v_t = F_t f v_w .
\end{equation}
Expressing the tether force by Eq.~(\ref{eq:tetherforce}) results in
%
%\begin{equation}\label{eq:power}
%P =   f(\sin\theta \cos\phi - f)^2  K P_w
%\end{equation}
%
%without kiteconstant $K$:
\begin{equation}\label{eq:power}
\zeta = \frac{P}{P_w S} =   C_R \left[ 1 + \left( \frac{L}{D}\right) ^2\right] f(\sin\theta\cos\phi  - f)^2 ,
\end{equation}
where $P_w$ denotes the wind power density at the altitude of the kite
\begin{equation}\label{eq:powerdensity}
P_w = \frac{1}{2} \rho  v_w^3 .
\end{equation}
Equation~(\ref{eq:power}) defines the instantaneous power harvesting factor $\zeta$ as the normalised traction power per wing surface area.

\subsection{Relative Flow Conditions at the Kite}
\label{sec:relative_flow}

\begin{figure}[h]
	\centering
	\def\svgwidth{200pt}
	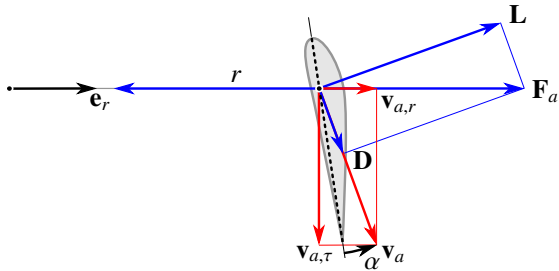
	\caption{Relative flow components $\vec{v}_{a,\tau}$ and $\vec{v}_{a,r}$ as well as force components $\vec{L}$ and $\vec{D}$ acting on the kite which is represented by the centre airfoil. The chord line is indicated by dots.}\label{fig:velocity_force_diagram}
\end{figure}

The aerodynamic coefficients used in Eqs.~(\ref{eq:lift}) and (\ref{eq:drag}) depend on the relative flow conditions that the kite experiences along its flight path.
For rigid and flexible membrane wings the key influencing parameter is the angle of attack $\alpha$, defined as the angle between the chord line of the wing and the apparent wind velocity vector $\vec{v}_a$.
The sketch in Fig.~\ref{fig:velocity_force_diagram} illustrates this, implying that the heading of the wing is in plane with the radial and tangential velocity components.
This is generally the case if the wing is not asymmetrically deformed due to steering actuation and sideslip velocity components can be neglected.

It can be shown from Fig.~\ref{fig:velocity_force_diagram} that the angle of attack does not vary along the flight path of a massless kite if the angle between wing and tether is constant. For flexible membrane wings this angle is generally controlled by the bridle line system which has the function of transferring the aerodynamic load to the tether. On the other hand, Eq.~(\ref{eq:triangle_similarity}) links the angle between the velocity vector $\vec{v}_a$ and the tether to the lift-to-drag ratio $L/D$. A constant $L/D$ thus ensures a constant $\alpha$, and vice versa.
While the relative flow angle is constant along the flight path, the magnitude of the relative flow velocity changes according to Eq.~(\ref{eq:Vapp_final}).

The effect of gravity induces variations of the flow angle along the flight path because the aerodynamic force $\vec{F}_{a}$ is not aligned anymore with the radial direction. In the following section this framework will be extended to include gravitational forces.

\subsection{Effect of Gravity on the Tether Force}
\label{sec:tether_gravity}
Equations~(\ref{eq:windshear}) to (\ref{eq:powerdensity}) provide an analytic modelling framework for the operation of a kite in pumping cycles for the ideal case of negligible gravity.
However, a real system is subject to gravitational and inertial forces which affect the flight behaviour and consequently also the traction power. 
%The effect of mass has been quantified in \cite[Eq. (7.8)]{Argatov2012} considering an arbitrary predefined trajectory for constant tether length.

In the present modelling framework we assume that the tether is long compared to the geometrical dimensions of the kite.
Accordingly, the kite is represented by a point mass $m$ and its gravitational force $m\vec{g}$ directly contributes to the quasi-steady force equilibrium at point $\vec{K}$.
Because of the long tether the angular velocities $\dot{\theta}$ and $\dot{\phi}$ are relatively small and the effect of inertial forces can be neglected.
The tether, on the other hand, is suspended between the ground station and the kite, its mass $m_t$ is continuously distributed over its length and the distributed loading by gravity and aerodynamic drag leads to sagging.

\begin{figure}[h]
	\centering
	\includegraphics[width=\linewidth]{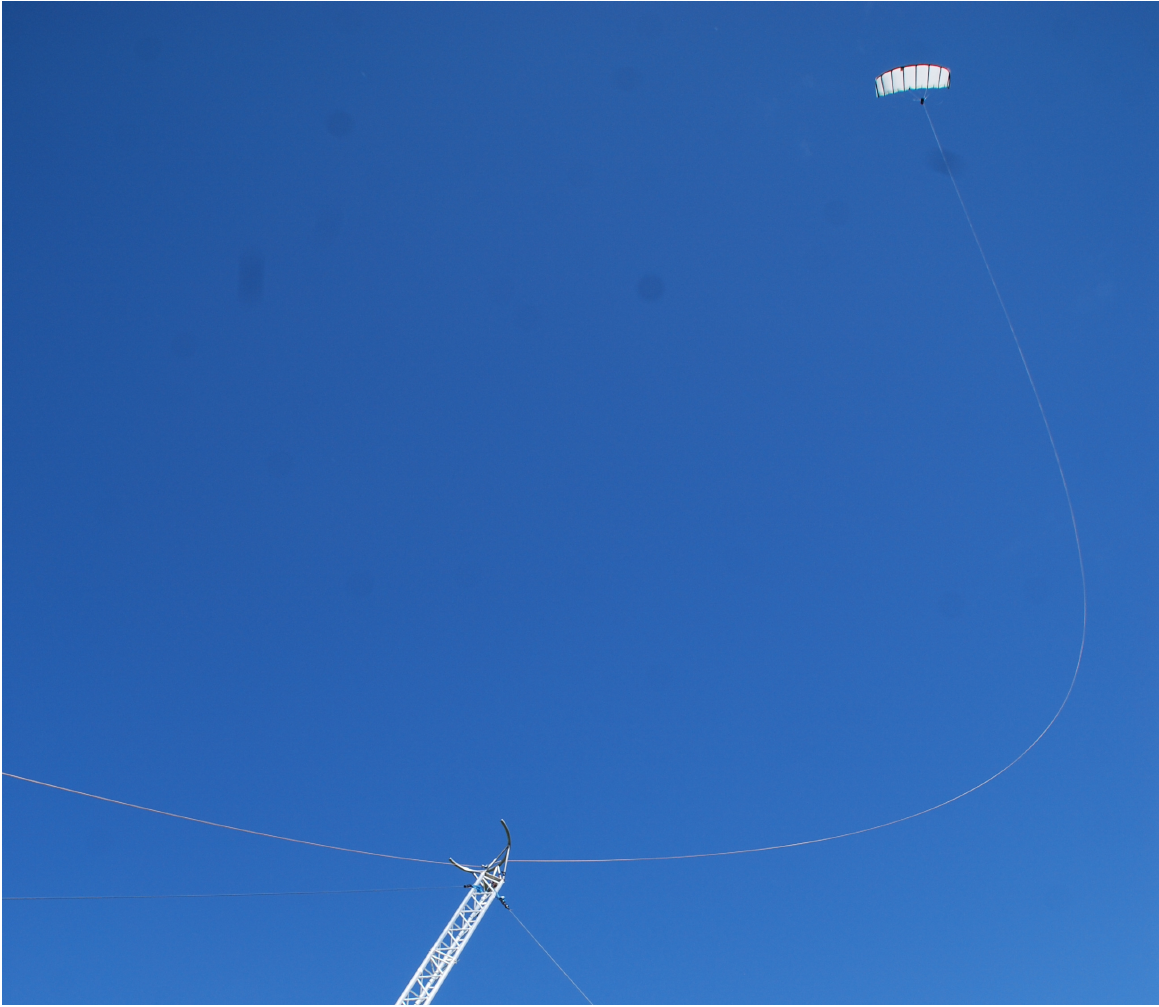}
	\caption{Strong sagging of the tether at low wind speed and static kite position on 23 August 2012. The kite has a surface area of 25 m$^2$.}
	\label{fig:sagging}
\end{figure}
\begin{figure}[h]
	\centering
	\def\svgwidth{250pt}
	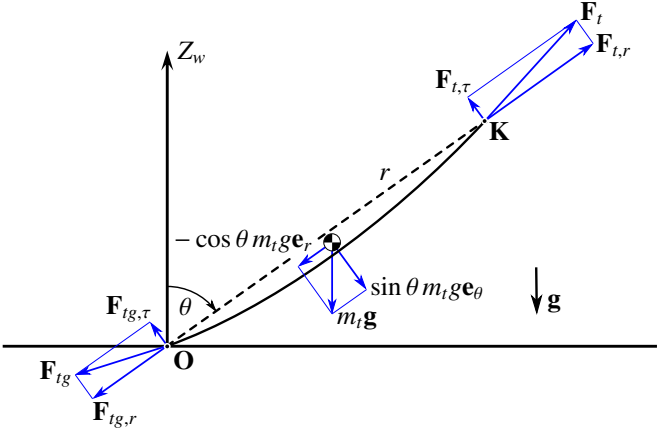
	\caption{Free body diagram of the deformed tether in the $\phi_k$-plane. The reaction forces $\vec{F}_{tg}$ and $\vec{F}_t$ acting at the suspension points $\vec{O}$ and $\vec{K}$, respectively, are decomposed into radial and tangential components. The idealised straight tether, which coincides with the radial coordinate, is included as dashed line.}\label{fig:tethermass}
\end{figure}

The photo shown in Fig.~\ref{fig:sagging} captures a moment of a particularly pronounced effect of gravity and aerodynamic drag.
This specific case was the combined result of low wind velocity and low reel-in speed, both contributing to a reduced tension in the tether.
To calculate the force $\vec{F}_t$ that the kite exerts on the tether and the force $\vec{F}_{tg}$ that the ground station exerts on the tether we use the free body diagram illustrated in Fig.~\ref{fig:tethermass}.

Because of its flexibility the tether can support only tensile forces and no bending moment and as consequence the tether force is always locally aligned with the tether, following its curvature.
This holds also for the tether suspension points, as indicated by the corresponding reaction forces included in Fig.~\ref{fig:tethermass}.
The sketch illustrates how the sagging induces the tangential reaction force components $\vec{F}_{t,\tau}$ and $\vec{F}_{tg,\tau}$ which are balancing the resultant tangential component of the gravitational loading. 

The reaction forces are calculated from the force and moment equilibria of the deformed tether.
For small to moderate sagging the centre of gravity of the tether is located halfway between the suspension points in terms of ground plane distances. 
The tether force vector can be resolved in spherical coordinates $(r, \theta, \phi)$ as a function of the tensile force $F_t$ at the kite, the tether mass $m_t$ and its orientation $\theta$
%
%\begin{align}
%   \label{eq:radialtetherforce}
%   F_{tg,r}    & = F_{t,r} - \cos\theta\, m_t g, \\
%   \label{eq:tangentialtetherforce}
%   F_{tg,\tau} & = F_{t,\tau} = \frac{1}{2} \sin\theta\, m_t g, \\
%   \label{eq:tetherforceatkite}
%   F_t         & = \sqrt{F_{t,r}^2 + F_{t,\tau}^2}, \\
%   \label{eq:tetherforceatground}
%   F_{tg}      & = \sqrt{F_{tg,r}^2 + F_{tg,\tau}^2}.
%\end{align}
%
\begin{equation}\label{eq:tether_force}
   \vec{F}_t =
   \begin{bmatrix}
      \sqrt{F^2_t-F^2_{t,\tau}} \\[0.3em]
     F_{t,\tau} \\[0.3em]
      0
   \end{bmatrix}
   =
   \begin{bmatrix}
   \sqrt{F^2_t-\frac{1}{4}\sin^2\theta\, m^2_t g^2} \\[0.3em]
   -\frac{1}{2}\sin\theta\, m_t g \\[0.3em]
   0
   \end{bmatrix}.
\end{equation}
%
%Once the tensile force $F_t$ in the tether is known, Eqs.~(\ref{eq:radialtetherforce}) to (\ref{eq:tetherforceatground}) can be used to calculate the force components and the angular deviations at the suspension points.
%
The tensile force $F_{tg}$ at the ground station can be calculated as
\begin{equation}
%   F_{tg} = \sqrt{\left( \sqrt{F^2_t - \frac{1}{4}\sin^2\theta\, m^2_t g^2} - \cos\theta\, m_t g \right)^2} \\ 
%   \overline{ + \frac{1}{4}\sin^2\theta\, m^2_t g^2 \raisebox{7mm}{}},
    F_{tg} = \sqrt{\left( \sqrt{F^2_t - F^2_{t,\tau}} - \cos\theta\, m_t g \right)^2 + F^2_{t,\tau}}\, ,
\end{equation}
with the sagging-induced tangential force component given by $F_{t,\tau} = \sfrac{1}{2} \sin\theta\, m_t g$.
For strong sagging the ground plane distance between the centre of mass and the kite decreases and as a result $F_{t,\tau}$ increases while $F_{tg,\tau}$ decreases.
If the tether mass is small compared to the tensile force the sagging will be small and the tensile forces at both suspension points will differ very little. 
We can introduce the relative gravitational force
\begin{equation}\label{eq:rel_gravitational_Force}
\gamma = \frac{m_t g}{F_t}
\end{equation}
to quantify the relative importance of gravity.
For small values of $\gamma$ the effect of gravitational forces will only be minor.

\subsection{Analytic Model Including Effect of Gravity}
\label{sec:analytic_model_gravity}
\begin{figure}[h]
	\centering
	\def\svgwidth{250pt}
	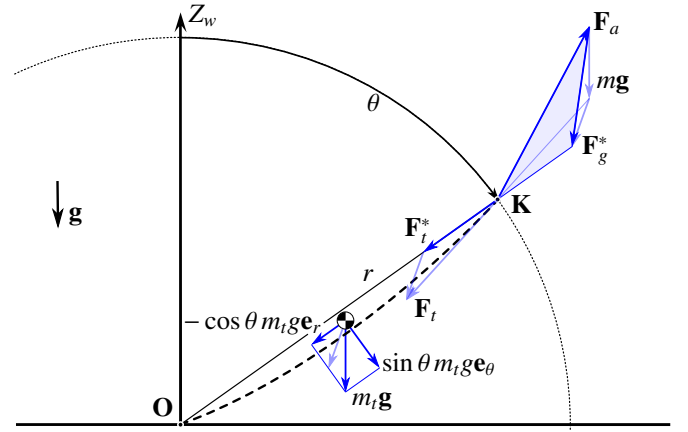
	\caption{Steady force equilibrium of the kite $\vec{K}$ in the $\phi=const.$ plane showing the original force triangle, $\vec{F}_t\vec{F}_am\vec{g}$, and the triangle resulting from the lumped approach, $\vec{F}^*_t\vec{F}_a\vec{F}^*_g$ (shaded in blue).}\label{fig:forcegravity}
\end{figure}

The tether force $\vec{F}_t$ given by Eq.~(\ref{eq:tether_force}) describes the effect of the kite on the tether.
It reversely acts, but with opposite sign, on the kite and implicitly includes the sagging-induced effect of gravity, the tangential force component $F_{t,\tau}$.
The quasi-steady force equilibrium is extended to 
\begin{equation}\label{eq:force_equilibrium_gravity}
    \vec{F}_t + m\vec{g} + \vec{F}_a = \vec{F}^*_t + \vec{F}^*_g + \vec{F}_a = 0,
\end{equation}
with
\begin{equation}\label{eq:force_gravity}
\vec{F}^*_{g} =
\begin{bmatrix}
-\cos\theta \\[0.3em]
\sin\theta \\[0.3em]
0
\end{bmatrix} m g 
+
\begin{bmatrix}
-\cos\theta   \\[0.3em]
\frac{1}{2} \sin\theta \\[0.3em]
0
\end{bmatrix}  m_{t} g ,
\end{equation}
and
\begin{equation}\label{eq:force_tether}
\vec{F}^*_{t} =
\begin{bmatrix}
-\sqrt{F^2_t-\frac{1}{4}\sin^2\theta\, m^2_t g^2} + \cos\theta\, m_t g \\[0.3em]
0  \\[0.3em]
0 
\end{bmatrix} .
\end{equation}
With the starred versions of the forces we have formally removed the gravitational contribution of the tether from the internal structural force $\vec{F}_t$ and lumped it to the gravitational force $m\vec{g}$ of the kite. A similar approach was used with Eq.~(\ref{eq:totaldrag}) to lump the aerodynamic drag of the tether to the drag of the kite.
Regarding Eq.~(\ref{eq:force_gravity}) it should be noted that unlike the contribution of the kite the contribution of the tether is not vertical because of the sagging of the tether and the fact that it is attached to the ground station. The resulting tether force $\vec{F}^*_{t}$ acts in radial direction.
Figure~\ref{fig:forcegravity} illustrates the described lumping approach and the effect on the steady force equilibrium of the kite.

\begin{figure}[h]
	\centering
	\def\svgwidth{250pt}
	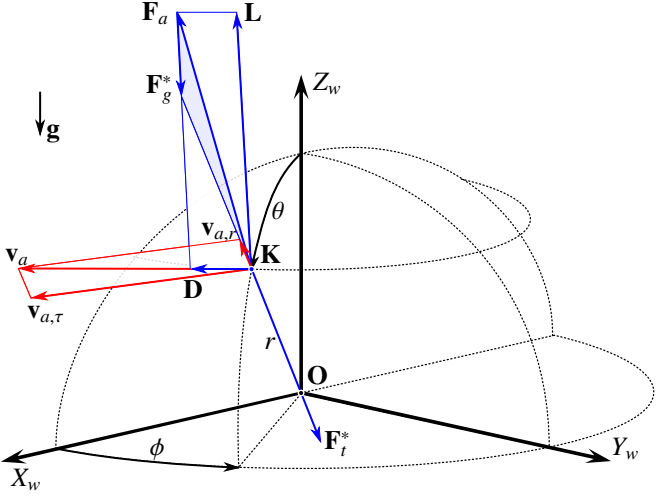
	\caption{Steady force equilibrium considering the effect of gravity. Adapted from \cite{Schmehl2013}.}\label{fig:trianglesimilarityweight}
\end{figure}

The apparent wind velocity and the decomposition of the aerodynamic force into lift and drag components is illustrated in Fig.~\ref{fig:trianglesimilarityweight}.
Because the gravitational force $\vec{F}^*_g$ causes a disalignment of the aerodynamic force $\vec{F}_a$ and the tether force $\vec{F}^*_t$, the geometric similarity of the force and velocity diagrams does not hold anymore. Consequently, the kinematic ratio $\kappa = v_{a,\tau}/v_{a,r}$ can not be expressed by the lift-to-drag ratio $L/D$, as stated by Eq.~(\ref{eq:triangle_similarity}) which is valid for the limiting case of vanishing mass. 
Starting from Eq.~(\ref{eq:v_a3}) the nondimensional apparent wind velocity can be formulated as
\begin{equation}\label{eq:v_a_inertial}
\frac{v_a}{v_w} = (\sin\theta\cos\phi  - f) \sqrt{1 + \left( \frac{v_{a,\tau}}{v_{a,r}}\right) ^2}
\end{equation}
and the tangential kite velocity factor now takes the form
\begin{equation}\label{eq:lambda}
\lambda = a + \sqrt{a^2 + b^2 - 1 + \left(\frac{v_{a,\tau}}{v_{a,r}}\right)^2 \left( b -f \right)^2} 
\end{equation}
with the trigonometric coefficients $a$ and $b$ defined by Eqs.~(\ref{eq:a_trigcoeff}) and (\ref{eq:b_trigcoeff}). 
The magnitude of the resultant aerodynamic force $F_a$ can be formulated by using Eq.~(\ref{eq:v_a_inertial}) 
\begin{equation}\label{eq:F_a_magnitude2}
\frac{F_a}{q S} =  C_R \left[ 1 + \left( \frac{v_{a,\tau}}{v_{a,r}}\right) ^2\right]  (\sin\theta\cos\phi  - f)^2 .
\end{equation}
\subsection{Iterative Solution Procedure}
\label{sec:iterative_solution}
In the following we describe an iterative procedure to solve for the unknown kinematic ratio  $\kappa$.
Maintaining a quasi-steady motion requires a kinematic ratio for which the aerodynamic force balances the tangential components of the gravitational force. This is expressed by
\begin{equation}\label{eq:F_a_theta}
F_{a,\theta} =  - F^*_{g,\theta} = - \left(\frac{1}{2} m_t + m \right) g \sin{\theta} .
\end{equation}  
%%
%\begin{equation}\label{eq:F_a_phi}
%F_{a,\phi} = -F_{i,\phi} .
%\end{equation} 
%%
The radial component is determined by 
\begin{equation}\label{eq:F_a_r}
F_{a,r} =   \sqrt{F^2_a - F^2_{a,\theta}} ,
\end{equation} 
using Eqs.~(\ref{eq:F_a_magnitude2}) and (\ref{eq:F_a_theta}) to resolve the forces on the right hand side.
Finally, the definition of the aerodynamic drag force 
\begin{equation}\label{eq:F_a_projection}
D = \frac{\vec{F}_a \cdot \vec{v}_a}{v_a} 
\end{equation}
is rewritten to obtain the following expression for the lift-to-drag ratio
\begin{equation}\label{eq:L_D_iterate}
\frac{L}{D}  = \sqrt{\left( \frac{F_a v_a}{\vec{F}_a \cdot \vec{v}_a}\right)^2 - 1 } .
\end{equation}
This equation can be employed to iteratively determine the kinematic ratio \cite{Noom2013}.

The process starts with setting the target value $G^*$ to the given lift-to-drag ratio $L/D$ of the kite and setting the initial guess $\kappa_1=G^*$ based on Eq.~(\ref{eq:triangle_similarity}).
The following steps are performed to update $\kappa_i$ in $i=1, \ldots, n$ iterations: First, the spherical components of the apparent wind velocity are computed from Eqs.~(\ref{eq:v_a_inertial}) and (\ref{eq:lambda}), then the respective components of the resultant aerodynamic force from Eqs.~(\ref{eq:F_a_magnitude2}), (\ref{eq:F_a_theta}) and  (\ref{eq:F_a_r}).
Using Eq.~(\ref{eq:L_D_iterate}) the value of the lift-to-drag ratio $G_i$ corresponding to the current value of $\kappa_i$ is computed.
From this we determine the updated value of the kinematic ratio as 
\begin{equation}\label{eq:kappa_iteration}
   \kappa_{i+i} = \kappa_i \sqrt{\frac{G^*}{G_i}}.
\end{equation}
This iteration loop is repeated until the lift-to-drag ratio $G_i$ calculated from Eq.~(\ref{eq:L_D_iterate}) is sufficiently close to the target value $G^*$.

\begin{figure}[h]
	\centering
	\includegraphics[width=\linewidth]{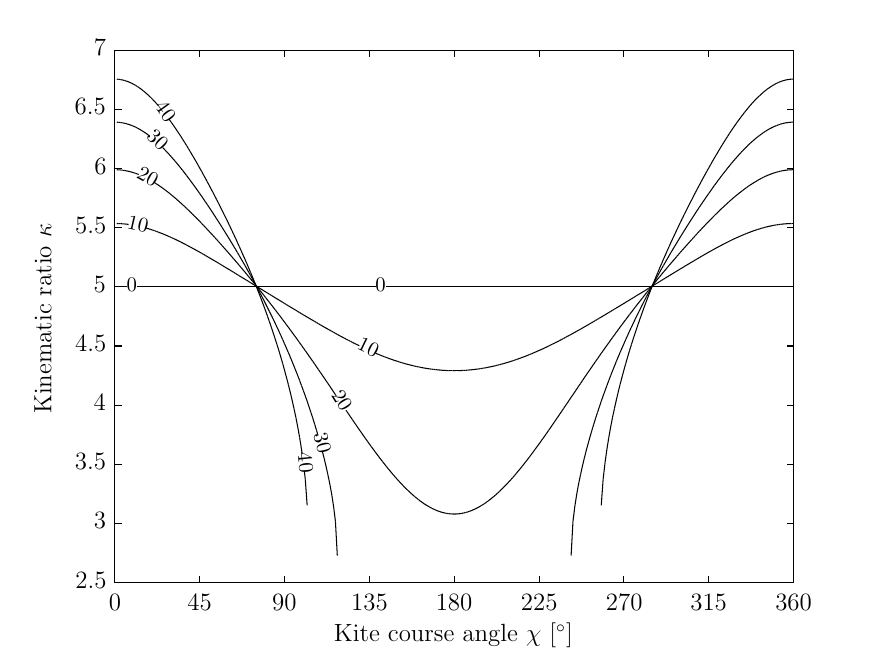}
	\caption{Kite mass $m$ as function of course angle $\chi$  and kinematic ratio $\kappa$ for $\beta=25^\circ$, $f=0.37$, $L/D=5$, $C_L=1$, $S=16.7$ m$^2$, $v_w=7$ m/s and $\rho=1.225$ kg/m$^3$ \cite{Schmehl2013}.}
	\label{fig:kinematiclifttodragratio}
\end{figure}

The effect of gravity on the instantaneous traction power can be significant depending on the kite course angle. Figure~\ref{fig:kinematiclifttodragratio} shows computed isolines of the kite mass as functions of the kite course angle and the kinematic ratio for a representative example.
For horizontal or upward flight ($90^{\circ}\leq\chi\leq 270^{\circ}$) the kinematic ratio is always smaller than the lift-to-drag ratio. The kinematic ratio can become zero when the component of the gravitational force opposing the flight direction is larger than the component of the aerodynamic force in flight direction. In this case the  forces in flight direction can not be in a quasi-steady equilibrium and the algorithm fails to identify a physical solution.
 
For downwards flight ($\chi<90^{\circ}$ or $\chi > 270^{\circ}$)  the kinematic ratio can become larger than the lift-to-drag ratio and increases with increasing mass of the kite. In specific cases, like an exceptionally heavy kite flying vertically downward while reeling out fast, the kinematic ratio starts to approach infinity. Also in this situation the model fails to identify a quasi-steady equilibrium. It is recommended to further investigate this situation in future research. In the current work these extreme situations do not occur and the kinematic ratio does not approach these limits.  

The effect of gravity on the average traction power generation can be significant, as for the upward flying regions where the kinematic ratio becomes smaller, the quasi-steady flight velocity of the wing reduces. This means that the upward flying regions of a closed-loop trajectory require more time than the downward flying regions. As a result the time average course angle can be expected to have an upward component as a result of the mass.

In the following sections we adapt the developed theoretical framework to the specific flight manoeuvres in the different phases of the pumping cycle illustrated in Fig.~\ref{fig:trajectory}. We start the cycle with the retraction phase because at the start of this phase is the only fix point of the trajectory determined by given problem parameters $\beta_o$ and $r_{max}$.

\subsection{Retraction Phase}
\label{sec:retraction}
The objective of the retraction phase is to pull the kite back to the minimum tether length $r_{min}$ at a minimal cost of energy, while ensuring stable flight throughout this manoeuvre. The retraction energy is calculated as the integral of the instantaneous traction power $P_i$ over the retraction time $\Delta t_i$. This is conventionally achieved by reducing the angle of attack of the wing, which reduces the aerodynamic coefficients but not the wing reference area. A more aggressive, but also more risky manoeuvre, such as sideways flagging, substantially decreases also the wing area \cite{Kitegen2009}. Within the scope of the present analysis the aerodynamic force is modified solely by means of the aerodynamic coefficients.

It is assumed that the aerodynamic coefficients $C_{L,i}$ and $C_{D,i}$ are constant during the retraction phase. At the start of the phase the tether is at its maximum length $r_{max}$ and the elevation angle has still the constant value $\beta_o$ of the traction phase. The course angle is set to a constant value of $\chi_i = 180^\circ$ in order to fly in upwards direction during the complete retraction phase.

The trajectory described by the kite is located in the $\phi_i = 0$ plane. 
The position of the kite is updated by a finite difference scheme
\begin{equation}\label{eq:retraction_trajectory}
\vec{r}(t + \Delta t) = \vec{r}(t) + \vec{v}_k(t) \Delta t.
\end{equation}
We define the characteristic time of the traction phase as
\begin{equation}\label{eq:charsystemtime}
t^* = \frac{r_{max} - r_{min}}{v_{w,ref}}
\end{equation}
and use this together with a given nondimensional time step $\Delta T$ to scale the integration time step $\Delta t$ to the physical dimension of the system
\begin{equation}\label{eq:nondimentionaltime}
\Delta t = t^* \Delta T.
\end{equation} 

As the kite describes its path through the retraction phase, a control strategy needs to be defined to determine the reel-out factor $f$. Three principal strategies can be applied: velocity control, force control and power control. We will use a constant force $F_{t,i}$ over the entire retraction phase, because this minimizes the total retraction time for a tether with a given tensile strength, which, in the first instance, also maximizes the net power output of the system.
This requires for each retraction step the solution of Eq.~(\ref{eq:tetherforce}) for the reel-out factor 
\begin{equation}\label{eq:reeloutfactor_fixedforce}
f = \sin\theta\cos\phi \pm \sqrt{\frac{F_{t,i}}{q S C_R \left[ 1 + \left( \frac{L}{D}\right) ^2\right]} } .
\end{equation}
The larger value of $f$ can be excluded because it describes the unphysical case of compressive loading of the tether. Tensile loading requires a radially outward pointing aerodynamic force $\vec{F}_a$ which is linked to a positive value of $v_{a,r}$. According to Eq.~(\ref{eq:v_a3}) this is only possible for $f \leq \sin\theta \cos\phi$ which can only be fulfilled if the root is subtracted from $\sin\theta\cos\phi$.

For a constant and uniform wind velocity $v_w$ and constant reeling factor the kite would asymptotically approach a steady flight state which is characterised by a constant elevation angle $\beta_{i,\infty}$. This radial retraction state is generally not reached before the minimum tether length $r_{min}$ is reached and the retraction phase is terminated at $t_A$.

\subsection{Transition Phase}
\label{sec:transition}

As shown in Fig.~\ref{fig:trajectory}, the retraction phase generally ends at an elevation angle that is substantially larger than the constant elevation angle $\beta_o$ of the traction phase. On the other hand, the tether force $F_{t,i}$ during retraction is much lower than the force $F_{t,o}$ during the traction phase. The objective of the transition phase is to fly the kite back to the lower angle $\beta_o$ and to safely increase the force in the tether to $F_{t,o}$. 

To initiate the transition flight manoeuvre at $t_A$ the aerodynamic coefficients are set to the values $C_{L,o}$ and $C_{D,o}$ of the traction phase, i.e. the kite is powered. At the same time the course angle is set to $\chi=0^\circ$, which means the kite is flying in a downward direction. The control algorithm generally aims to keep the tether at constant length, but takes corrective action to ensure that the tether tension stays within a limited range during the transition phase.

Because the kite can overfly the ground station during the retraction phase the described flight manoeuvre can result in a sudden drop of the tether tension below the required minimum value for the kite to ensure a stable operation. In such situation the tether is reeled in further to restore the minimum tension. For the simulation and the operation of the real system the target force $F_{t,i}$ of the retraction phase is used as lower limit. As consequence, the parameter $r_{min}$ can only be regarded as a target value and the true minimum tether length can be less as a result of the described minimum tension requirement. 

On the other hand, flying to a lower elevation angle into the so-called wind window increases the tether tension which could exceed the value $F_{t,o}$ set for the traction phase. In this situation the reeling velocity is increased to stay below the value $F_{t,o}$.
The transition phase ends when the required elevation angle $\beta_{o}$ for the traction phase is reached. 

\subsection{Traction Phase}
\label{sec:traction}

During the traction phase the kite is operated in crosswind motion to maximise the traction force and thus also the generated mechanical power. A variety of different flight manoeuvres are in use, of which circular and figure eight trajectories are most frequently described in literature. 

Instead of resolving the tangential motion component of the manoeuvre we use a constant representative flight state to describe the average traction force and power of the kite. As consequence the angular coordinates $\beta$ and $\phi$ as well as the course angle $\chi$ have constant values during the traction phase. The proposed approach has the advantage to not only reduce the simulation times substantially but also to keep the model generally applicable for a range of different crosswind manoeuvres.
We hypothesise that this constant representative flight state is best determined as a time average of the real flight state, taking into account that it is the predicted traction power that should match the average traction power of the crosswind manoeuvre.  
The constant representative flight state is a predefined experience-based setting and can be evaluated on the basis of experimental data or by means of a dynamic kite model. 

According to Eq.~(\ref{eq:power}) the traction power depends on the product of $\cos\beta$ and $\cos\phi$ and for this reason the time average of the trigonometric functions is used to define the representative angular positions $\phi_o$ and $\beta_o$ by
\begin{equation}\label{eq:averagepositions}
\cos\phi_o = \overline{\cos\phi} \qquad \text{and} \qquad \cos\beta_o = \overline{\cos\beta}.
\end{equation}
This averaging implies a weighting factor that decreases from $1$ from the centre of the wind window, when the tether is aligned with the $X_w$-axis, to $0$ at the side of the wind window, when the tether is perpendicular to the $X_w$-axis. For a figure eight trajectory the averaging results in $\phi_o$ and $\beta_o$ at the centre of one of the figure eight lobes.
Because the kite flies slower in upward than in downward direction the average course angle $\chi_o$ is expected to be larger than 90$^\circ$. We leave it for further research to find a relation between $\chi_o$ and the mass and aerodynamic properties of the kite.
The traction phase is terminated when the maximum tether length $r_{max}$ is reached. 

\subsection{Complete Pumping Cycle}
\label{sec:cycle}

The mean mechanical power production during one pumping cycle is computed from the mean traction power and time duration of each phase 
%
% \lsymb[f]{$P_m$}{mean mechanical power during a pumping cycle}{W}{sortsymbol}\lsymb[f]{$P_o$}{mean power produced during reel-out phase}{W}{sortsymbol}\lsymb[f]{$P_i$}{mean power consumed during reel-in phase}{W}{sortsymbol}\lsymb[f]{$P_{tr}$}{mean power produced/consumed during transition phase}{W}{sortsymbol}\lsymb[f]{$t_{tr}$}{duration of transition phase}{s}{sortsymbol}\lsymb[f]{$t_{o}$}{duration of reel-out phase}{s}{sortsymbol}\lsymb[f]{$t_{i}$}{duration of reel-in phase}{s}{sortsymbol}
%
\begin{equation}\label{eq:pc_P_general}
P_m = \frac{\overline{P}_o \Delta t_o + \overline{P}_i \Delta t_i + \overline{P}_x \Delta t_x}{\Delta t_o + \Delta t_i + \Delta t_x} ,
\end{equation}
where the indices $o$, $i$ and $x$ denote the reel-out, reel-in and transition phases, respectively.
Using Eq.~(\ref{eq:pc_P_general}), an average power harvesting factor 
\begin{equation}\label{eq:meanharvestfact}
\zeta_m = \frac{P_m}{P_w S} , 
\end{equation}
can be defined for the complete pumping cycle. To account for the varying atmospheric conditions along the cycle trajectory, the wind power density is evaluated at an average  traction altitude
\begin{equation}\label{eq:meantractionaltitude}
z_{mt} = \frac{1}{2} \cos\theta\, (r_{min}+r_{max}). 
\end{equation}
The equivalent for a horizontal axis wind turbine would be the hub height.

%%%%%%%%%%%%%%%%%%%%%%%%%%%%%%%%%%%%%%%%%%%%%%%%%%%%%%%%%%%%%%%%%%%%%%%%%%%%%%%%%
\section{Experimental Setup}
\label{sec:experimentalsetup}
%%%%%%%%%%%%%%%%%%%%%%%%%%%%%%%%%%%%%%%%%%%%%%%%%%%%%%%%%%%%%%%%%%%%%%%%%%%%%%%%%
%
The quality of the presented quasi-steady model is assessed on the basis of measurement data retrieved from comprehensive tests of a pumping kite power system. In this section we outline the key features of the technology demonstrator and select two specific representative test cases for comparison.
Because the aerodynamic characteristics of the kite in the different phases of the cycle have a decisive influence on the computed power output particular attention is devoted to this subject. 

The common approach to determine the aerodynamic characteristics of rigid wings under controlled conditions are wind tunnel measurements of scaled models or computational fluid dynamics. Although these techniques have been applied to tethered flexible membrane wings \cite{Wachter2008,Bungart2009} the practical usability of the data is limited. On the one hand, windtunnel measurements are costly and because of the strong fluid-structure coupling the aero-elastic behaviour of scale models can generally not be extrapolated to the size of the real system. On the other hand, reasonably accurate aerodynamic simulations of deforming membrane wings are still a major challenge for currently available computational methods.

Tow testing of kites has developed as an interesting alternative to determine the aerodynamic performance of kites \cite{Wood2017}. Although cost-effective, this technique also imposes a clear limit on the wing size. 

We describe a procedure for estimating the aerodynamic properties directly from available flight data. This approach has the advantage that the aerodynamic loading and structural​ deformation of the wing during the specific flight manoeuvres of the pumping cycle is taken into account.

\subsection{Technology Demonstrator}
\label{sec:technology_demonstrator}

\begin{figure}[h]
	\centering
	\includegraphics[height=3.295cm]{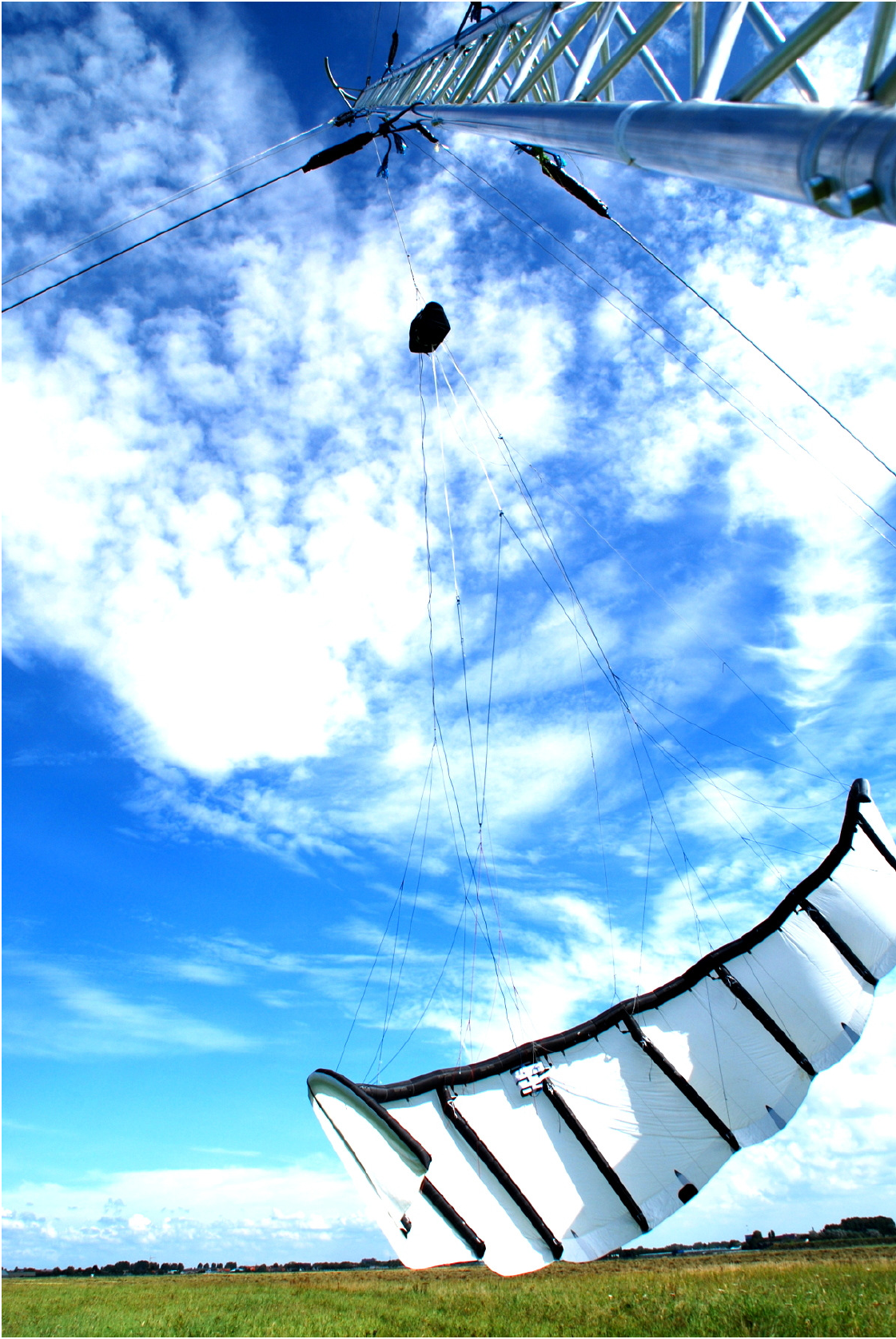} \hspace{-6pt}
	\includegraphics[height=3.295cm]{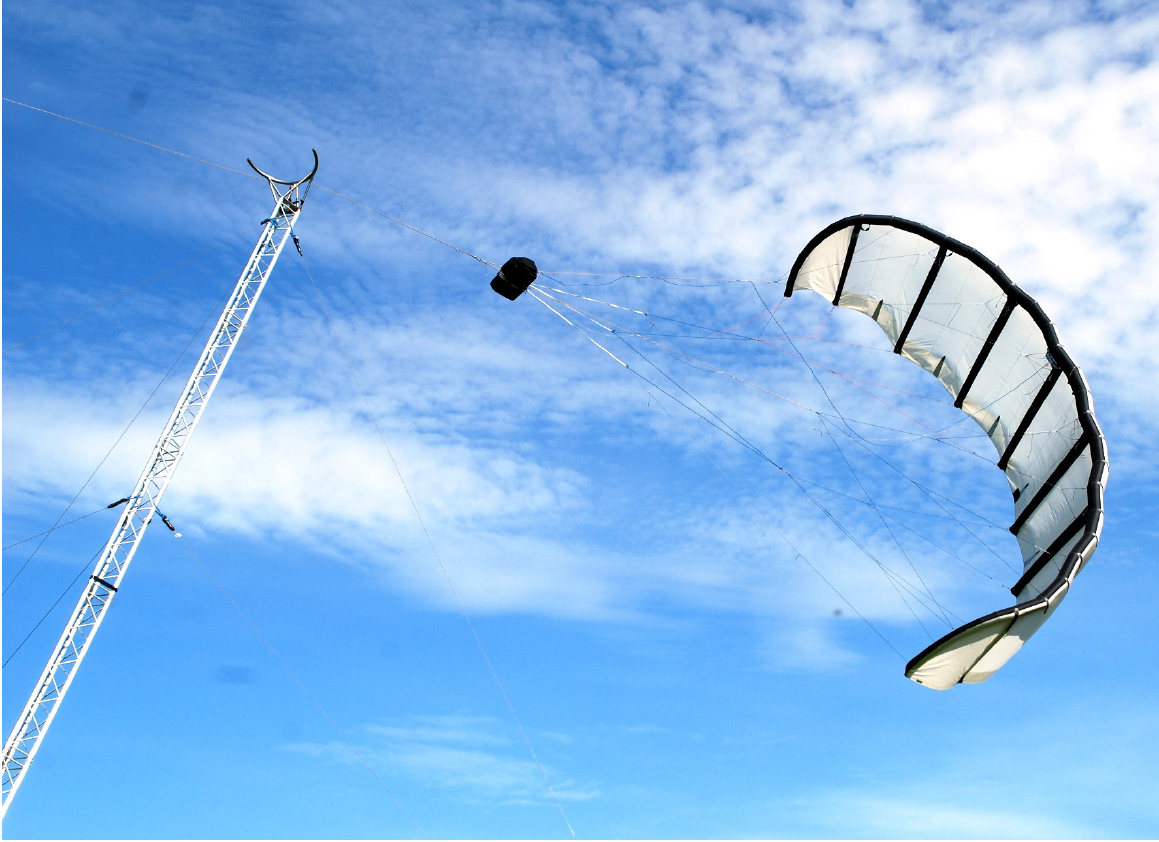} \hspace{-6pt}
	\includegraphics[height=3.295cm]{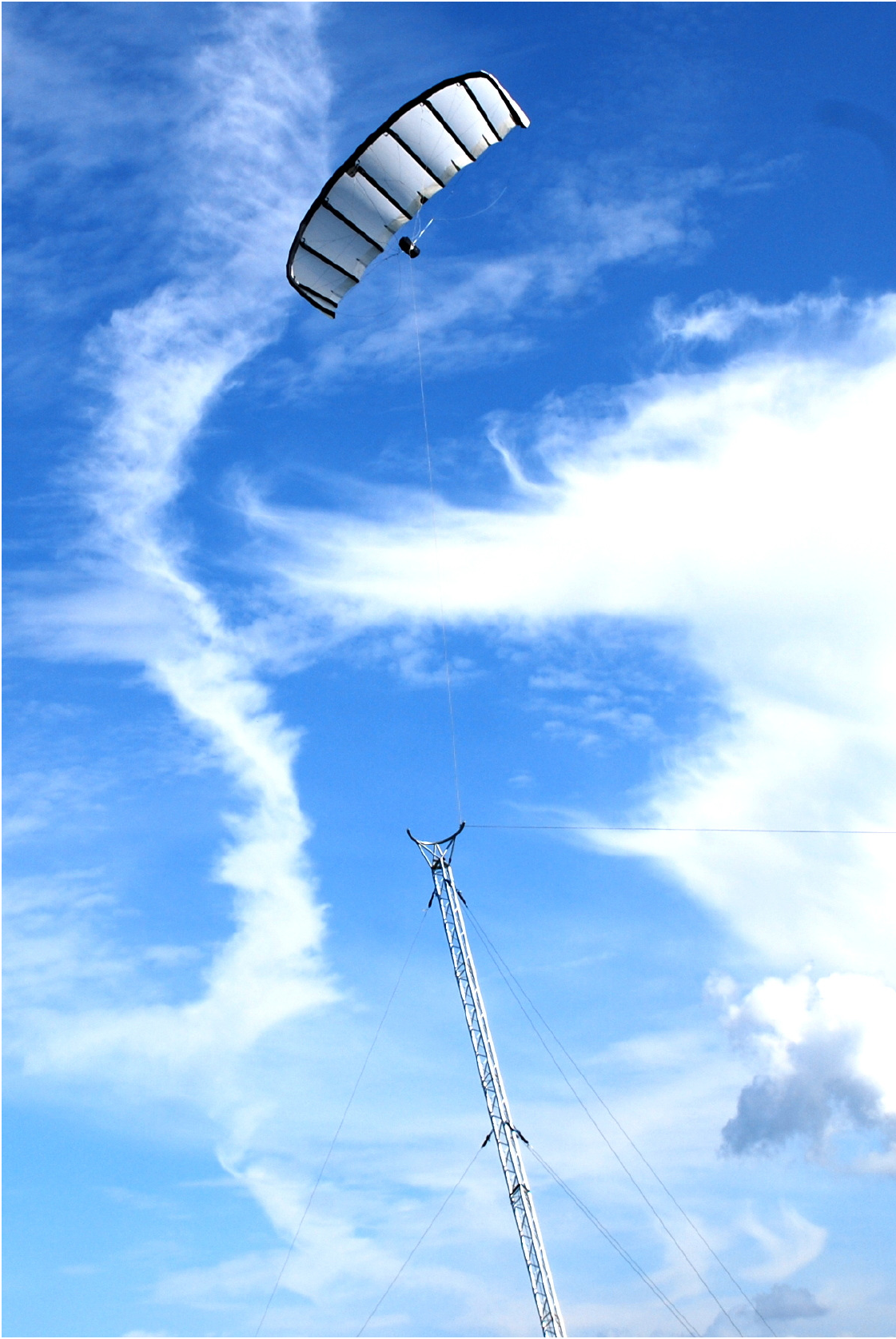}
	\caption{Experimental kite launch from upside-down hanging position on 23 August 2012.}
	\label{fig:kitelaunchsequence}
\end{figure}

The 20 kW technology demonstrator employed for the present study is in periodical test operation since January 2010. 
\begin{figure}[h]
\centering
\scriptsize
\def\svgwidth{\linewidth}
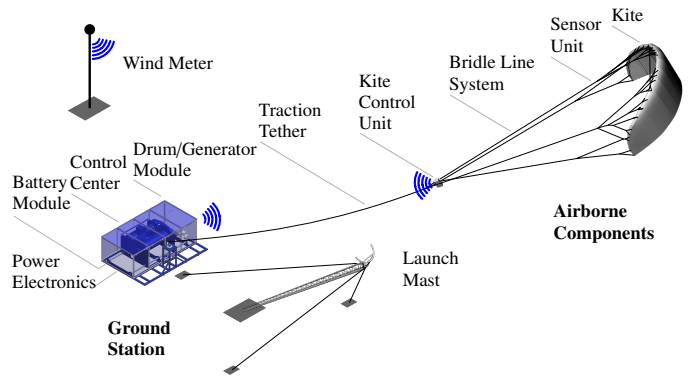
\caption{Kite power system with optional launch mast. Adapted from \cite{VanDerVlugt2013}.}
\label{fig:systemcomponents}
\end{figure}
Figure~\ref{fig:systemcomponents} shows an overview of the system and its major components. 
A detailed description of the hard- and software components, the installed measurement equipment and statistical performance data is provided in \cite{VanDerVlugt2013}.
The retrofitted experimental launch setup is described in \cite{Haug2012,Schmehl2014}.
A photographic sequence of the launch procedure is shown in Fig.~\ref{fig:kitelaunchsequence} with video footage available from \cite{Schmehl2014b}.
Starting in 2016, the spin-off company Kitepower B.V. is developing a commercial 100 kW version of the technology demonstrator \cite{TUDelftTV2018,KitepowerBV2018}.

\subsection{Selected Test Cases}
\label{sec:test_cases}

\begin{figure}[h]
   \centering
   \includegraphics[width=\linewidth]{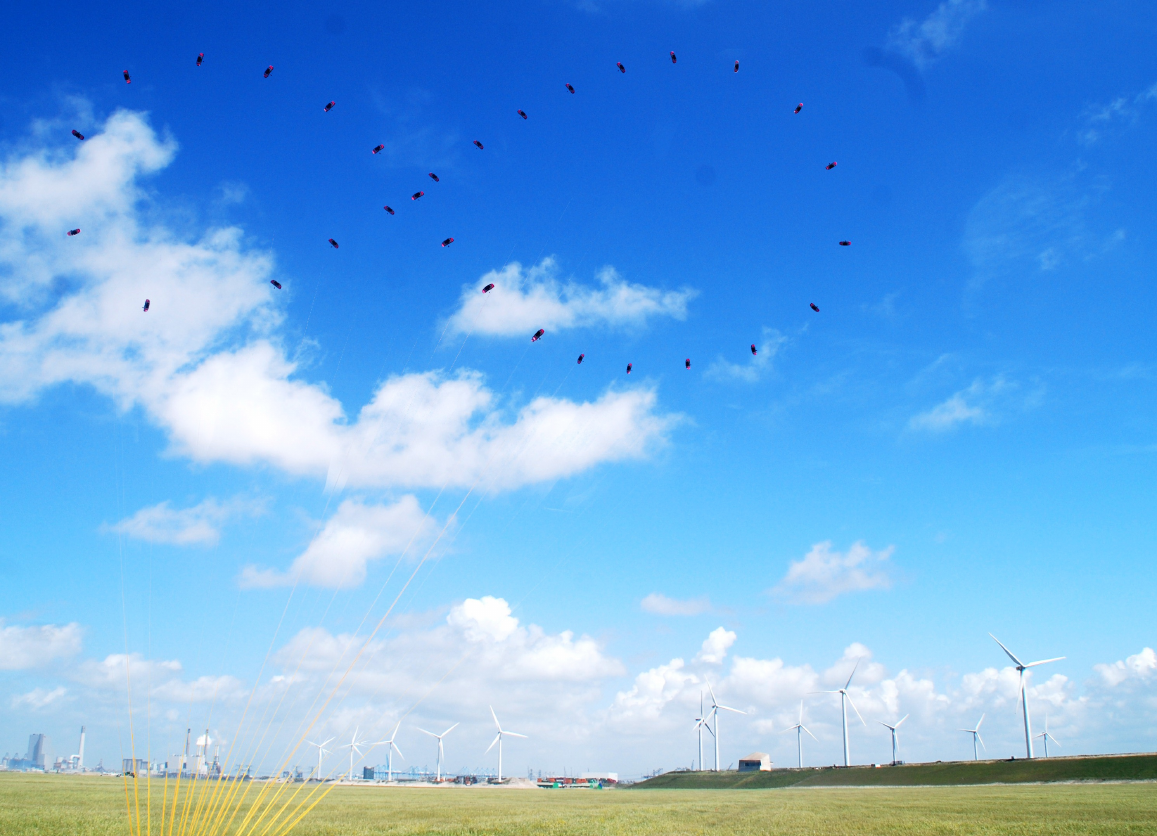}
   \caption{Composite photo of the 14 m$^2$ kite flying a figure of eight manoeuvre ($\Delta t=1 s$) on 23 June 2012 at the Maasvlakte 2 of Rotterdam Harbour \cite{Fechner2013a}.}
   \label{fig:cycle33sec}
\end{figure}

Two different test cases have been selected to assess the quality of the derived modelling framework.
Firstly, for the strong wind analysis a single representative cycle was selected randomly from a dataset recorded on 23 June 2012. The experiment was performed on the Maasvlakte 2 of the Rotterdam Harbour in The Netherlands, on an open field near the beach (see \cite{Kitepower2012a} and Fig.~\ref{fig:cycle33sec}). 
The test conditions were favourable with an undisturbed wind approaching from the sea at an average velocity of 9.9 m/s. For this test a reinforced production kite was used, a Genetrix Hydra 14 m$^2$ with a projected surface area of $S=10.2$ m$^2$ modified to withstand the high wing loading occurring in this experiment. 
The mass of the kite is 5 kg and the control unit including the used sensor unit has a mass of 10 kg such that the total mass of the airborne system components is set to $m=15$ kg.

Secondly, the presented moderate wind data was obtained at Valkenburg, a former military airfield in The Netherlands located at 3 km inland. A steady 5.9 m/s north-eastern wind, blowing parallel to the coastline on 7 May 2013, provided good testing conditions. During this test a scaled up and redesigned version of the Genetrix Hydra with a wing surface area of $A=25$ m$^2$, a projected area of $S=18.6$ m$^2$ and a mass of $m=10.6$ kg was used. This kite is shown in Figs.~\ref{fig:sagging} and \ref{fig:kitelaunchsequence}.

\subsection{Resultant Aerodynamic Coefficient}
\label{sec:kite_res_aero_coeff}

To estimate the aerodynamic force coefficient $C_R$ of the kite from available experimental data we start with the tether force $F_{tg}$ measured at the ground station.  This value is then used to derive the aerodynamic force components at the kite in radial and tangential directions, $F_{a,r}$ and $F_{a,\tau}$, respectively.
The radial component is calculated from the radial force equilibrium of the tether illustrated in Fig.~\ref{fig:tethermass} as
\begin{equation}\label{eq:aeroforcer_fromexp}
	F_{a,r} = \sqrt{F^2_{tg}-\frac{1}{4}\sin^2\theta\, m^2_t g^2} + \cos\theta \left(m_t + m \right) g ,
\end{equation}
assuming that the tether is only moderately sagging. Combining this with the tangential component defined by Eq.~(\ref{eq:F_a_theta}) we can compute the total aerodynamic force according to
\begin{equation}\label{eq:aeroforce_fromexp}
	F_a = \sqrt{F_{a,r}^2 + F_{a,\tau}^2} 
\end{equation}
as a function of system parameters and the measured tether force at the ground.

Next to the aerodynamic force the estimation process also requires information about the apparent wind velocity. To measure $v_a$ directly some flights of the test campaign were equipped with a Pitot tube mounted in the bridle line system between the wing and the kite control unit. However, the quality of this data was insufficient and for this reason we resorted to use Eq.~(\ref{eq:Vapp_definition}) to determine $\vec{v}_a$ as difference of the wind velocity vector $\vec{v}_w$ at the kite position and the kite velocity vector $\vec{v}_k$. To determine $\vec{v}_w$ the wind speed $v_{w,ref}$ measured at the reference altitude was extrapolated to the kite position using Eq.~(\ref{eq:windshear}), whereas $\vec{v}_k$ was determined using the GPS sensor attached to the kite.

The resultant aerodynamic force coefficient can then be derived from Eq.~(\ref{eq:tetherforce_a}) as
\begin{equation}\label{eq:cr_fromexp}
C_R = \frac{2 F_a}{\rho v_{a}^2 S} ,
\end{equation}
which implicitly contains the drag contribution of the tether. 
Because this is small it has not been taken into account.

\begin{figure}[h!]
    \centering
    \fontsize{6.5pt}{10pt}\selectfont %matches font size in PDF
	\def\svgwidth{224pt}
	\vspace{-5pt}
	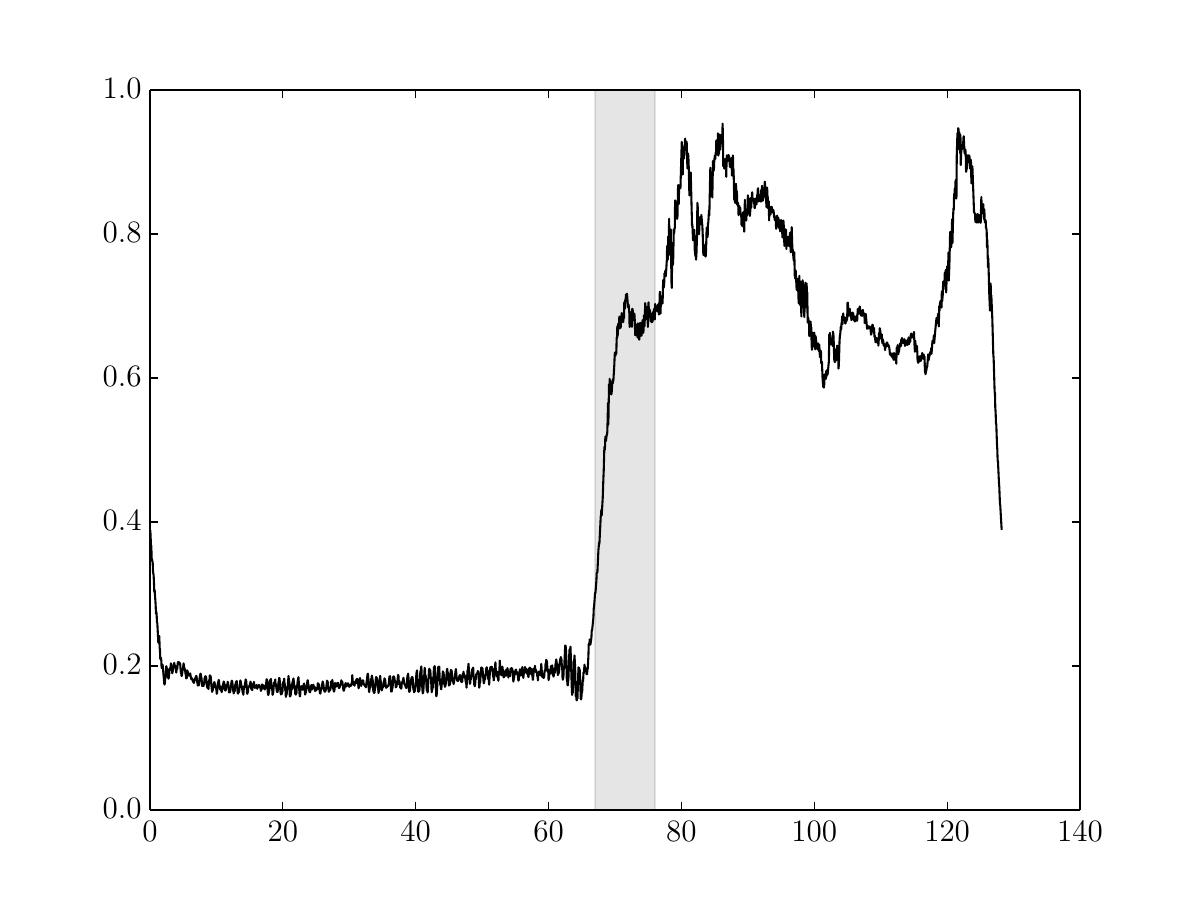 \\[-22pt]
	\def\svgwidth{224pt}
	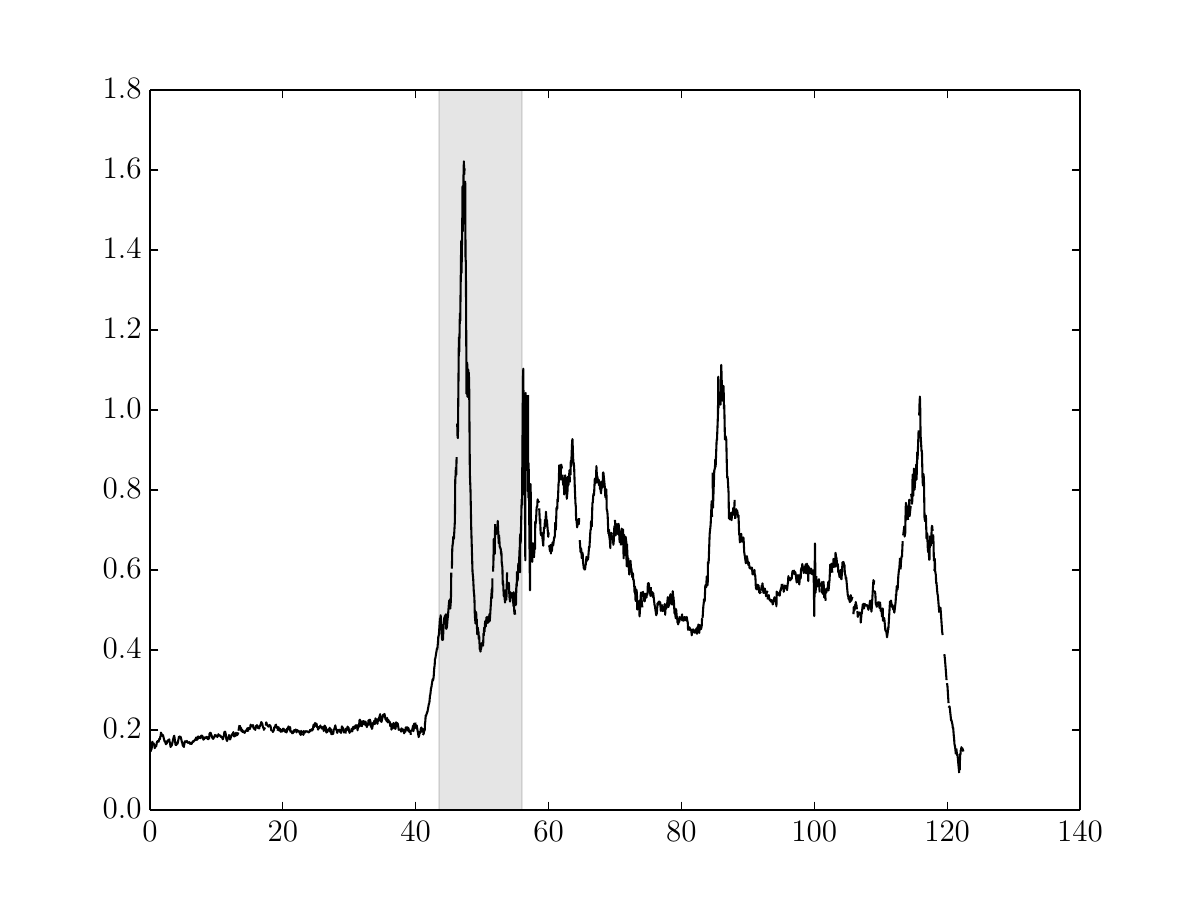 \\[-5pt]
    \caption{Estimated resultant force coefficient $C_R$ over a full pumping cycle. The retraction phase starts at $t = 0$, the grey regions indicate the transition phases and the cycle is completed with the end of the traction phase.}
    \label{fig:cr}
\end{figure}

Figure~\ref{fig:cr} shows the variation of the resultant force coefficient $C_R$ over representative pumping cycles. During the retraction phase $C_R$ is low and varies only within a narrow band, while during the traction phase the value is three to four times higher, showing also substantially larger variations. These variations can be explained as follows. 

Firstly, we have shown in Sect.~\ref{sec:relative_flow} that the angle of attack is constant along the flight path of an idealised massless kite. However, the effect of gravity on a real kite induces variations of the angle of attack which in turn lead to variations of $C_R$. Secondly, by extrapolating wind data that is measured at ground level it is not possible to account for local wind gusts and leads to over- or underestimation of the instantaneous value of $C_R$. We account for this effect by determining $C_{R,i}$ and $C_{R,o}$ as time averages over the retraction and traction phases, respectively, as specified in Table~\ref{tab:parameters}.

%%
%\subsubsection{Method 1}
%%
%Using the work of \citet{noom2013} we can find the lift-to-drag ratio $L/D$. First the aerodynamic force has to be dissected into its components. In the global RF $(xw, yw, zw)$, the vertical component of $F_a$ $F_{az}$ is defined as follows:
%%
%\begin{equation} 
%	F_{az} = F_{ar} \sin{\beta} - F_{at} \cos{\beta}
%\end{equation}
%%
%The horizontal projection of $F_a$ is then defined as:
%%
%\begin{equation}
%	F_{a,hor} = \sqrt{F_a^2 - F_{a,z}^2}
%\end{equation}
%%
%The horizontal distance $r_{hor}$ is defined as:
%%
%\begin{equation}
%	r_{hor} = \sqrt{x^2 + y^2}
%\end{equation}
%%
%With these distances it is now possible to compute the separate components of $\vec{F_a}$:
%%
%\begin{align}
%	F_a,x &= F_{a,hor} \frac{r_\mr{x}}{r_\mr{hor}}\\
%	F_a,y &= F_{a,hor} \frac{r_\mr{y}}{r_\mr{hor}}
%\end{align}
%%
%\citet[p.38]{noom2013} then supplies us with the lift-to-drag ratio:
%%
%\begin{equation}
%	L/D = \sqrt{ \frac{\Fa \Vapp}{\vFa \cdot \vVapp}^2 - 1}
%\end{equation}
%%
%\subsubsection{Method 2}
%%

%%
%%
\subsection{Lift-to-Drag Ratio}
\label{sec:kite_lift-to-drag-ratio}

\begin{figure}[h]
	\centering
	\def\svgwidth{250pt}
	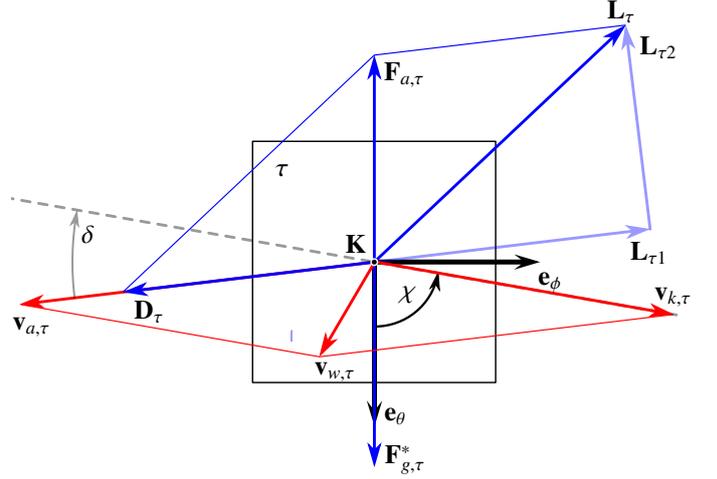
	\caption{Tangential velocity and force components acting on the kite. The placement of the local tangential plane $\tau$ is shown in Fig.~\ref{fig:coordinatesystem}. The tangential flight direction is given by the vector $\vec{v}_{k,\tau}$ and indicated by the dashed line.}\label{fig:tangential_equilibrium}
\end{figure}

To estimate the lift-to-drag ratio $L/D$ of the kite we analyse the forces in the tangential plane. A similar, but more simplified approach has been proposed in \cite{Argatov2009}. Starting point is the quasi-steady force equilibrium given by Eq.~(\ref{eq:force_equilibrium_gravity}). Noting that the tether force $\vec{F}^*_t$ defined by Eq.~(\ref{eq:force_tether}) has only a radial component the equilibrium in the tangential plane $\tau$ reduces to
\begin{equation}\label{eq:force_equilibrium_tangential}
    \vec{F}^*_{g,\tau} + \vec{F}_{a,\tau} = \vec{F}^*_{g,\tau} + \vec{D}_{\tau} + \vec{L}_{\tau} = 0 ,
\end{equation}
which is illustrated in Fig.~\ref{fig:tangential_equilibrium} together with the tangential velocity components.
For two specific flight modes the tangential force equilibrium can be reduced to a scalar equation relating force contributions in the tangential flight direction.

In the traction phase the kite is operated in crosswind manoeuvres. To generate a high tether force the kite needs to fly substantially faster than the wind speed ($v_k \gg v_w$), which is the case for a high lift-to-drag ratio ($L/D \gg 1$). This is quantitatively described by the tangential velocity factor defined by Eq.~(\ref{eq:tangentialkitevelocityfactor}). As consequence, the alignment of the velocity components $\vec{v}_{k,\tau}$ and $\vec{v}_{a,\tau}=\vec{v}_{w,\tau}-\vec{v}_{k,\tau}$ increases with the flight speed, the angle $\delta$ shown in Fig.~\ref{fig:tangential_equilibrium} decreases until it practically vanishes for $L/D \gg 1$.
For this limiting case we consider the tangential force equilibria in flight direction $\tau_1$ and in orthogonal direction $\tau_2$
\begin{align}
   \label{eq:tau1_force_balance}
   L_{\tau1} & + g \left(\frac{1}{2} m_t + m \right) \sin\theta \cos\chi - D_{\tau} = 0, \\
   \label{eq:tau2_force_balance}
   L_{\tau2} & - g \left(\frac{1}{2} m_t + m \right) \sin\theta \sin\chi = 0.
\end{align}
The gravitational contributions are orthogonal projections of $\vec{F}^*_{g,\theta}$ defined by Eq.~(\ref{eq:force_gravity}) onto the $\tau_1$- and $\tau_2$-directions, using the course angle $\chi$.

Because for fast crosswind manoeuvres the lift force $L$ is by far larger than the gravitational force $F^*_{g}$ we can conclude from Eq.~(\ref{eq:tau2_force_balance}) that $L \gg L_{\tau2}$ and accordingly also $\vec{L}\approx\vec{L}_r+\vec{L}_{\tau1}$. To determine $L_{\tau1}$ and $D_{\tau1}$ in Eq.~(\ref{eq:tau1_force_balance}) we orthogonally project $\vec{L}$ and $\vec{D}$ onto the tangential plane using Eq.~(\ref{eq:triangle_similarity}) and following the illustration in Fig.~\ref{fig:velocity_force_diagram}. This projection is possible because for fast crosswind manoeuvres the deviation of the resultant aerodynamic force $\vec{F}_a$ from the radial direction $\vec{e}_r$ can be neglected.
The resulting force equilibrium in $\tau_1$-direction is as follows
\begin{equation}\label{eq:tangential_force_balance}
    \frac{L}{\sqrt{1+\kappa^2}} + g \left(\frac{1}{2} m_t + m \right) \sin\theta \cos\chi - \frac{\kappa D}{\sqrt{1+\kappa^2}} = 0.
\end{equation}
We use Eq.~(\ref{eq:v_a_inertial}) to determine the kinematic ratio from measured data 
\begin{equation}\label{eq:v_a_inertial_solved}
\kappa = \sqrt{ \left( \frac{v_{a}}{v_w (\sin\theta\cos\phi  - f)} \right)^2 - 1}.
\end{equation}

In the retraction phase the kite moves in the $\phi=0$ plane with a course angle of $\chi=180^{\circ}$. Accordingly, the force components $\vec{F}^*_{g,\tau}$, $\vec{L}_{\tau}$ and $\vec{D}_{\tau}$ are all aligned with $\vec{v}_{k,\tau}$. Similar to fast crosswind flight we can use Eqs.~(\ref{eq:v_a_inertial_solved}) and (\ref{eq:tangential_force_balance}) to estimate the lift-to-drag ratio.

Starting from an initial estimate $G_1=\kappa$, which is based on Eq.~(\ref{eq:triangle_similarity}), the lift-to-drag ratio $G$ is determined iteratively using the following equation 
\begin{equation}\label{eq:tangential_force_balance_iterate}
	G_{i+1}= \kappa - \sqrt{1 + \kappa^2} g \left(\frac{1}{2} m_t + m \right) \sin{\theta} \cos\chi 
		\frac{\sqrt{1 + G_{i}^2}}{F_a} ,
\end{equation}
with $i=1, \dots , n$ and $F_{a}$ calculated from Eqs.~(\ref{eq:F_a_theta}), (\ref{eq:aeroforcer_fromexp}) and (\ref{eq:aeroforce_fromexp}) as a function of system parameters and the tether force at the ground station. Equation~(\ref{eq:tangential_force_balance_iterate}) is derived from Eq.~(\ref{eq:tangential_force_balance}) by solving for $L/D$ and substituting the remaining drag force $D$ by $F_a/\sqrt{1+G^2}$. Because the lift-to-drag ratio does not significantly change anymore after two iterations we use $L/D=G_3$ as solution.
%
%\begin{equation}
%	L/D = \left(\kr - \sqrt{1 + kr^2} \cos{\beta} \cos{\theta} 9.81 \left(\mk + \mkcu \right) \right)
%		\frac{\sqrt{1 + (L/D)^2}}{\Ft}
%\end{equation}
%

This estimate still includes the effect of tether drag $D_t$ according to Eqs.~(\ref{eq:totaldrag}) and (\ref{eq:tetherdrag}). To eliminate this we first recalculate the total aerodynamic drag 
\begin{equation}
	D = \frac{\Fa}{\sqrt{1 + \left(\frac{L}{D}\right)^2}}
\end{equation}
and from this determine the lift-to-drag ratio of the kite without the tether

\begin{equation}
    \frac{L}{D_k}  = \frac{L}{D} ~ \frac{D}{D - D_t}.
\end{equation}

\begin{figure}[h]
    \centering
    \fontsize{6.5pt}{10pt}\selectfont %matches font size in PDF
    \def\svgwidth{224pt}
	\vspace{-5pt}
    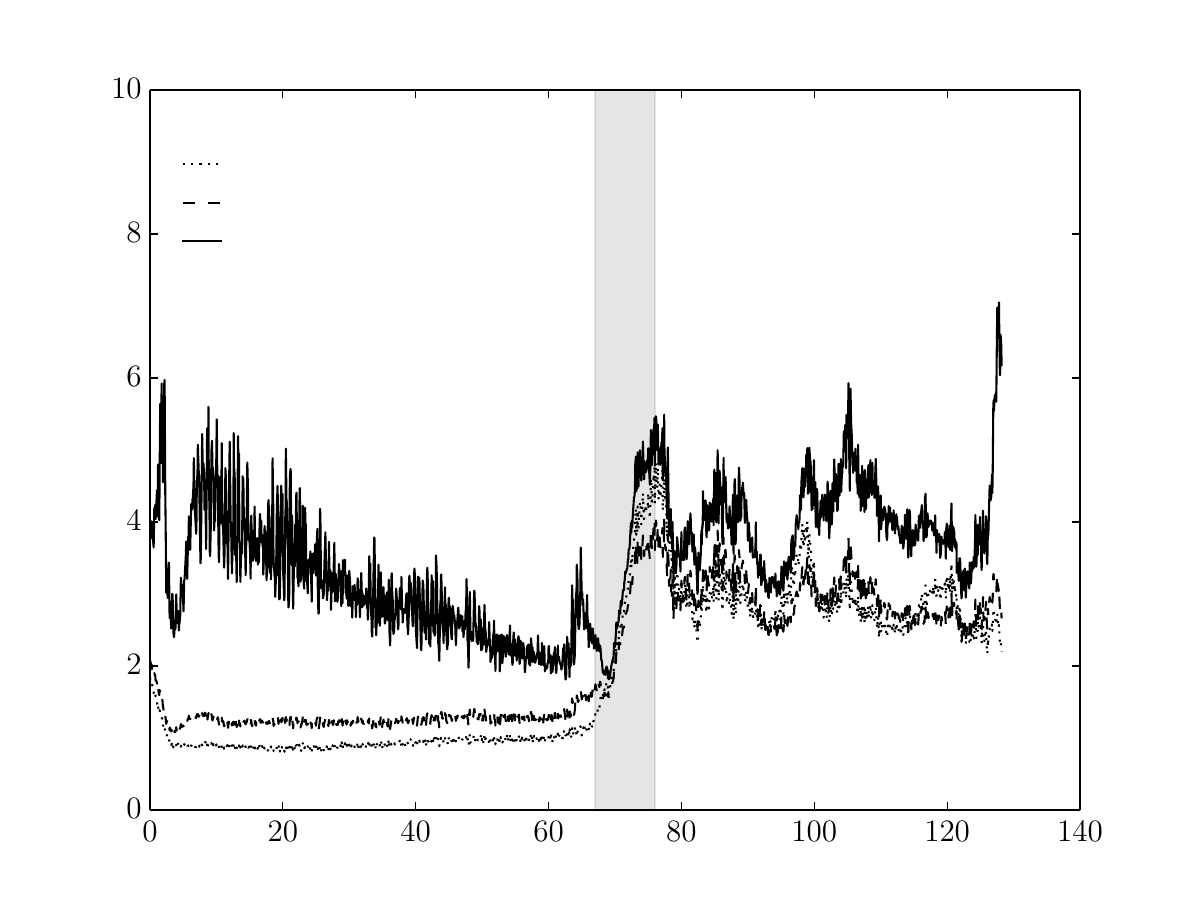 \\[-22pt]
    \def\svgwidth{224pt}
    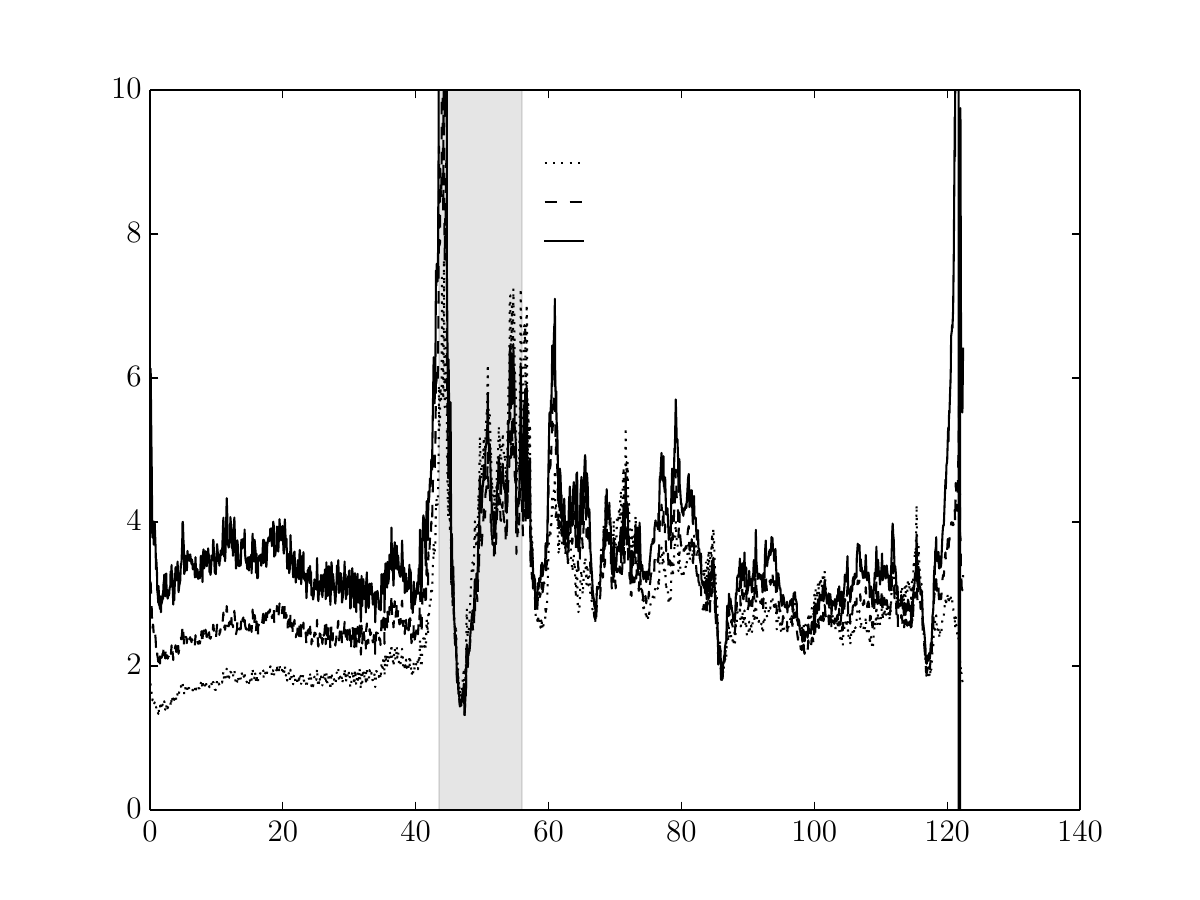 \\[-5pt]
    \caption{Estimated lift-to-drag ratio over a full pumping cycle. The grey regions indicate the transition phases.}
    \label{fig:loverd_HYDRA}
\end{figure}

Figure~\ref{fig:loverd_HYDRA} shows how the lift-to-drag ratio at the different stages of the described estimation process varies over a representative pumping cycle. Time averages for $L/D_k$ can be determined for each phase of the cycle, as specified in Table~\ref{tab:parameters}.
It is important to note that the estimation quality crucially depends on the accuracy at which the wind velocity at the altitude and time of flight can be determined. 
%The available experimental setup only allowed to measure the wind velocity at 6 m high, this data was extrapolated with the use of Eq.~(\ref{eq:windshear}). 
%
%
% How to fix text/math in Rolf's PDF diagrams
% - Load a copy of original PDF of diagram in Inkscape
% - Replace or modify text elements that should be overloaded  in Latex
% - Export PDF+Latex from Inkscape
% - Leave in pdf_tex file only those text elements that should be overloaded in Latex, i.e. remove all others.
% - remove the replaced text elements in original PDF of diagram and load this via \includegraphics
%

%%%%%%%%%%%%%%%%%%%%%%%%%%%%%%%%%%%%%%%%%%%%%%%%%%%%%%%%%%%%%%%%%
\section{Results and Validation}
\label{sec:results}
%%%%%%%%%%%%%%%%%%%%%%%%%%%%%%%%%%%%%%%%%%%%%%%%%%%%%%%%%%%%%%%%%

The presented modelling framework is suitable to derive a fast estimate of the system performance. Optionally, the mass of the kite and tether can be taken into account at the the expense of extra calculation time required to iteratively determine the force equilibrium. In this section we compare simulation results and measured data for two representative test cases, one for moderate and one for strong wind speed. The pumping cycles are calculated on the basis of simulation parameters that are as close as possible to the conditions of the experiment. 
The comparison is based on kite position and velocity, tether tension and generated mechanical power.

\begin{table}[h]
\setlength{\tabcolsep}{0pt}
%\begin{tabular}{lrr} 
\begin{tabular}{@{}p{0.6\linewidth}>{\raggedleft}p{0.2\linewidth}>{\raggedleft}p{0.2\linewidth}}
\toprule
Environmental parameters & \tabularnewline
\midrule
Wind condition & moderate & strong \tabularnewline 
Reference wind speed $v_{w,ref}$ & 5.9 m/s & 9.9 m/s \tabularnewline
Reference height $h_{ref}$ & 6 m & 6 m\tabularnewline
Roughness length $z_0$ & 0.07 m & 0.07 m \tabularnewline
Average traction altitude $z_{mt}$ & 139 m & 252 m \tabularnewline
Wind speed at $z_{mt}$ & 10.1 m/s & 18.2 m/s  \tabularnewline
\toprule
Operational parameters  &  \tabularnewline 
\midrule
Reel-out azimuth angle $\phi_{o}$ & 10.6$^\circ$ & 10.5$^\circ$\tabularnewline
Reel-out elevation angle $\beta_{o}$ & 26.6$^\circ$ & 27.0$^\circ$\tabularnewline
Reel-out course angle $\chi_{o}$ & 96.4 $^\circ$ & 100.9$^\circ$\tabularnewline
Min. tether length $r_{min}$ & 234 m & 390 m\tabularnewline
Max. tether length $r_{max}$& 385 m & 720 m\tabularnewline
Reel-out tether force $F_{t,o}$ & 3069 N & 3008 N \tabularnewline 
Reel-in tether force $F_{t,i}$ & 750 N & 749 N \tabularnewline
\toprule
Kite and tether parameters & \tabularnewline
\midrule
Kite surface area $A$ & 25 m$^2$ & 14 m$^2$ \tabularnewline 
Projected kite area $S$ & 19.8 m$^2$ & 10.2 m$^2$ \tabularnewline 
Mass kite incl. control unit $m$ & 19.6 kg & 15.0 kg \tabularnewline
Traction phase $L/D_k$ & 3.6 & 4.0 \tabularnewline 
Retraction phase $L/D_k$   & 3.5 & 3.1 \tabularnewline
Traction phase res. coefficient $C_{R,o}$ & 0.61 & 0.71 \tabularnewline
Retraction phase res. coefficient $C_{R,i}$ & 0.20 & 0.18 \tabularnewline
Traction phase lift coefficient $C_{L,o}$ & 0.59 & 0.69 \tabularnewline
Retraction phase lift coefficient $C_{L,i}$ & 0.15 & 0.17 \tabularnewline
Tether drag coefficient $C_{D,t}$ & 1.1 & 1.1 \tabularnewline
Tether diameter $d_t$ & 4 mm & 4 mm\tabularnewline
Tether density $\rho_t$ & 724 kg/m$^3$ & 724 kg/m$^3$\tabularnewline
\toprule  
Simulation parameters  & &   \tabularnewline 
\midrule
Nondimensional time step $\Delta T$ & 0.01 & 0.01 \tabularnewline
\bottomrule 
\end{tabular}
\caption{Model input parameters, representative for the two experimental datasets.}
\label{tab:parameters}
\end{table}

Table~\ref{tab:parameters} shows the modelling parameters that are used for this comparison.
The required temporal discretisation of the cycle by means of a nondimensional time step $\Delta T$ is determined in Sect.~\ref{sec:convergence}. The angular positions $\phi_o$ and $\beta_o$ and the course angle  $\chi_o$ during the traction phase are determined by time averaging the data as explained in Sect.~\ref{sec:traction}. The minimum and maximum tether lengths are also determined from the data. Both, in the experiment and in the model, the tether force during the traction phase was controlled to a set value of $F_{t,o}=3000$ N and during the retraction phase to a set value of $F_{t,i}=750$ N.
The reference wind speed is measured at an altitude of 6 m above the ground. The surface roughness length is estimated to be 0.07 m. The material density of the tether listed in Table~\ref{tab:parameters} is lower than the material density of Dyneema\regtm{} as a result of the braiding process.  

\subsection{Convergence Study}
\label{sec:convergence}

\begin{figure}[h]
	\centering
	\includegraphics[width=224pt]{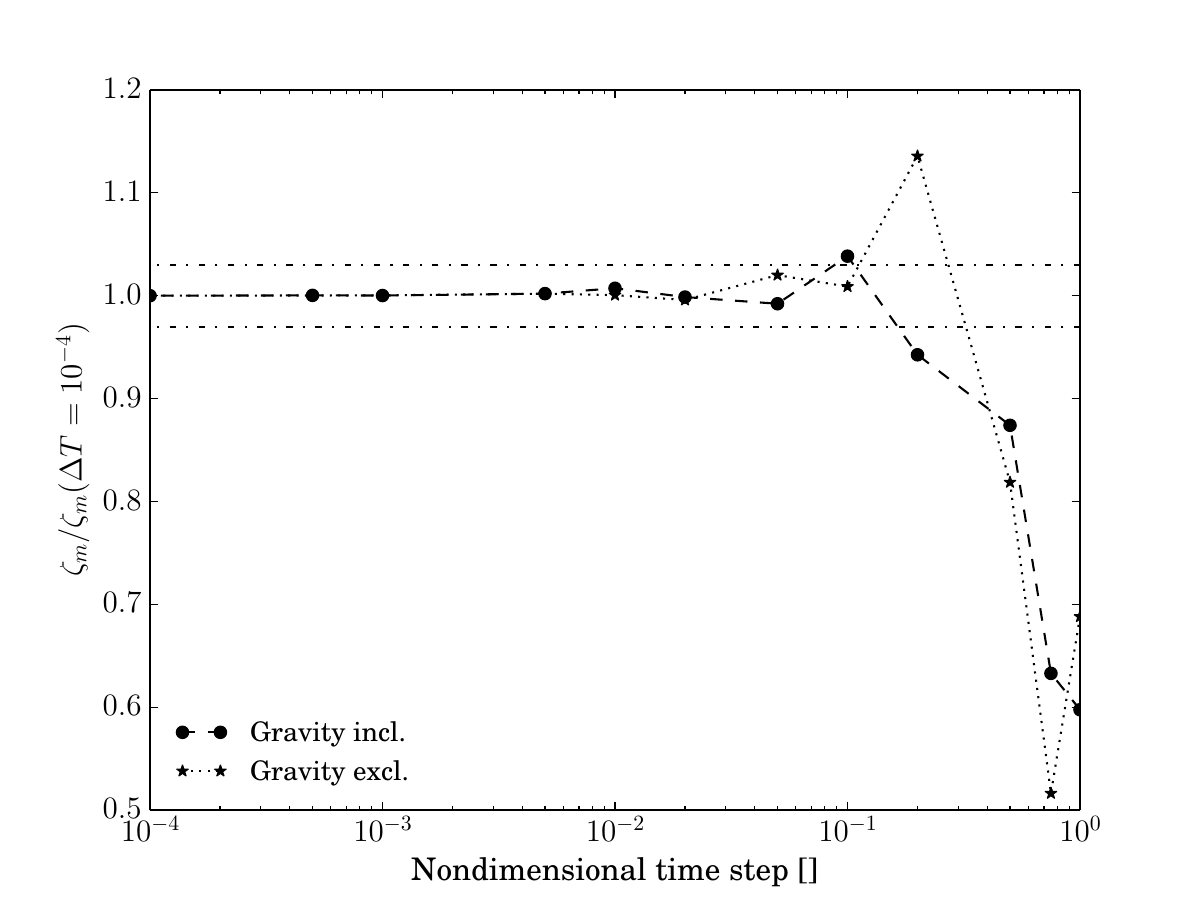}
	\caption{
		Convergence of the average power harvesting factor $\zeta_m$ normalised by the power harvesting factor $\zeta_{m}(\Delta T = 10^{-4})$ for the smallest nondimensional time step used in this convergence study. The dash-dotted lines indicate the 3\% convergence range. The convergence study is for the strong wind case.}
	\label{fig:convergence}
\end{figure}

As explained in Sect.~\ref{sec:model} the model equations are numerically integrated  in time. Figure~\ref{fig:convergence} shows how the accuracy of the integration result, the average power harvesting factor, is influenced by the constant integration time step. For a time step of $\Delta T < 0.1$, the simulations converge to less than 3\% deviation from the reference solution. This holds for simulations excluding and including the effect of gravity as well as for the strong and moderate wind cases. For this reason we use an integration step size of $\Delta T = 0.01$. As a result, the gravity-including simulation of the strong wind case requires 534 time steps to complete an entire pumping cycle.   

\subsection{Flight Trajectory}

\begin{figure}[h]
	\centering
	\fontsize{6.5pt}{10pt}\selectfont %matches font size in PDF
	\def\svgwidth{224pt}
	\vspace{-5pt}
	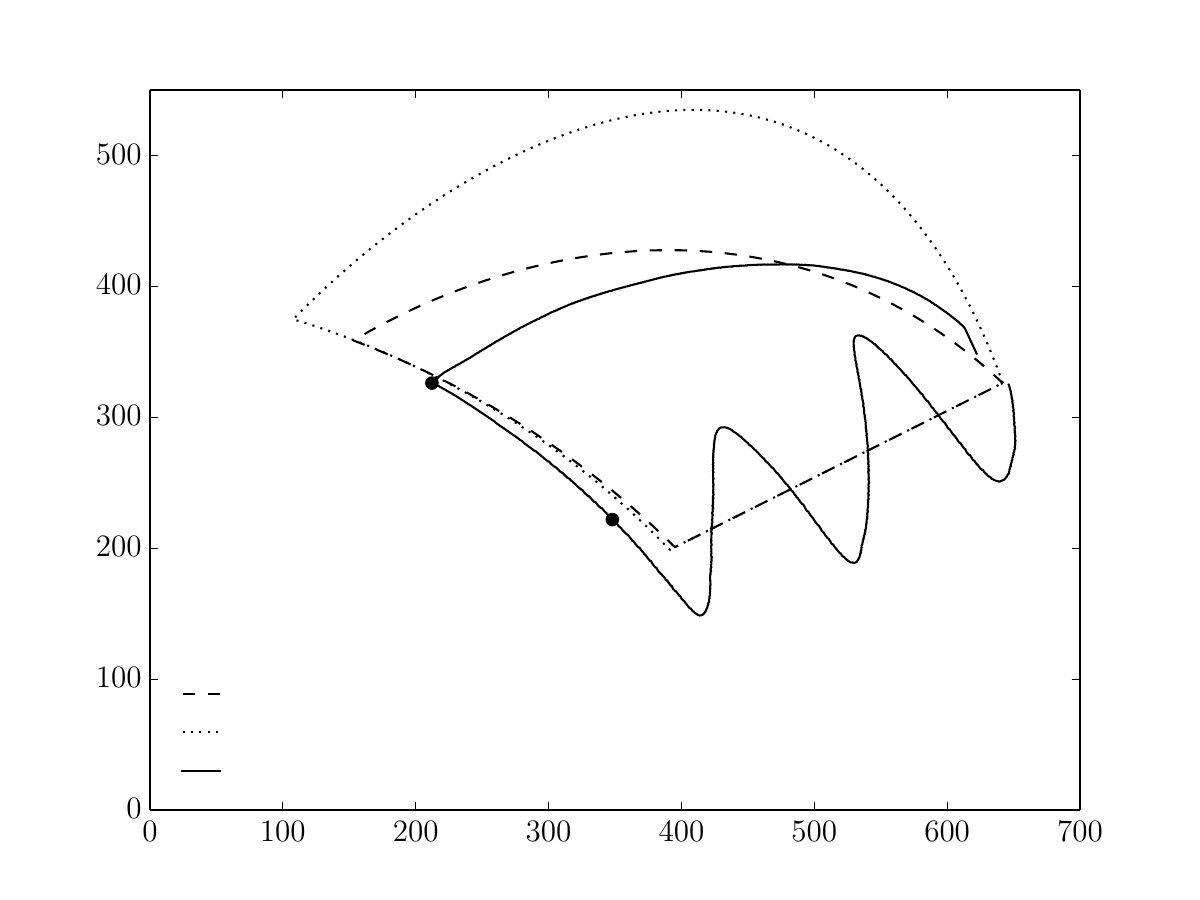 \\[-22pt]
	\def\svgwidth{224pt}
	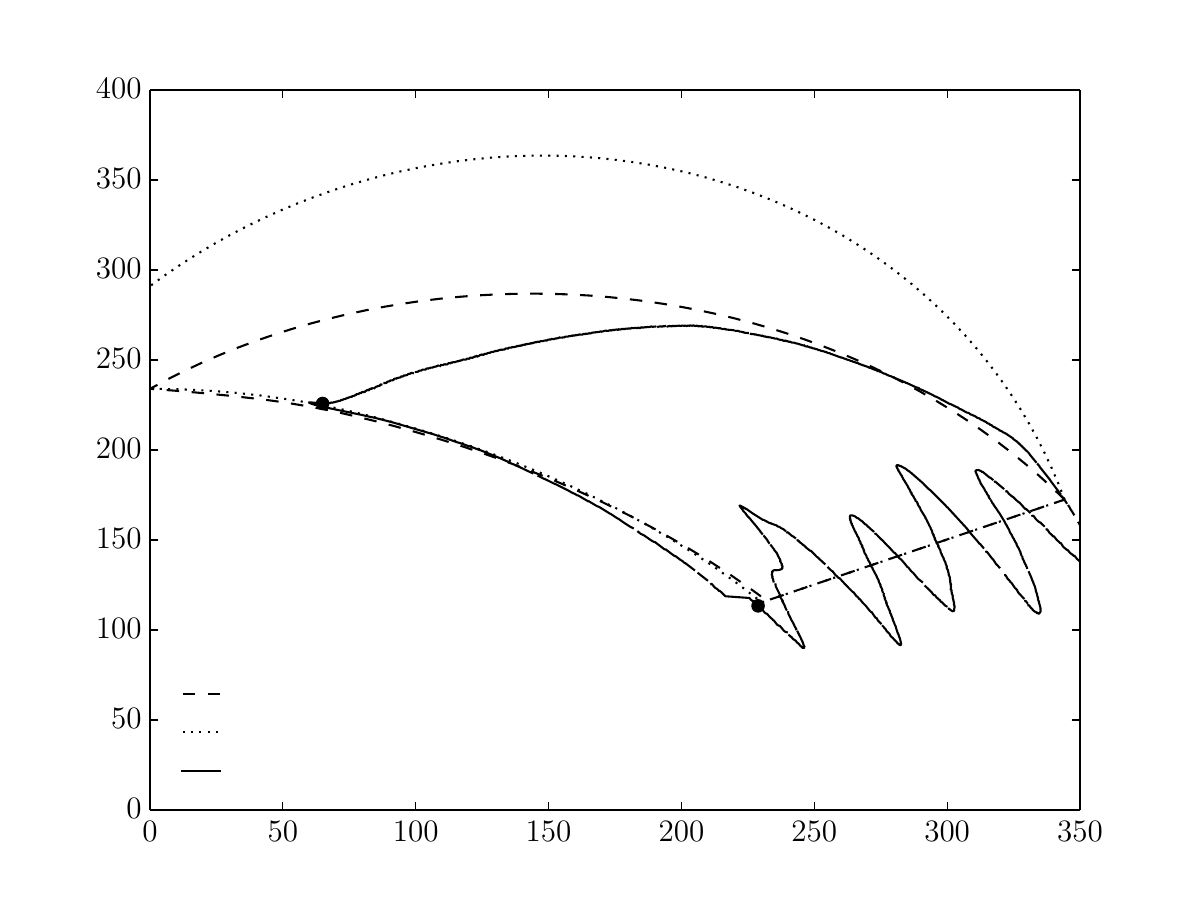 \\[-5pt]
	\caption{Position of the kite over a full pumping cycle. The dotted line represents the computed path neglecting gravity, the dashed line the computed path accounting for gravity and the solid line the measured path.}
	\label{fig:kitepathgrnddistheight}
\end{figure}

The computed and measured flight paths of the kite are depicted in the side views shown in Fig.~\ref{fig:kitepathgrnddistheight}. The horizontal distance is measured from the ground station while the height is measured from the ground.
Most obvious are the differences in the retraction phase which indicates how important the consideration of the gravitational effect is.
The lower flight path due to gravity is the result of two different mechanisms.

Firstly, we note that during retraction the gravitational force acting on the kite is of the same order of magnitude as the tether force.
As consequence, the radial component of the gravitational force significantly contributes to counterbalancing the resultant aerodynamic force of the kite and by that alleviates the tensile loading of the tether. 
In the extreme case of gliding flight towards the ground station the tensile loading can be reduced to a very low value. 

Secondly, the tangential component of the gravitational force exerts a particularly strong effect during the first part of the retraction phase.
In this period the kite flies upwards from a low elevation angle, the tether tension is low and the tangential component of the gravitational force adds up to the drag force to decelerate the kite.
This force effect keeps the kite from reaching a high velocity and by that limits the generated traction force. 

Because we adjust the reeling velocity to achieve a constant set value of the tether force, $F_{t,i} = 750$ N, the kite can be retracted faster in the simulation accounting for gravity.
This analysis is quantitatively supported by the reeling velocities shown in Sect.~\ref{sec:kinematic_properties}

During the transition phase the flight paths are all very similar.
Because the aerodynamic coefficients of the traction phase are used and the kite is flying a downward crosswind manoeuvre, a positive reeling velocity is required to not exceed the constant set value of the tether force, $F_{t,o} = 3000$ N, in this phase.
This can be seen from the data presented in Figs.~\ref{fig:kitepathgrnddistheight} and \ref{fig:reeloutvel}.

\begin{figure}[h]
	\centering
	\fontsize{6.5pt}{10pt}\selectfont %matches font size in PDF
	\def\svgwidth{224pt}
	\vspace{-5pt}
	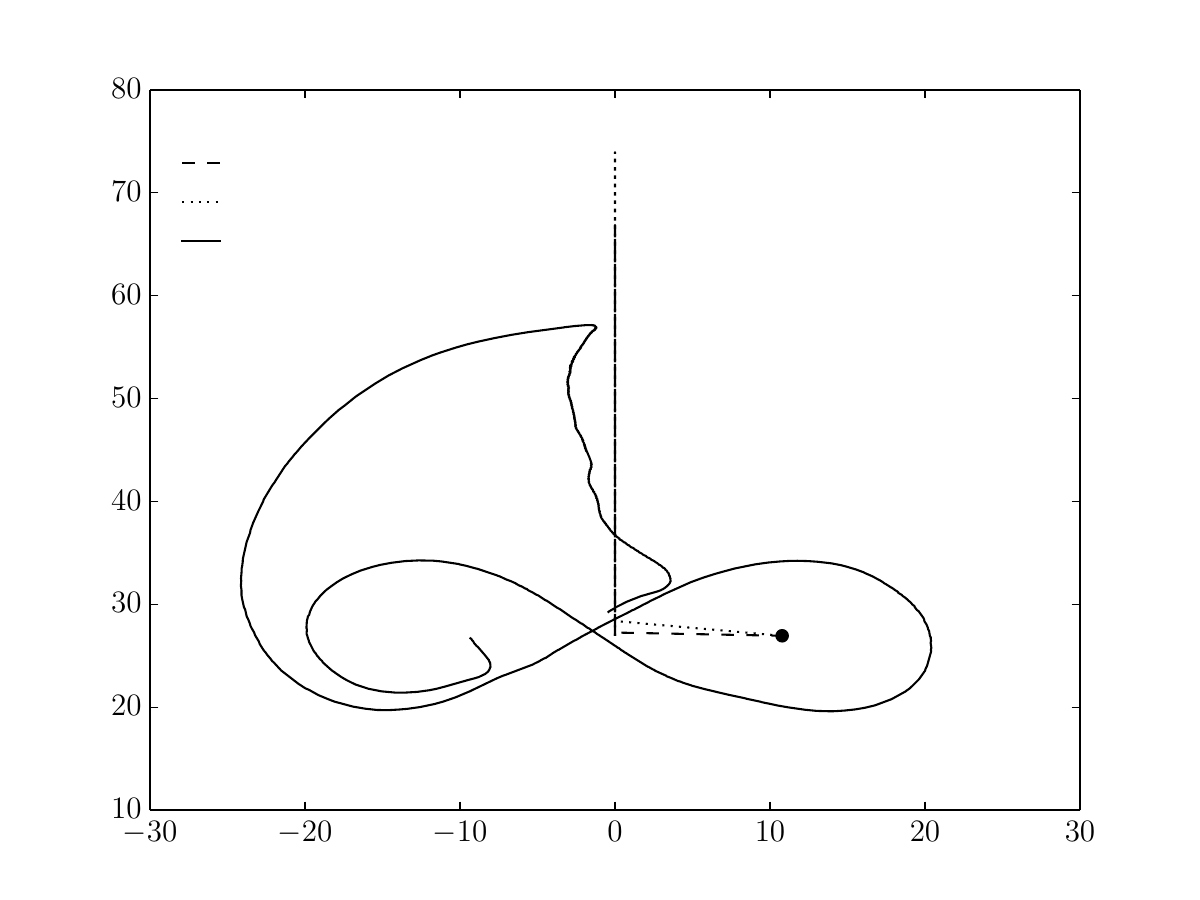 \\[-22pt]
	\def\svgwidth{224pt}
	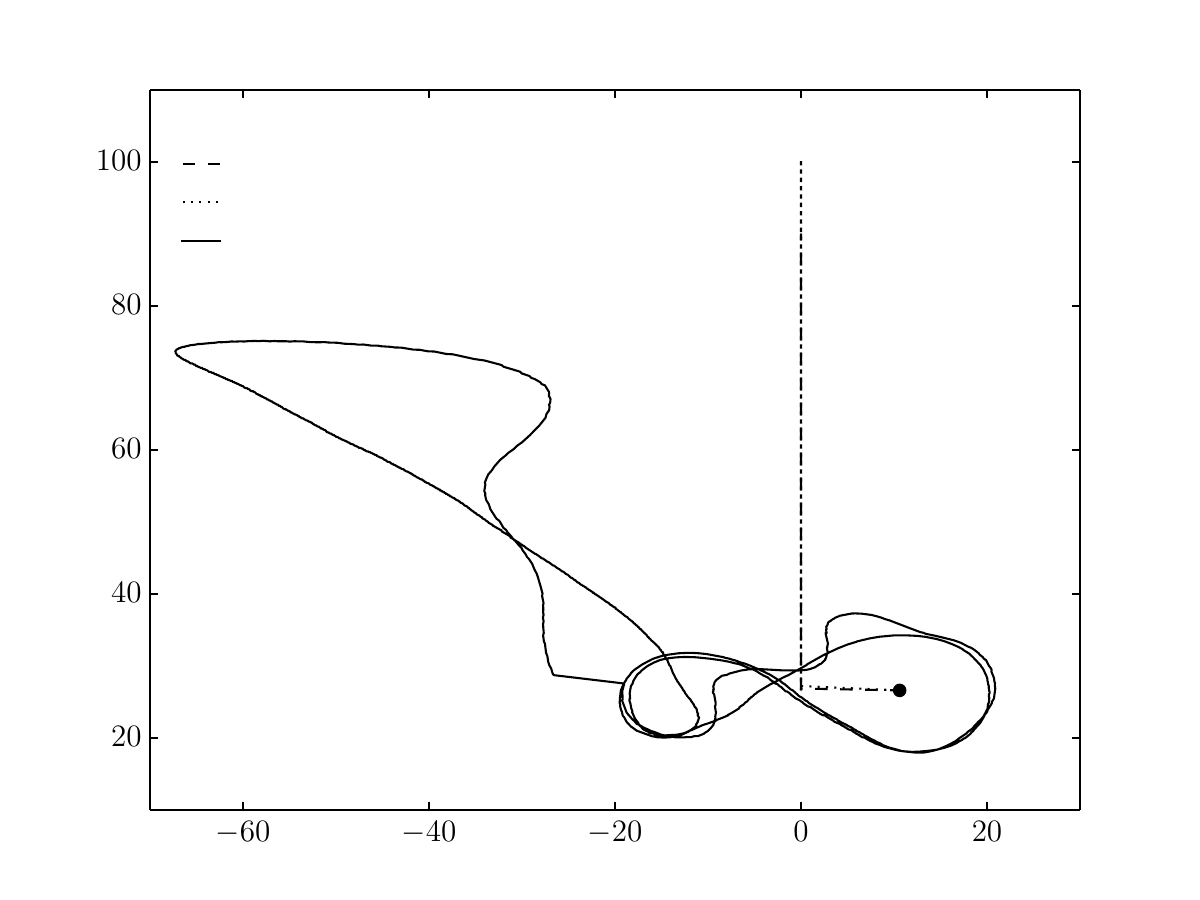 \\[-5pt]
	\caption{Angular coordinates of the kite over a full pumping cycle. The dot at the centre of the figure eight lobe indicates the constant values $\phi_o$ and $\beta_o$ that are used during the traction phase. The vertical line represents the computed flight path during the retraction phase.}
	\label{fig:kitepathelevazi}
\end{figure}

During the traction phase the computed flight paths do not resolve the measured figure eight manoeuvres but only the average motion of the kite along the straight line segment defined by the constant elevation angle $\beta_o$ and azimuth angle $\phi_o$.
This is also visible from the diagrams in Fig.~\ref{fig:kitepathelevazi} which complement the side views of the flight paths.
As discussed in Sect.~\ref{sec:traction} this constant average flight state during the traction phase coincides with the centre of one of the figure eight lobes. 

In terms of angular coordinates $\beta$ and $\phi$ the computed retraction and transition paths are straight and centred line segments.
However, the measured retraction paths show a significant deviation from the central line.
For the strong wind case the kite reaches an azimuth angle of $\phi = 25^\circ$ during the transition phase to smoothly connect to the first figure eight manoeuvre of the traction phase.
For the moderate wind case the measured retraction and transition paths go far through the side of the wind window.
This is an alternative technique of decreasing the traction force, which was used in this specific flight test.

\begin{figure}[h]
    \centering
    \fontsize{6.5pt}{10pt}\selectfont %matches font size in PDF
    \def\svgwidth{224pt}
    \vspace{-5pt}
    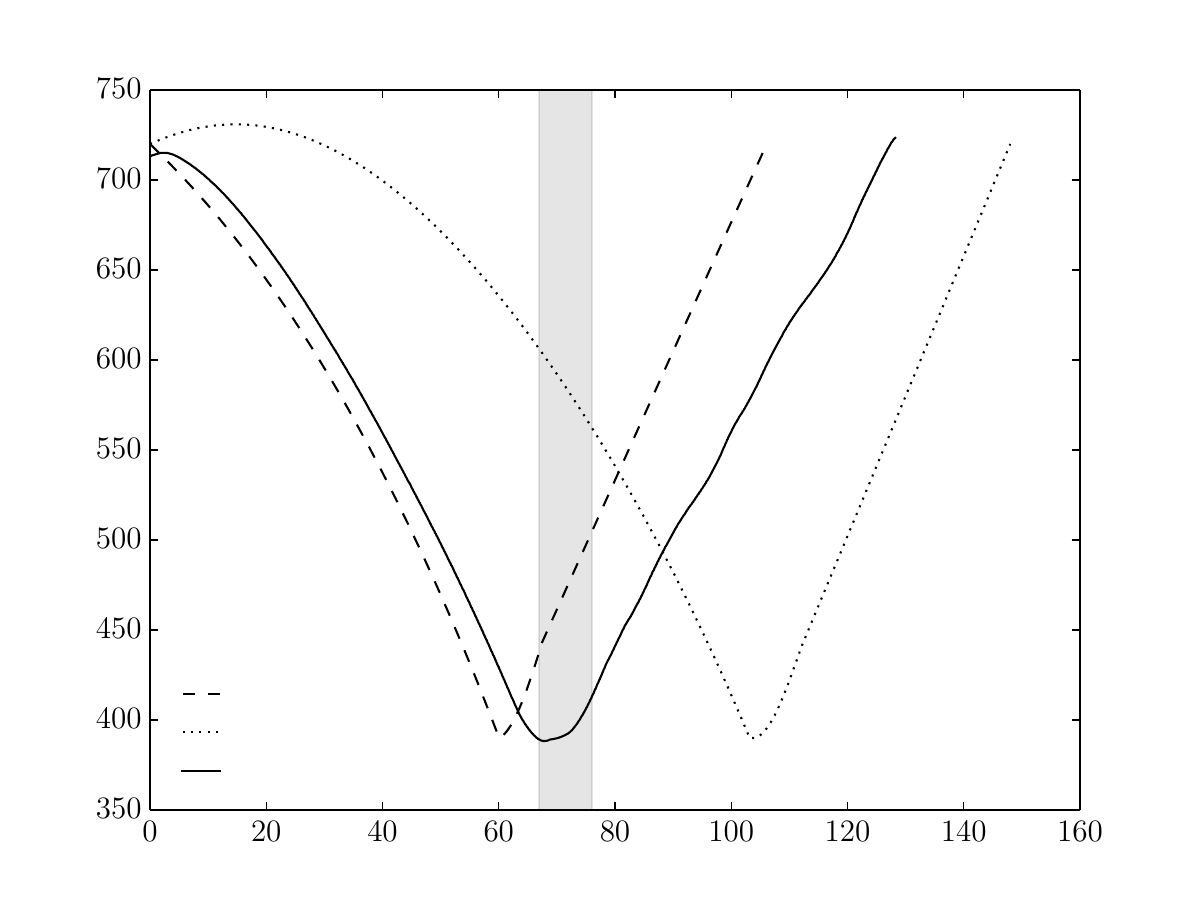 \\[-22pt]
    \def\svgwidth{224pt}
    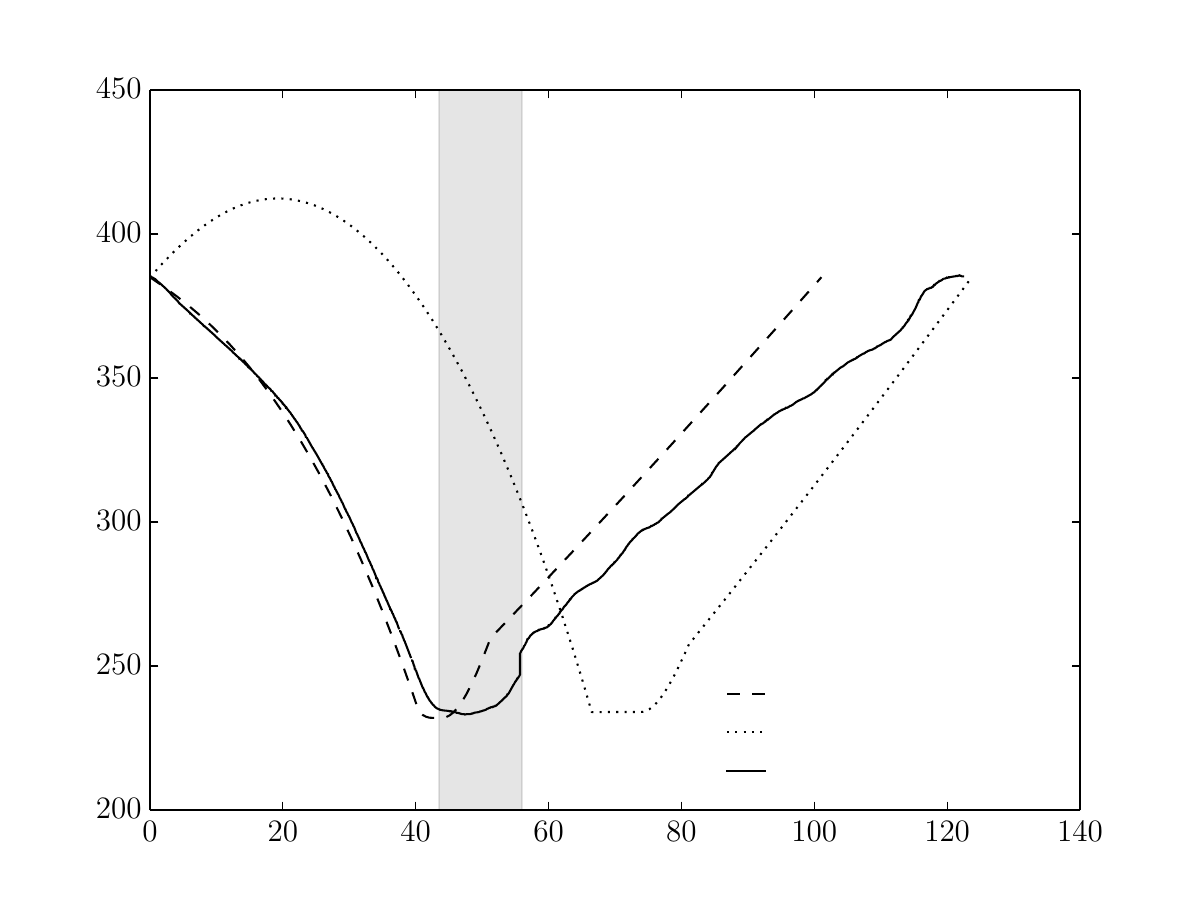 \\[-5pt]
    \caption{Tether length over a full pumping cycle.  The grey regions represent the transition phases in the experiment.}
    \label{fig:tetherlength}
\end{figure}

Plotting the tether length over time puts the comparison into a time perspective. From Fig.~\ref{fig:tetherlength} it can be seen that the real system and the simulation model accounting for gravity immediately start reeling in the tether, while the simulation model neglecting gravity initially continues to reel out the tether. At the same time the kite flies to a higher elevation angle which allows retracting the tether at the set value of the tether force, $F_{t,i} = 750$ N. As consequence, the retraction phase ends at a higher elevation angle which means that the flight path in the transition phase is longer. It can be concluded that the simulation of the retraction and transition phases takes substantially longer when neglecting the effect of gravity. 

\subsection{Kinematic Properties}
\label{sec:kinematic_properties}

\begin{figure}[h]
	\centering
	\fontsize{6.5pt}{10pt}\selectfont %matches font size in PDF
	\def\svgwidth{224pt}
	\vspace{-5pt}
	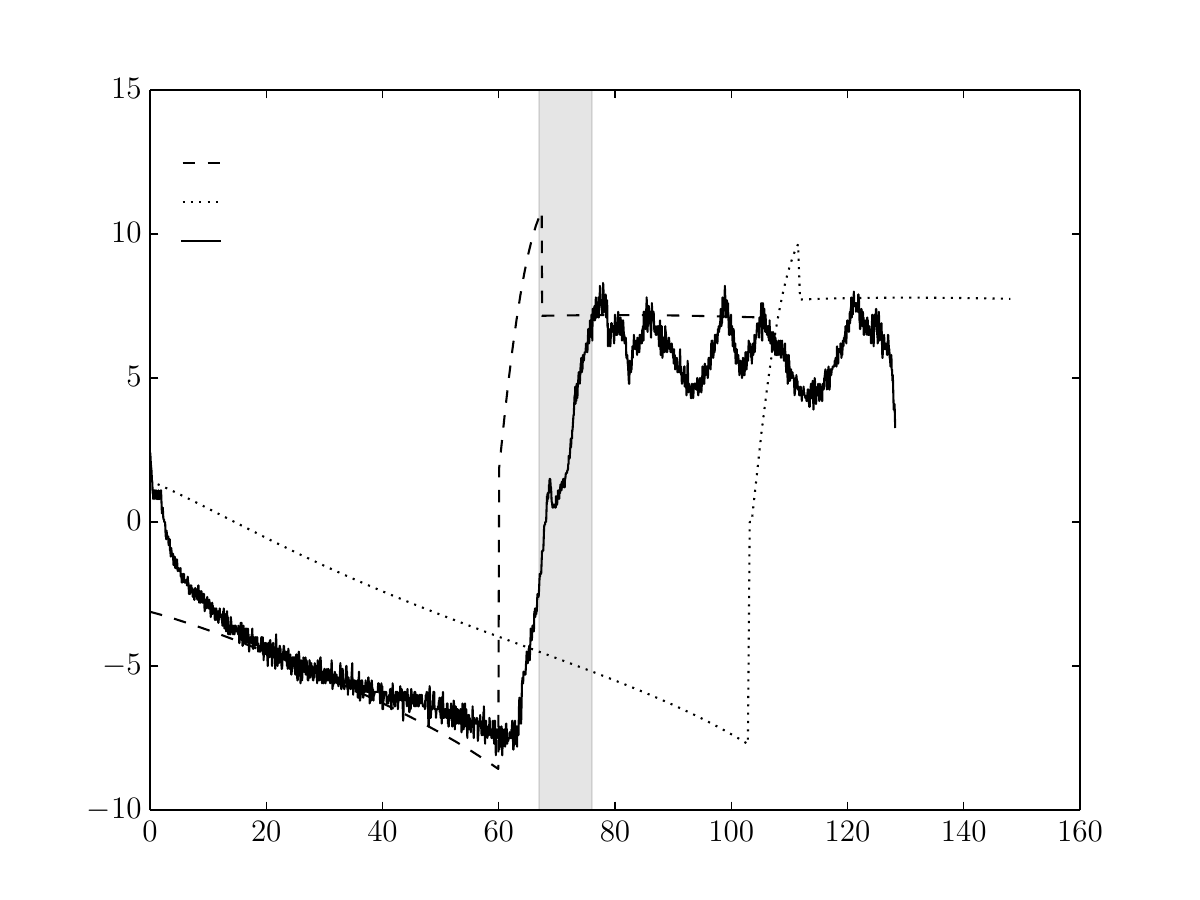 \\[-22pt]
	\def\svgwidth{224pt}
	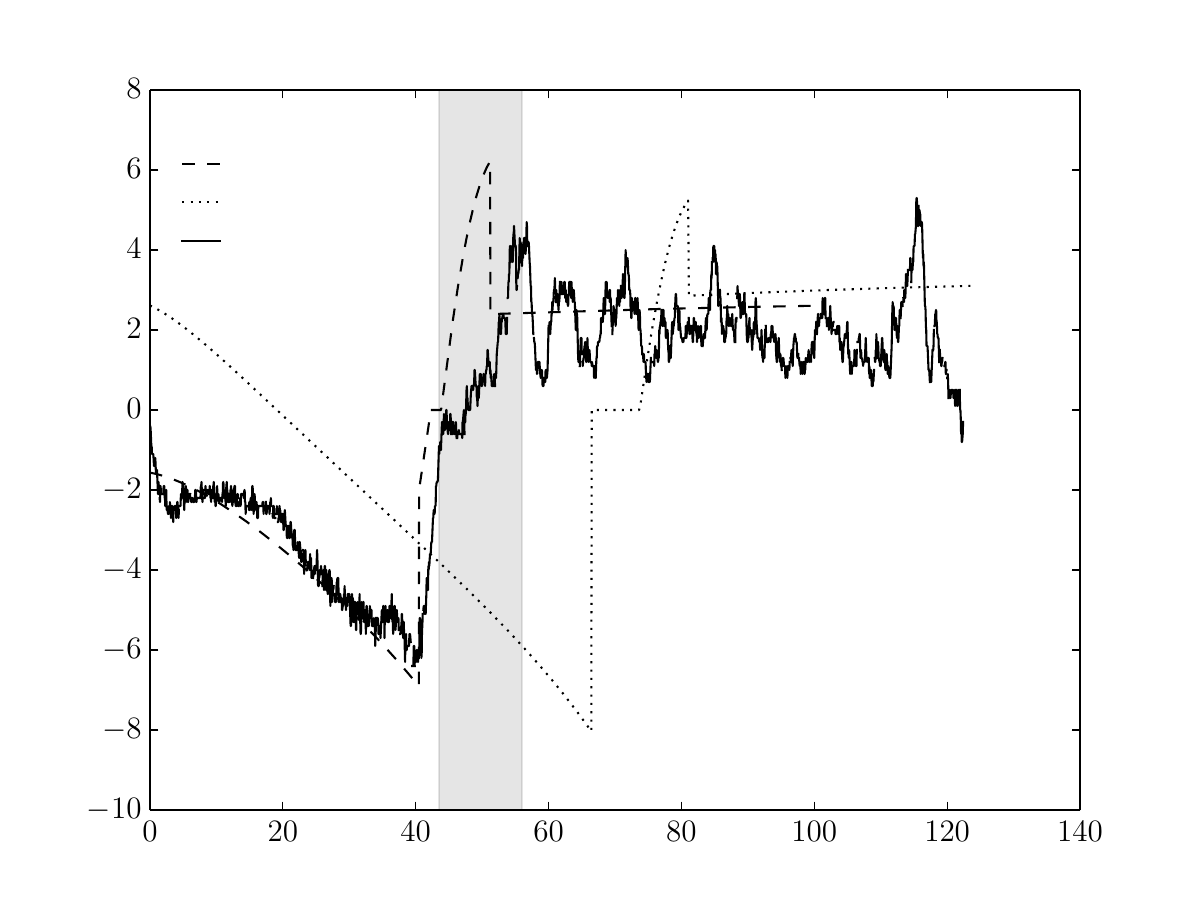 \\[-5pt]
	\caption{Tether reeling velocity over a full pumping cycle.}
	\label{fig:reeloutvel}
\end{figure}

Comparing the reeling velocity of the tether $v_t$, the flight velocity of the kite $v_k$, the wind velocity $v_w$ and the apparent wind velocity $v_a$ provides additional insight into the behaviour of the quasi-steady model and the effect of gravity. The reeling velocity is illustrated in Fig.~\ref{fig:reeloutvel}. The diagrams show that during the retraction phase the tether is reeled in with continuously increasing speed which is a consequence of the constant force control. From Eqs.~(\ref{eq:tetherforce}) and (\ref{eq:F_a_magnitude2}) it can be seen that as the elevation angle $\beta$ increases the aerodynamic force $F_a$ decreases. As consequence the retraction velocity can be increased continuously to keep the tether force at its set value.

\begin{figure}[h]
    \centering
    \fontsize{6.5pt}{10pt}\selectfont %matches font size in PDF
    \def\svgwidth{224pt}
    \vspace{-5pt}
    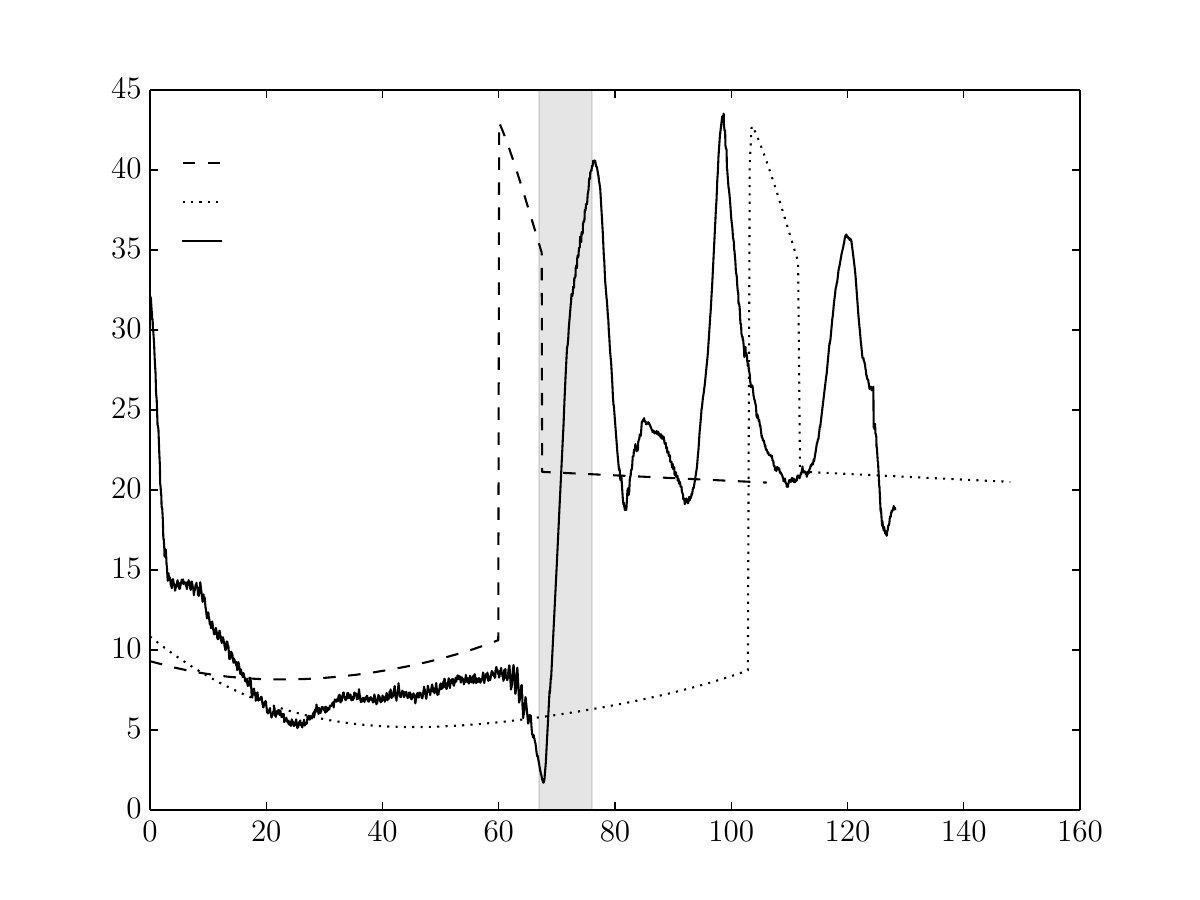 \\[-22pt]
    \def\svgwidth{224pt}
    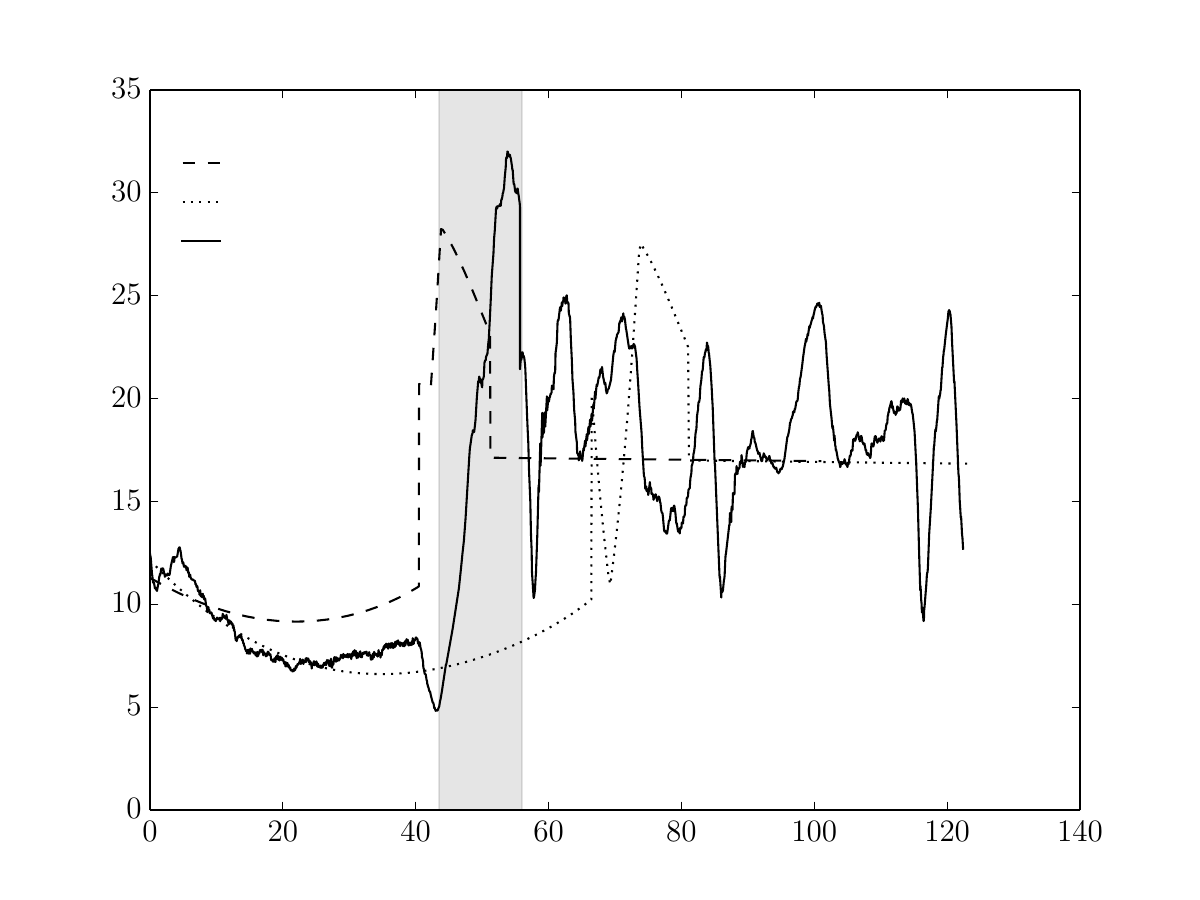 \\[-5pt]
    \caption{Flight velocity of the kite over a full pumping cycle.}
    \label{fig:kitevel}
\end{figure}

The flight velocity of the kite is illustrated in Fig.~\ref{fig:kitevel}. In the traction phase both simulation models exhibit a velocity that is slightly decreasing during the traction phase. This behaviour is a result of the competing effects of wind velocity and tether drag. On the one hand the wind velocity increases with the flight altitude which for itself would lead to an increase of the flight velocity according to Eqs.~(\ref{eq:lambda_definition}), (\ref{eq:tangentialkitevelocityfactor}) and (\ref{eq:lambda}). On the other hand the aerodynamic drag of the tether increases with the tether length. For the specific case the effect of tether drag predominates such that the flight velocity slightly decreases.

\begin{figure}[h]
    \centering
    \fontsize{6.5pt}{10pt}\selectfont %matches font size in PDF
    \def\svgwidth{224pt}
    \vspace{-5pt}
    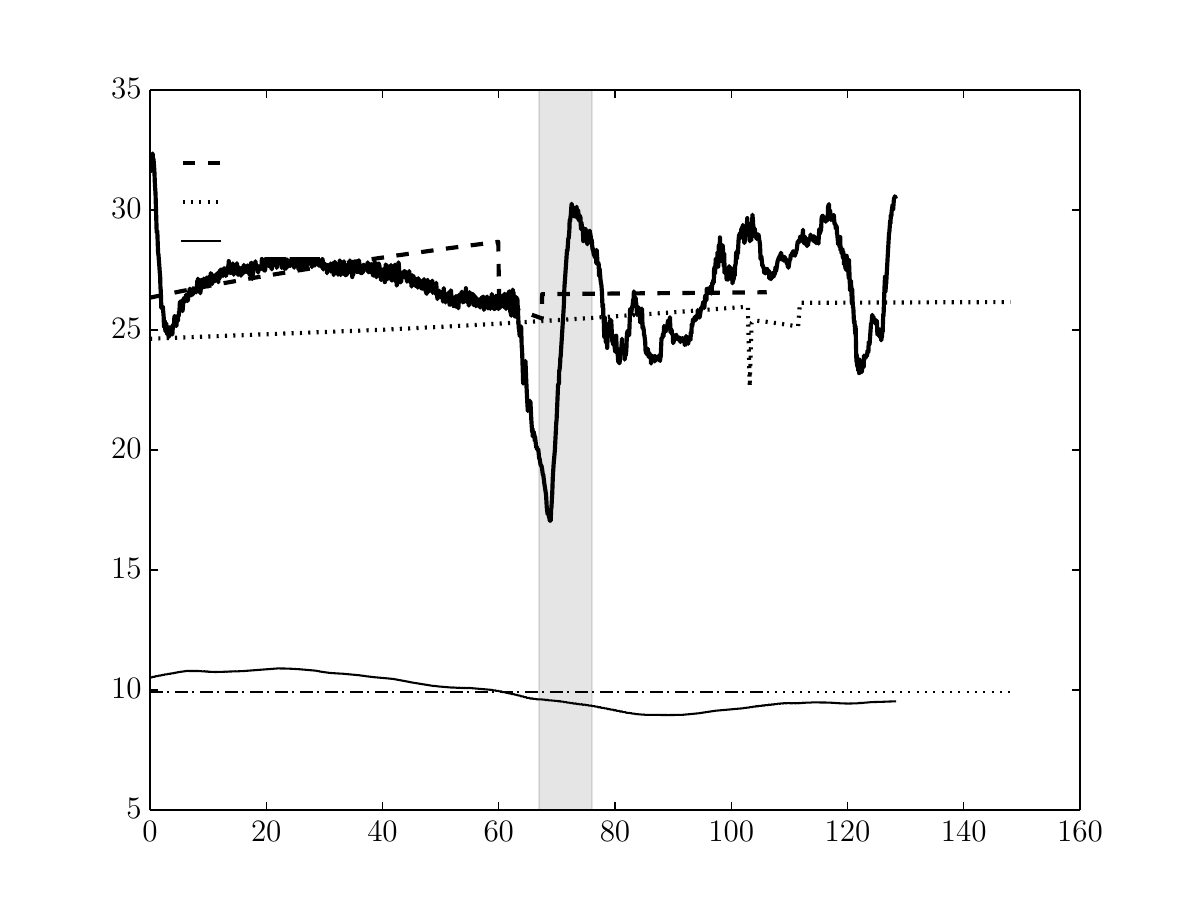 \\[-22pt]
    \def\svgwidth{224pt}
    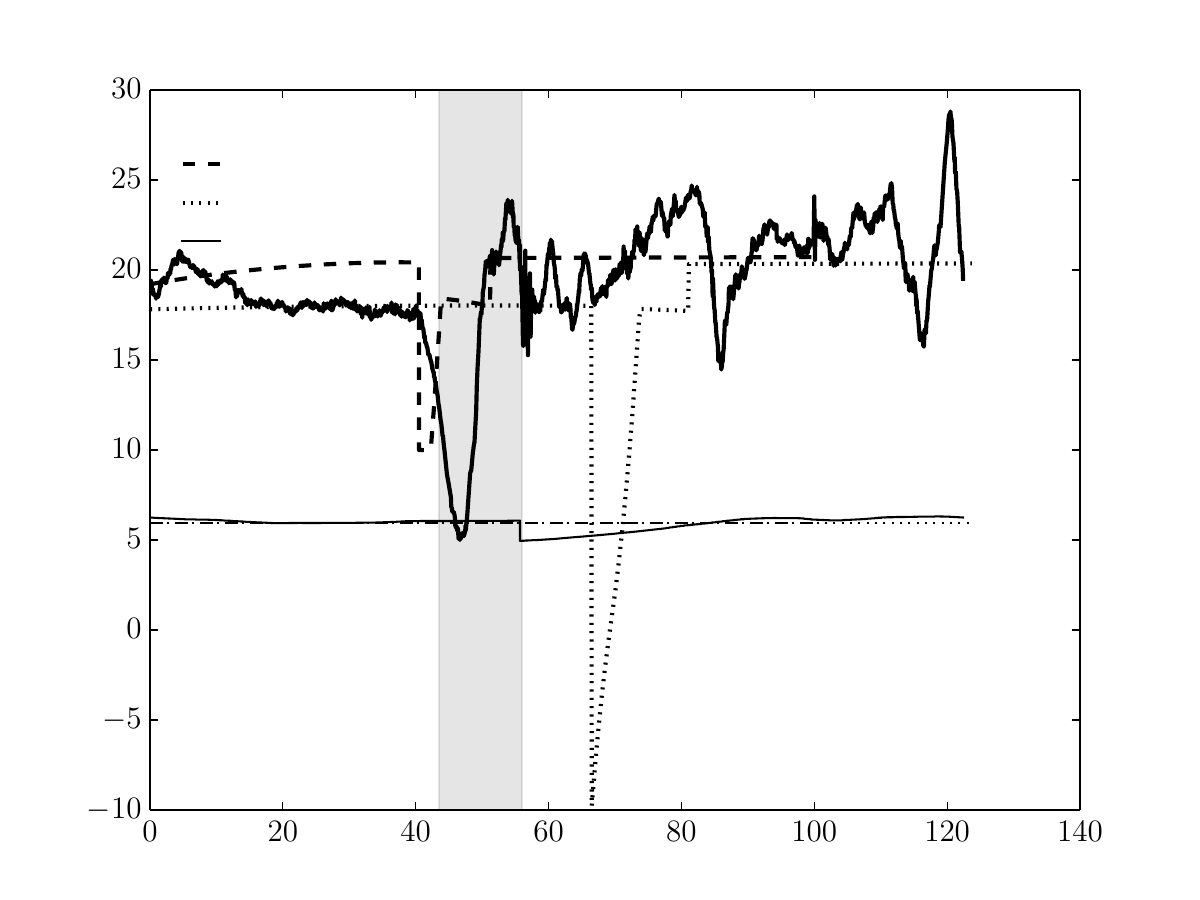 \\[-5pt]
    \caption{True wind velocity at 6 m high (thin lines) and apparent wind velocity (thick lines) for a full pumping cycle.}
    \label{fig:apparentvel}
\end{figure}

The apparent wind velocity experienced by the kite and the reference wind speed at 6 m altitude are depicted in Fig.~\ref{fig:apparentvel}. Both simulations use a constant value of the reference wind speed and Eq.~(\ref{eq:windshear}) to extrapolate to the wind velocity at the flight altitude. The apparent wind velocity is evaluated according to Eqs.~(\ref{eq:Vapp_final}) and (\ref{eq:v_a_inertial}). We can recognize that during the retraction phase the computed apparent wind velocity increases slightly while it levels to a constant value in the traction phase. This is caused by the constant force control and the fact that the tether force and the apparent wind velocity are directly linked by the quadratic relationship given by Eq.~(\ref{eq:tetherforce_a}). 

A side effect in this equation is the resultant aerodynamic coefficient $C_R$ which increases slightly with the tether length as a result of the increasing drag contribution, as quantified by Eq.~(\ref{eq:aerocoeff}) and (\ref{eq:totaldragcoeff}). Because of this, imposing a constant tether force during the retraction phase leads to a slightly increasing apparent wind velocity. Gravity will enhance this effect. During the traction phase these side effects are negligible and imposing a constant tether force directly translates into a constant apparent wind velocity.

\subsection{Traction Force}

\begin{figure}[h]
	\centering
	\fontsize{6.5pt}{10pt}\selectfont %matches font size in PDF
	\def\svgwidth{224pt}
	\vspace{-5pt}
	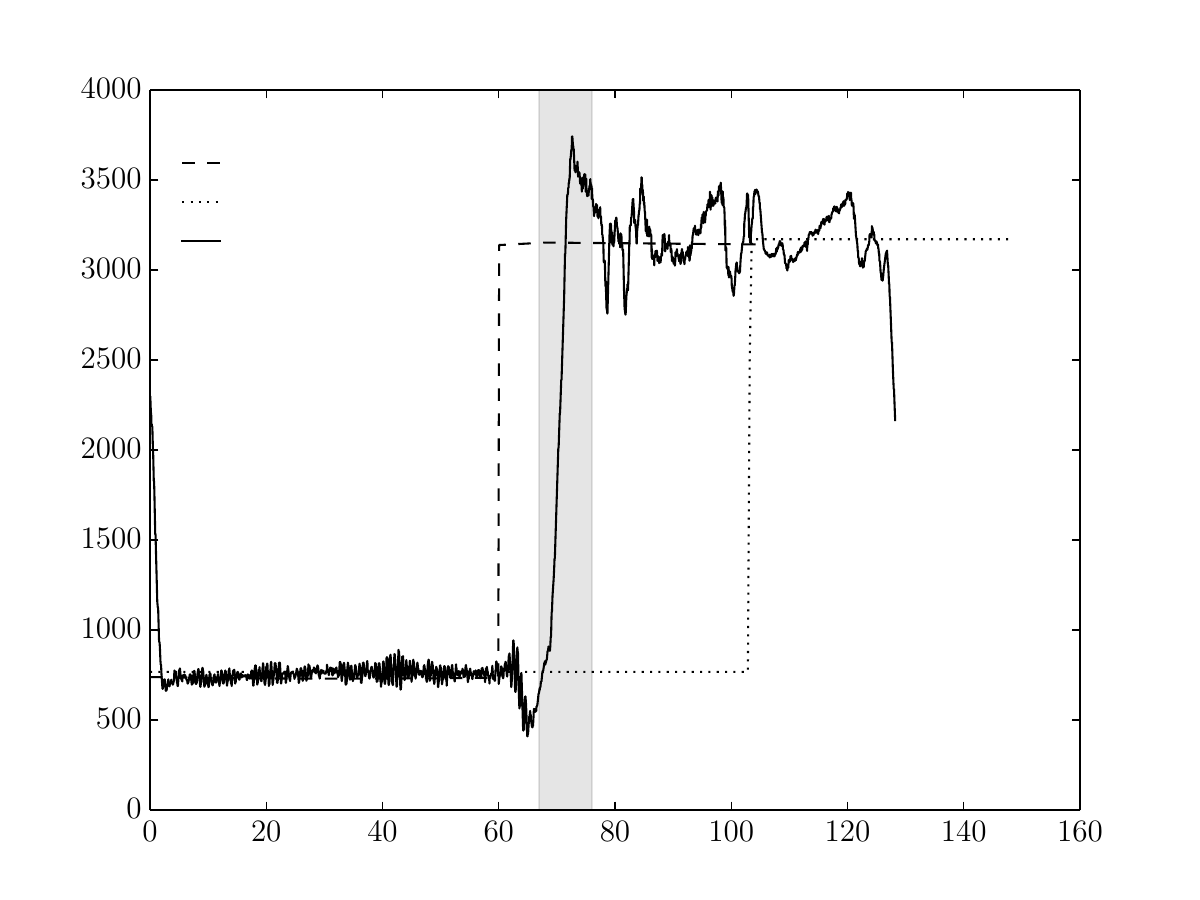 \\[-22pt]
	\def\svgwidth{224pt}
	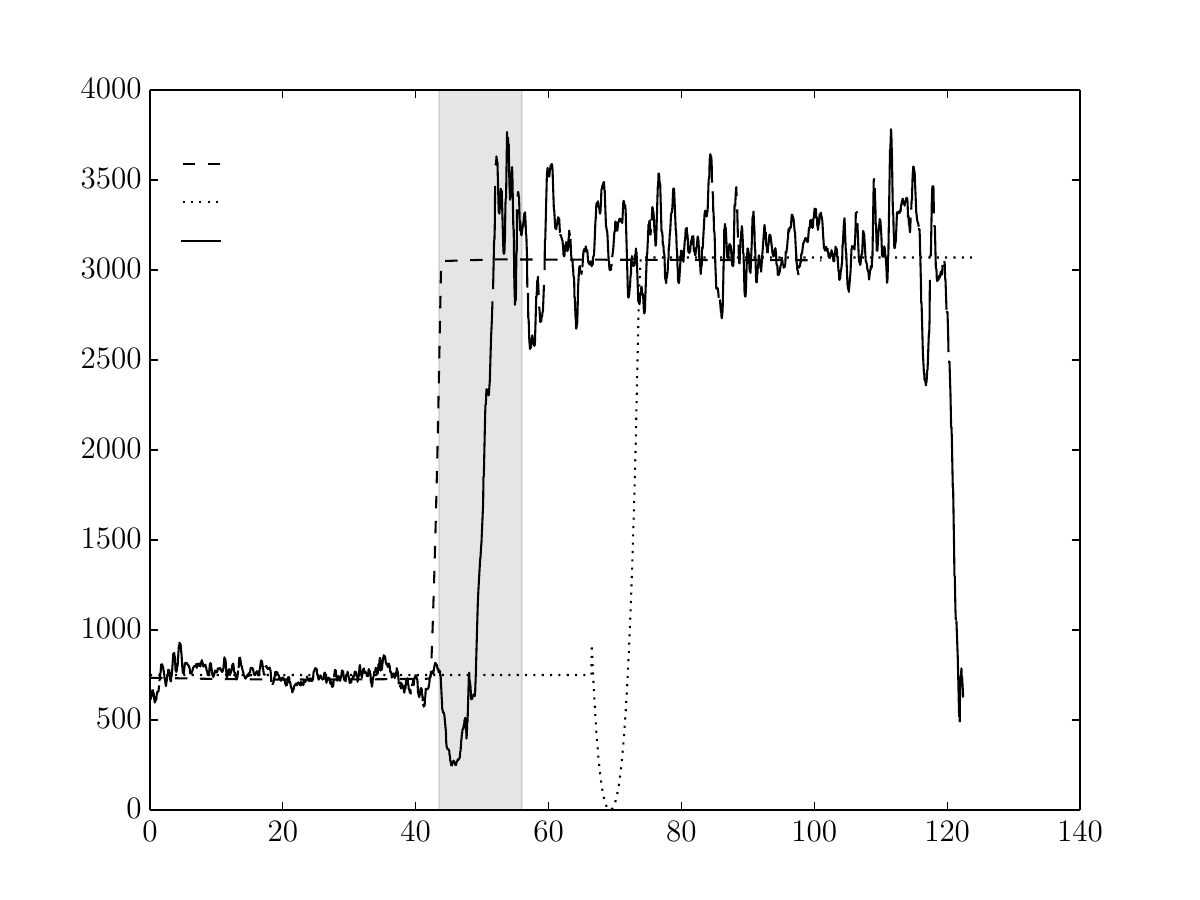 \\[-5pt]
	\caption{Tether force at the ground end of the tether over a full pumping cycle.}
	\label{fig:cabletension}
\end{figure}

Figure~\ref{fig:cabletension} shows the development of the tether force at the ground, $F_{t,g}$, over a full pumping cycle. Because we apply force control the computed tether force fits the measurements quite accurately, as expected. The largest deviation between simulations and experiment occurs in the traction phase. As the kite manoeuvres through the figure eight loops it is confronted with turbulence and wind gusts as well as motion-induced variations of the apparent wind velocity, as described by Eqs.~(\ref{eq:Vapp_final}) and (\ref{eq:v_a_inertial}).  As a result the tether tension experiences variations which the control mechanism of the ground station can not fully compensate anymore. This leads to instantaneous force overshoots of the set value by about 20\% which is taken into account in the system design of the technology demonstrator by defining the set value of the tether force with a safety margin.

\subsection{Traction Power}
\label{sec:power}

\begin{figure}[h]
	\centering
	\fontsize{6.5pt}{10pt}\selectfont %matches font size in PDF
	\def\svgwidth{224pt}
	\vspace{-5pt}
	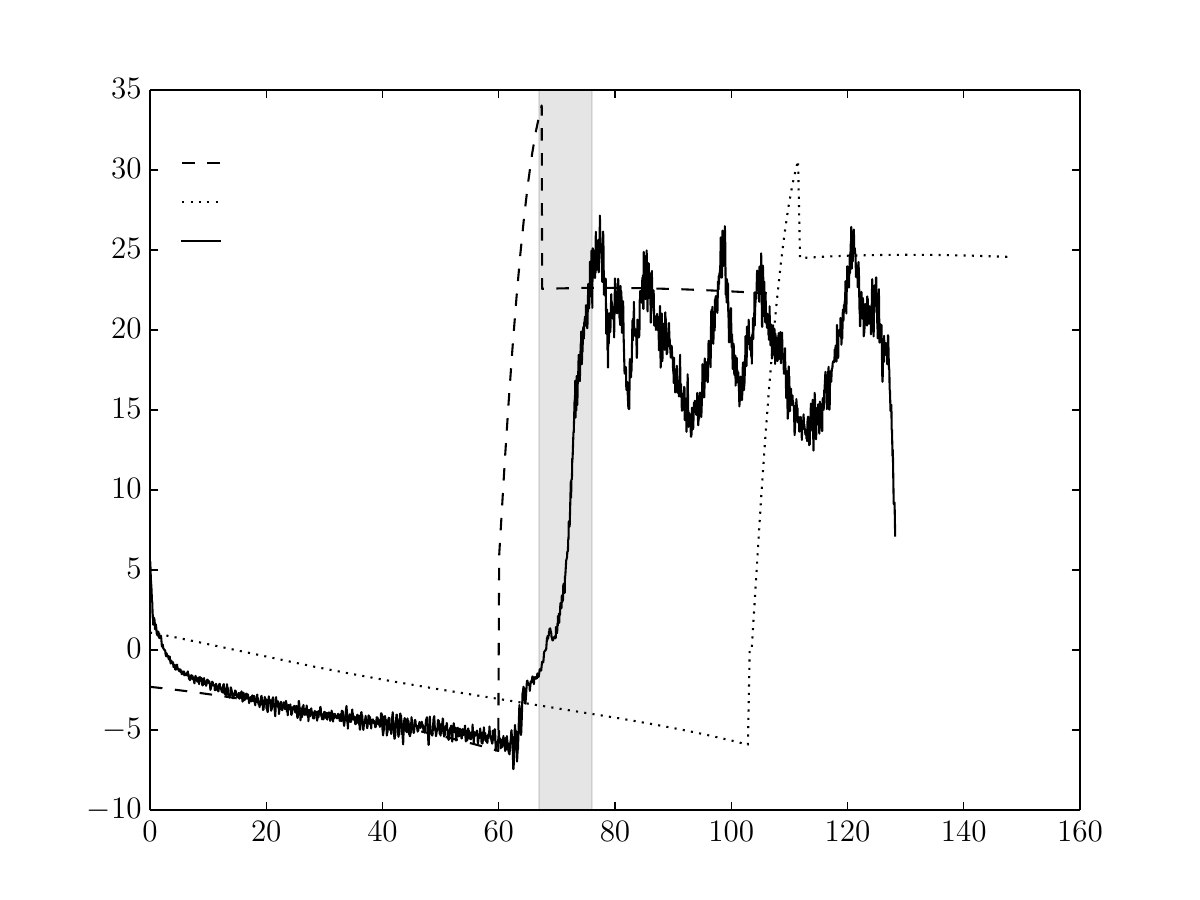 \\[-22pt]
	\def\svgwidth{224pt}
	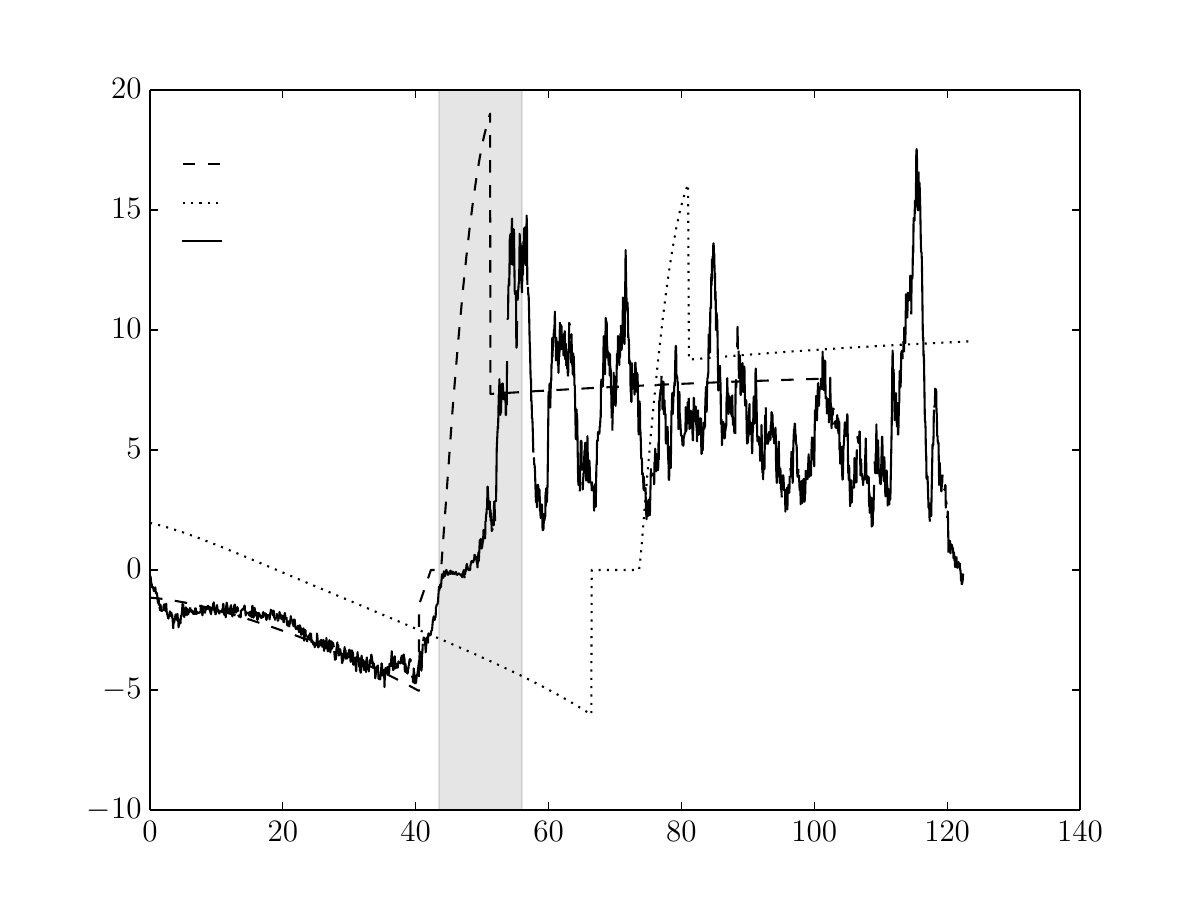 \\[-5pt]
	\caption{Mechanical power over a full pumping cycle.}
	\label{fig:power}
\end{figure}

Figure~\ref{fig:power} shows the instantaneous value of the traction power delivered to the ground station over a full pumping cycle. Tables~\ref{tab:results_highwind} and \ref{tab:results_lowwind} list the mean values for the cycle and its three phases. Considering the retraction phase we note that the consumed power and the time duration are within 10\% of the measured values when gravity is taken into account. We further note that the effect of gravity reduces the retraction time from a significant overestimation to a slight underestimation of the measured value. This underestimation can be explained by noting that the measured flight path is not perfectly straight as the computed paths, resulting in less efficient and thus slower retraction.

Also for the traction phase the generated power and time duration are closer to the measured values when accounting for gravity. The effect of gravity is however not as strong as in the retraction phase. Yet, even when accounting for gravity the simulation overestimates the generated power and underestimates the time duration of the phase. A possible reason could be an overprediction of the computed wind velocity at the operational altitude of the kite. Future research with more accurate wind measurements \cite{Oehler2017} and a comparison with a dynamic model \cite{Fechner2015} is necessary to better understand the reason for this difference.  

We finally compare the computed and measured performance characteristics for the complete pumping cycle. For strong wind conditions the measured data is between the two simulation results. For moderate wind conditions the simulation neglecting gravity is closer to the experiment. The close match can be traced back to a coincidental, mutual compensation of the modelling errors occurring in the different cycle phases. However, because the modelling errors per phase are generally lower when accounting for gravity we recommended to further improve this more advanced modelling option.
\begin{table}[h] 
\begin{tabular}{@{}llrrr@{}} 
\hline\noalign{\smallskip} 
& & Gravity  & Gravity &  \\ 
Phase &  Parameter &  excluded &  included & Experiment \\ 
\hline\noalign{\smallskip}
Retraction & $P_m$ [kW] & -2.46 & -4.03 & -3.64 \\ 
& Time [s] & 103 & 60 & 67 \\ 
\hline 
Transition & $P_m$  [kW] & 17.90 & 23.67 & 8.50 \\ 
& Time [s] & 8 & 7 & 9 \\ 
\hline 
Traction & $P_m$  [kW] & 24.72 & 22.57 & 19.12 \\ 
& Time [s] & 36 & 38 & 52 \\ 
\hline 
Complete & $P_m$  [kW] & 5.37 & 7.59 & 6.48 \\ 
cycle & Time [s] & 148 & 106 & 128 \\ 
\hline\noalign{\smallskip}
\end{tabular} 
\caption{Simulated and measured performance characteristics of the pumping cycle for strong wind conditions.} 
\label{tab:results_highwind} 
\end{table}

% HIGH WIND RESULT
%Cycle experimental result: 
%Cycle mean power: 6326.97141825
%Cycle mean retraction    //Power: -3690.71814179        //Time: 67.0    //Energy: 
%Cycle mean transition    //Power: 11254.2623944  		//Time: 9.0     //Energy: 
%Cycle mean traction      //Power: 17637.7160298  		//Time: 55.45   //Energy: 
%Retraction phase
%Transition phase
%Traction phase
%Cycle result: 
%Cycle mean power: 6535.69163101
%Cycle mean retraction    //Power: -3681.31572253        //Time: 64.2335981788   //Energy: -236464.15489
%Cycle mean transition    //Power: 18684.2968419  		//Time: 5.24749916762   //Energy: 98045.8321253
%Cycle mean traction      //Power: 19487.5612298  		//Time: 45.7482485203   //Energy: 891521.794193
%Retraction phase
%Transition phase
%Traction phase
%Cycle result: 
%Cycle mean power: 4435.60441615
%Cycle mean retraction    //Power: -2065.26437906         //Time: 120.444172298   //Energy: -248749.058713
%Cycle mean transition    //Power: 13756.6702131  //Time: 6.61329155221   //Energy: 90976.8709066
%Cycle mean traction      //Power: 21817.2362164  //Time: 41.5006395174   //Energy: 905429.255482

\begin{table}[!htb] 
\begin{tabular}{@{}llrrr@{}} 
\hline\noalign{\smallskip}
& & Gravity  & Gravity &  \\ 
Phase &  Parameter &  excluded &  included & Experiment \\ 
\hline\noalign{\smallskip}
Retraction & $P_m$ [kW] & -1.73 & -2.73 & -2.60 \\ 
& Time [s] & 66 & 40 & 43 \\ 
\hline 
Transition & $P_m$  [kW] & 5.08 & 8.11 & 3.44 \\ 
& Time [s] & 14 & 10 & 12 \\ 
\hline 
Traction & $P_m$  [kW] & 9.20 & 7.67 & 6.23 \\ 
& Time [s] & 42 & 49 & 66 \\ 
\hline 
Complete & $P_m$  [kW] & 2.84 & 3.55 & 2.79 \\ 
cycle & Time [s] & 123 & 101 & 122 \\ 
\hline\noalign{\smallskip}
\end{tabular} 
\caption{Simulated and measured performance characteristics of the pumping cycle for moderate wind conditions.} 
\label{tab:results_lowwind} 
\end{table}

% LOW WIND RESULT
%Cycle experimental result: 
%Cycle mean power: 2791.30316269
%Cycle mean retraction    //Power: -2597.5744994  //Time: 43.5    //Energy: 
%Cycle mean transition    //Power: 3440.55917769  //Time: 12.4500000477   //Energy: 
%Cycle mean traction      //Power: 6226.91135252  //Time: 66.4500000477   //Energy: 
%Retraction phase
%Transition phase
%Traction phase
%C_R
%Simulated Cycle result: mass 19.6 kg
%Cycle mean power: 2370.93206528
%Cycle mean retraction    //Power: -948.692268353         //Time: 115.735084489   //Energy: -109796.979832
%Cycle mean transition    //Power: 11273.3051675  //Time: 7.76796339215   //Energy: 87570.6218494
%Cycle mean traction      //Power: 14562.3343856  //Time: 25.8414648362   //Energy: 376312.051958
%Retraction phase
%Transition phase
%Traction phase
%C_R
%Simulated Cycle result: mass None kg
%Cycle mean power: 649.544020372
%Cycle mean retraction    //Power: -223.938186778         //Time: 510.643350358   //Energy: -114352.54597
%Cycle mean transition    //Power: 8949.47286858  //Time: 8.11485503922   //Energy: 72623.675006
%Cycle mean traction      //Power: 15976.3840934  //Time: 24.7073212413   //Energy: 394733.65407

%%%%%%%%%%%%%%%%%%%%%%%%%%%%%%%%%%%%%%%%%%%%%%%%%%%%%%%%%%%%%%%%%%%%%%%%%%%%%%%%%
\section{Conclusion}
\label{sec:conclusion}
%%%%%%%%%%%%%%%%%%%%%%%%%%%%%%%%%%%%%%%%%%%%%%%%%%%%%%%%%%%%%%%%%%%%%%%%%%%%%%%%%

The present study comprises a quasi-steady modelling framework for a pumping kite power system and a comprehensive validation of this framework based on experimental data.
The objective of the model is to estimate the mechanical power output as a function of the wind conditions, the system design and operational parameters.
Part of the study is a technique to estimate the aerodynamic properties of the kite using available measurement data.
The validation reference data is derived from two separate test campaigns of a technology demonstrator using kites of 14 and 25 m$^2$ surface area to generate 20 kW of nominal traction power.
The data for moderate and strong wind conditions comprises instantaneous values, average values for each of the three phases and for the complete cycle.

The computational effort to numerically integrate the flight path over a pumping cycle substantially reduces by not explicitly resolving the transverse crosswind manoeuvres. Using a two-dimensional idealisation of the cycle we find that three operational phases have to be distinguished: retraction, transition and traction. When accounting for this partitioning the modelling framework provides valuable insight into the energy conversion mechanisms which can be used as a starting point for systematic optimisation.

Per cycle phase the simulation is generally closer to the experimental data when accounting for gravitational effects.
Especially during the retraction phase, when the gravitational force is of the same order of magnitude as the other forces governing the flight motion of the kite, the effect of gravity is substantial and neglecting these contributions leads to pronounced deviations between simulation and experiment.
We thus recommended to always take the mass of the airborne components into account.

The analysis clearly indicates that additional information about the aerodynamic properties of the airborne system components and the atmospheric conditions will greatly improve the prediction quality. The current calculation of the apparent wind velocity from an extrapolated measured ground surface wind velocity and the GPS-velocity of the kite should be regarded only as a first step. As consequence, we recommend to include in future test campaigns also separate measurements of the wind velocity at several altitudes, for example, by statically positioning the kite at these altitudes and using the onboard wind sensor. Similarly, this wind sensor should be used for direct measurement of the apparent wind velocity \cite{Oehler2017}.

The presented modelling framework is perfectly suited as a basis for optimisation and scaling studies, also to predict the power generation potential and the achievable cost of energy at a specific deployment site \cite{Grete2014b}.
The framework has been used for designing and predicting the power output of a kite wind park \cite{Faggiani2018}.
The quasi-steady analysis is not suited, for example, for investigating peak loading during crosswind manoeuvres or for fully dynamic flight behaviour. For such analyses a dynamic system model needs to be used \cite{Fechner2015}.

\section*{Acknowledgements}
\label{sec:acknowledgements}

The authors would like to thank Johannes Oehler for proofreading the manuscript. Anna Bley and Roland Schmehl have received financial support by the project REACH (H2020-FTIPilot-691173), funded by the European Union's Horizon 2020 research and innovation programme under grant agreement No. 691173, and AWESCO (H2020-ITN-642682) funded by the European Union's Horizon 2020 research and innovation programme under the Marie Sk{\l}odowska-Curie grant agreement No. 642682.



%% References


\bibliographystyle{model3-num-names}
\bibliography{bibliography}

%% not want to use model3-num-names.bst.





% \end{thebibliography}

\end{document}

%% file: trajectory.pdf_tex
%% Creator: Inkscape inkscape 0.91, www.inkscape.org
%% Accompanies image file 'trajectory.pdf' (pdf, eps, ps)
\begingroup%
  \makeatletter%
  \providecommand\color[2][]{%
    \errmessage{(Inkscape) Color is used for the text in Inkscape, but the package 'color.sty' is not loaded}%
    \renewcommand\color[2][]{}%
  }%
  \providecommand\transparent[1]{%
    \errmessage{(Inkscape) Transparency is used (non-zero) for the text in Inkscape, but the package 'transparent.sty' is not loaded}%
    \renewcommand\transparent[1]{}%
  }%
  \providecommand\rotatebox[2]{#2}%
  \ifx\svgwidth\undefined%
    \setlength{\unitlength}{248bp}%
    \ifx\svgscale\undefined%
      \relax%
    \else%
      \setlength{\unitlength}{\unitlength * \real{\svgscale}}%
    \fi%
  \else%
    \setlength{\unitlength}{\svgwidth}%
  \fi%
  \global\let\svgwidth\undefined%
  \global\let\svgscale\undefined%
  \makeatother%
  \begin{picture}(1,0.58064516)%
    \put(0,0){\includegraphics[width=\unitlength,page=1]{trajectory.pdf}}%
    \put(0.12914308,0.04898219){\makebox(0,0)[lb]{\smash{0}}}%
    \put(0,0){\includegraphics[width=\unitlength,page=2]{trajectory.pdf}}%
    \put(0.21183753,0.04898219){\makebox(0,0)[lb]{\smash{50}}}%
    \put(0,0){\includegraphics[width=\unitlength,page=3]{trajectory.pdf}}%
    \put(0.29552353,0.04898219){\makebox(0,0)[lb]{\smash{100}}}%
    \put(0,0){\includegraphics[width=\unitlength,page=4]{trajectory.pdf}}%
    \put(0.38697819,0.04898219){\makebox(0,0)[lb]{\smash{150}}}%
    \put(0,0){\includegraphics[width=\unitlength,page=5]{trajectory.pdf}}%
    \put(0.48035182,0.04898219){\makebox(0,0)[lb]{\smash{200}}}%
    \put(0,0){\includegraphics[width=\unitlength,page=6]{trajectory.pdf}}%
    \put(0.57245148,0.04898219){\makebox(0,0)[lb]{\smash{250}}}%
    \put(0,0){\includegraphics[width=\unitlength,page=7]{trajectory.pdf}}%
    \put(0.66396896,0.04898219){\makebox(0,0)[lb]{\smash{300}}}%
    \put(0,0){\includegraphics[width=\unitlength,page=8]{trajectory.pdf}}%
    \put(0.75639119,0.04898219){\makebox(0,0)[lb]{\smash{350}}}%
    \put(0,0){\includegraphics[width=\unitlength,page=9]{trajectory.pdf}}%
    \put(0.84863199,0.04898219){\makebox(0,0)[lb]{\smash{400}}}%
    \put(0.47130053,0.00593518){\makebox(0,0)[lb]{\smash{$X_w$ [m]}}}%
    \put(0.11226575,0.06432998){\makebox(0,0)[lb]{\smash{0}}}%
    \put(0,0){\includegraphics[width=\unitlength,page=10]{trajectory.pdf}}%
    \put(0.09700382,0.16205235){\makebox(0,0)[lb]{\smash{50}}}%
    \put(0,0){\includegraphics[width=\unitlength,page=11]{trajectory.pdf}}%
    \put(0.08308011,0.25977459){\makebox(0,0)[lb]{\smash{100}}}%
    \put(0,0){\includegraphics[width=\unitlength,page=12]{trajectory.pdf}}%
    \put(0.08243494,0.3574972){\makebox(0,0)[lb]{\smash{150}}}%
    \put(0,0){\includegraphics[width=\unitlength,page=13]{trajectory.pdf}}%
    \put(0.08272479,0.45521971){\makebox(0,0)[lb]{\smash{200}}}%
    \put(0,0){\includegraphics[width=\unitlength,page=14]{trajectory.pdf}}%
    \put(0.08207962,0.55294232){\makebox(0,0)[lb]{\smash{250}}}%
    \put(0,0){\includegraphics[width=\unitlength,page=15]{trajectory.pdf}}%
    \put(0.60068548,0.72300762){\color[rgb]{0,0,0}\makebox(0,0)[lb]{\smash{}}}%
    \put(0.04815447,0.2742502){\color[rgb]{0,0,0}\rotatebox{90}{\makebox(0,0)[lb]{\smash{$Z_w$ [m]}}}}%
    \put(0.68680343,0.48596799){\makebox(0,0)[lb]{\smash{retraction, $i$}}}%
    \put(0.28333495,0.35703741){\makebox(0,0)[lb]{\smash{transition, $x$}}}%
    \put(0.65319354,0.25069893){\makebox(0,0)[lb]{\smash{traction, $o$}}}%
    \put(0,0){\includegraphics[width=\unitlength,page=16]{trajectory.pdf}}%
    \put(0.28288596,0.2725669){\color[rgb]{0,0,0}\makebox(0,0)[lb]{\smash{$\vec{v}_w$}}}%
    \put(0,0){\includegraphics[width=\unitlength,page=17]{trajectory.pdf}}%
    \put(0.22302504,0.10768873){\makebox(0,0)[lb]{\smash{}}}%
    \put(0.22727107,0.13076718){\color[rgb]{0,0,0}\makebox(0,0)[lb]{\smash{$\beta$}}}%
    \put(0.27439632,0.11295053){\makebox(0,0)[lb]{\smash{}}}%
    \put(0.2901166,0.11066484){\color[rgb]{0,0,0}\makebox(0,0)[lb]{\smash{$\beta_o$}}}%
    \put(0,0){\includegraphics[width=\unitlength,page=18]{trajectory.pdf}}%
    \put(0.8489167,0.37965349){\makebox(0,0)[lb]{\smash{$t_0$}}}%
    \put(0.31305446,0.44929277){\makebox(0,0)[lb]{\smash{$t_A$}}}%
    \put(0.50505455,0.20609274){\makebox(0,0)[lb]{\smash{$t_B$}}}%
    \put(0.84860506,0.34284645){\makebox(0,0)[lb]{\smash{$t_C$}}}%
  \end{picture}%
\endgroup%

%% file: kinematics_b.pdf_tex
%% Creator: Inkscape inkscape 0.91, www.inkscape.org
%% Accompanies image file 'kinematics_b.pdf' (pdf, eps, ps)
\begingroup%
  \makeatletter%
  \providecommand\color[2][]{%
    \errmessage{(Inkscape) Color is used for the text in Inkscape, but the package 'color.sty' is not loaded}%
    \renewcommand\color[2][]{}%
  }%
  \providecommand\transparent[1]{%
    \errmessage{(Inkscape) Transparency is used (non-zero) for the text in Inkscape, but the package 'transparent.sty' is not loaded}%
    \renewcommand\transparent[1]{}%
  }%
  \providecommand\rotatebox[2]{#2}%
  \ifx\svgwidth\undefined%
    \setlength{\unitlength}{267.18120117bp}%
    \ifx\svgscale\undefined%
      \relax%
    \else%
      \setlength{\unitlength}{\unitlength * \real{\svgscale}}%
    \fi%
  \else%
    \setlength{\unitlength}{\svgwidth}%
  \fi%
  \global\let\svgwidth\undefined%
  \global\let\svgscale\undefined%
  \makeatother%
  \begin{picture}(1,0.65872898)%
    \put(0,0){\includegraphics[width=\unitlength,page=1]{kinematics_b.pdf}}%
    \put(0.01988711,0.00615993){\color[rgb]{0,0,0}\makebox(0,0)[lb]{\smash{$X_w$}}}%
    \put(0.91848373,0.0531381){\color[rgb]{0,0,0}\makebox(0,0)[lb]{\smash{$Y_w$}}}%
    \put(0.47287072,0.60381348){\color[rgb]{0,0,0}\makebox(0,0)[lb]{\smash{$Z_w$}}}%
    \put(0.46387789,0.16045761){\color[rgb]{0,0,0}\makebox(0,0)[lb]{\smash{$\vec{O}$}}}%
    \put(0.26598461,0.50567017){\color[rgb]{0,0,0}\makebox(0,0)[lb]{\smash{$\vec{e}_{r}$}}}%
    \put(0.59438732,0.31323734){\color[rgb]{0,0,0}\makebox(0,0)[lb]{\smash{$\vec{e}_{\phi}$}}}%
    \put(0.30406133,0.12844612){\color[rgb]{0,0,0}\makebox(0,0)[lb]{\smash{$\vec{e}_{\theta}$}}}%
    \put(0.22795231,0.0530576){\color[rgb]{0,0,0}\makebox(0,0)[lb]{\smash{$\phi$}}}%
    \put(0.34344571,0.4354995){\color[rgb]{0,0,0}\makebox(0,0)[lb]{\smash{$\vec{v}_{k,r}$}}}%
    \put(0.61549413,0.20219964){\color[rgb]{0,0,0}\makebox(0,0)[lb]{\smash{$\vec{v}_{k,\tau}$}}}%
    \put(0.58427227,0.28331828){\color[rgb]{0,0,0}\makebox(0,0)[lb]{\smash{$\vec{v}_{k}$}}}%
    \put(0.3995636,0.21274959){\color[rgb]{0,0,0}\makebox(0,0)[lb]{\smash{$r$}}}%
    \put(0.21157151,0.27259457){\color[rgb]{0,0,0}\makebox(0,0)[lb]{\smash{$\vec{v}_w$}}}%
    \put(0.00568123,0.35000626){\color[rgb]{0,0,0}\makebox(0,0)[lb]{\smash{$\vec{v}_a$}}}%
    \put(0.39316812,0.29274965){\color[rgb]{0,0,0}\makebox(0,0)[lb]{\smash{$\chi$}}}%
    \put(0.37883144,0.34411449){\color[rgb]{0,0,0}\makebox(0,0)[lb]{\smash{$\vec{K}$}}}%
    \put(0.32092711,0.37433039){\color[rgb]{0,0,0}\makebox(0,0)[lb]{\smash{$\tau$}}}%
    \put(0.13944558,0.38657597){\color[rgb]{0,0,0}\makebox(0,0)[lb]{\smash{$-\vec{v}_{k}$}}}%
    \put(0.4106599,0.40855374){\color[rgb]{0,0,0}\makebox(0,0)[lb]{\smash{$\theta$}}}%
  \end{picture}%
\endgroup%

%% file: kinematics_c.pdf_tex
%% Creator: Inkscape inkscape 0.91, www.inkscape.org
%% Accompanies image file 'kinematics_c.pdf' (pdf, eps, ps)
\begingroup%
  \makeatletter%
  \providecommand\color[2][]{%
    \errmessage{(Inkscape) Color is used for the text in Inkscape, but the package 'color.sty' is not loaded}%
    \renewcommand\color[2][]{}%
  }%
  \providecommand\transparent[1]{%
    \errmessage{(Inkscape) Transparency is used (non-zero) for the text in Inkscape, but the package 'transparent.sty' is not loaded}%
    \renewcommand\transparent[1]{}%
  }%
  \providecommand\rotatebox[2]{#2}%
  \ifx\svgwidth\undefined%
    \setlength{\unitlength}{267.18120117bp}%
    \ifx\svgscale\undefined%
      \relax%
    \else%
      \setlength{\unitlength}{\unitlength * \real{\svgscale}}%
    \fi%
  \else%
    \setlength{\unitlength}{\svgwidth}%
  \fi%
  \global\let\svgwidth\undefined%
  \global\let\svgscale\undefined%
  \makeatother%
  \begin{picture}(1,0.65872898)%
    \put(0,0){\includegraphics[width=\unitlength,page=1]{kinematics_c.pdf}}%
    \put(0.01988711,0.00615992){\color[rgb]{0,0,0}\makebox(0,0)[lb]{\smash{$X_w$}}}%
    \put(0.91848373,0.05313809){\color[rgb]{0,0,0}\makebox(0,0)[lb]{\smash{$Y_w$}}}%
    \put(0.47287072,0.60381347){\color[rgb]{0,0,0}\makebox(0,0)[lb]{\smash{$Z_w$}}}%
    \put(0.01166967,0.34611352){\color[rgb]{0,0,0}\makebox(0,0)[lb]{\smash{$\vec{v}_a$}}}%
    \put(0.04454833,0.25914861){\color[rgb]{0,0,0}\makebox(0,0)[lb]{\smash{$\vec{v}_{a,\tau}$}}}%
    \put(0.30479408,0.38432997){\color[rgb]{0,0,0}\makebox(0,0)[lb]{\smash{$\vec{v}_{a,r}$}}}%
    \put(0.2481727,0.60653988){\color[rgb]{0,0,0}\makebox(0,0)[lb]{\smash{$\vec{F}_a$}}}%
    \put(0.36304648,0.60552624){\color[rgb]{0,0,0}\makebox(0,0)[lb]{\smash{$\vec{L}$}}}%
    \put(0.27844288,0.28954114){\color[rgb]{0,0,0}\makebox(0,0)[lb]{\smash{$\vec{D}$}}}%
    \put(0.4106599,0.40855373){\color[rgb]{0,0,0}\makebox(0,0)[lb]{\smash{$\theta$}}}%
    \put(0.3995636,0.21274959){\color[rgb]{0,0,0}\makebox(0,0)[lb]{\smash{$r$}}}%
    \put(0.22795231,0.05305759){\color[rgb]{0,0,0}\makebox(0,0)[lb]{\smash{$\phi$}}}%
    \put(0.37150864,0.15972677){\color[rgb]{0,0,0}\makebox(0,0)[lb]{\smash{$\beta$}}}%
    \put(0.48983679,0.0617913){\color[rgb]{0,0,0}\makebox(0,0)[lb]{\smash{$\vec{F}_t$}}}%
    \put(0.46387789,0.1604576){\color[rgb]{0,0,0}\makebox(0,0)[lb]{\smash{$\vec{O}$}}}%
    \put(0.39330277,0.34219288){\color[rgb]{0,0,0}\makebox(0,0)[lb]{\smash{$\vec{K}$}}}%
  \end{picture}%
\endgroup%

%% file: velocity_force_diagram.pdf_tex
%% Creator: Inkscape inkscape 0.91, www.inkscape.org
%% Accompanies image file 'velocity_force_diagram.pdf' (pdf, eps, ps)
\begingroup%
  \makeatletter%
  \providecommand\color[2][]{%
    \errmessage{(Inkscape) Color is used for the text in Inkscape, but the package 'color.sty' is not loaded}%
    \renewcommand\color[2][]{}%
  }%
  \providecommand\transparent[1]{%
    \errmessage{(Inkscape) Transparency is used (non-zero) for the text in Inkscape, but the package 'transparent.sty' is not loaded}%
    \renewcommand\transparent[1]{}%
  }%
  \providecommand\rotatebox[2]{#2}%
  \ifx\svgwidth\undefined%
    \setlength{\unitlength}{176bp}%
    \ifx\svgscale\undefined%
      \relax%
    \else%
      \setlength{\unitlength}{\unitlength * \real{\svgscale}}%
    \fi%
  \else%
    \setlength{\unitlength}{\svgwidth}%
  \fi%
  \global\let\svgwidth\undefined%
  \global\let\svgscale\undefined%
  \makeatother%
  \begin{picture}(1,0.45454545)%
    \put(0,0){\includegraphics[width=\unitlength,page=1]{velocity_force_diagram.pdf}}%
    \put(0.7028182,0.28259261){\color[rgb]{0,0,0}\makebox(0,0)[lb]{\smash{$\vec{v}_{a,r}$}}}%
    \put(0.54445139,0.00016241){\color[rgb]{0,0,0}\makebox(0,0)[lb]{\smash{$\vec{v}_{a,\tau}$}}}%
    \put(0.70406207,-0.00009685){\color[rgb]{0,0,0}\makebox(0,0)[lb]{\smash{$\vec{v}_a$}}}%
    \put(0.94304592,0.44221407){\color[rgb]{0,0,0}\makebox(0,0)[lb]{\smash{$\vec{L}$}}}%
    \put(0.15842496,0.28577324){\color[rgb]{0,0,0}\makebox(0,0)[lb]{\smash{$\vec{e}_r$}}}%
    \put(0.65089955,0.16704335){\color[rgb]{0,0,0}\makebox(0,0)[lb]{\smash{$\vec{D}$}}}%
    \put(0.98608174,0.30027436){\color[rgb]{0,0,0}\makebox(0,0)[lb]{\smash{$\vec{F}_a$}}}%
    \put(0.4216103,0.33082953){\color[rgb]{0,0,0}\makebox(0,0)[lb]{\smash{$r$}}}%
    \put(0.67248144,-0.02224006){\color[rgb]{0,0,0}\makebox(0,0)[lb]{\smash{$\alpha$}}}%
  \end{picture}%
\endgroup%

%% file: tethermass.pdf_tex
%% Creator: Inkscape inkscape 0.91, www.inkscape.org
%% Accompanies image file 'tethermass.pdf' (pdf, eps, ps)
\begingroup%
  \makeatletter%
  \providecommand\color[2][]{%
    \errmessage{(Inkscape) Color is used for the text in Inkscape, but the package 'color.sty' is not loaded}%
    \renewcommand\color[2][]{}%
  }%
  \providecommand\transparent[1]{%
    \errmessage{(Inkscape) Transparency is used (non-zero) for the text in Inkscape, but the package 'transparent.sty' is not loaded}%
    \renewcommand\transparent[1]{}%
  }%
  \providecommand\rotatebox[2]{#2}%
  \ifx\svgwidth\undefined%
    \setlength{\unitlength}{1583.29130859bp}%
    \ifx\svgscale\undefined%
      \relax%
    \else%
      \setlength{\unitlength}{\unitlength * \real{\svgscale}}%
    \fi%
  \else%
    \setlength{\unitlength}{\svgwidth}%
  \fi%
  \global\let\svgwidth\undefined%
  \global\let\svgscale\undefined%
  \makeatother%
  \begin{picture}(1,0.59976723)%
    \put(0,0){\includegraphics[width=\unitlength,page=1]{tethermass.pdf}}%
    \put(0.26586287,0.15276722){\color[rgb]{0,0,0}\makebox(0,0)[lb]{\smash{$\theta$}}}%
    \put(0.50287324,0.1399927){\color[rgb]{0,0,0}\makebox(0,0)[lb]{\smash{$m_t \vec{g}$}}}%
    \put(0.55249702,0.18336513){\color[rgb]{0,0,0}\makebox(0,0)[lb]{\smash{$\sin\theta\, m_t g \vec{e}_{\theta}$}}}%
    \put(0.61070895,0.17515855){\color[rgb]{0,0,0}\makebox(0,0)[lb]{\smash{}}}%
    \put(0.2640759,0.54023504){\color[rgb]{0,0,0}\makebox(0,0)[lb]{\smash{$Z_w$}}}%
    \put(0.34156572,0.55214417){\color[rgb]{0,0,0}\makebox(0,0)[lb]{\smash{}}}%
    \put(0.89035737,0.55052253){\color[rgb]{0,0,0}\makebox(0,0)[lb]{\smash{$\vec{F}_{t,r}$}}}%
    \put(0.1367216,-0.00572549){\color[rgb]{0,0,0}\makebox(0,0)[lb]{\smash{$\vec{F}_{tg,r}$}}}%
    \put(0,0){\includegraphics[width=\unitlength,page=2]{tethermass.pdf}}%
    \put(0.81786264,0.16563025){\color[rgb]{0,0,0}\makebox(0,0)[lb]{\smash{$\vec{g}$}}}%
    \put(0.86719316,0.59816112){\color[rgb]{0,0,0}\makebox(0,0)[lb]{\smash{$\vec{F}_{t}$}}}%
    \put(0,0){\includegraphics[width=\unitlength,page=3]{tethermass.pdf}}%
    \put(0.7047621,0.49587868){\color[rgb]{0,0,0}\makebox(0,0)[rb]{\smash{$\vec{F}_{t,\tau}$}}}%
    \put(0.1062826,0.05678114){\color[rgb]{0,0,0}\makebox(0,0)[rb]{\smash{$\vec{F}_{tg}$}}}%
    \put(0.22563058,0.1551193){\color[rgb]{0,0,0}\makebox(0,0)[rb]{\smash{$\vec{F}_{tg,\tau}$}}}%
    \put(0.7298829,0.41756517){\color[rgb]{0,0,0}\makebox(0,0)[lb]{\smash{$\vec{K}$}}}%
    \put(0.25761114,0.07140357){\color[rgb]{0,0,0}\makebox(0,0)[lb]{\smash{$\vec{O}$}}}%
    \put(0.61510956,0.05305386){\color[rgb]{0,0,0}\makebox(0,0)[lb]{\smash{}}}%
    \put(0.56538799,0.35755073){\color[rgb]{0,0,0}\makebox(0,0)[lb]{\smash{$r$}}}%
    \put(0,0){\includegraphics[width=\unitlength,page=4]{tethermass.pdf}}%
    \put(0.46607299,0.25592173){\color[rgb]{0,0,0}\makebox(0,0)[rb]{\smash{$-\cos\theta\, m_t g \vec{e}_r$}}}%
  \end{picture}%
\endgroup%

%% file: forcesgravity.pdf_tex
%% Creator: Inkscape inkscape 0.91, www.inkscape.org
%% Accompanies image file 'forcesgravity.pdf' (pdf, eps, ps)
\begingroup%
  \makeatletter%
  \providecommand\color[2][]{%
    \errmessage{(Inkscape) Color is used for the text in Inkscape, but the package 'color.sty' is not loaded}%
    \renewcommand\color[2][]{}%
  }%
  \providecommand\transparent[1]{%
    \errmessage{(Inkscape) Transparency is used (non-zero) for the text in Inkscape, but the package 'transparent.sty' is not loaded}%
    \renewcommand\transparent[1]{}%
  }%
  \providecommand\rotatebox[2]{#2}%
  \ifx\svgwidth\undefined%
    \setlength{\unitlength}{1583.29130859bp}%
    \ifx\svgscale\undefined%
      \relax%
    \else%
      \setlength{\unitlength}{\unitlength * \real{\svgscale}}%
    \fi%
  \else%
    \setlength{\unitlength}{\svgwidth}%
  \fi%
  \global\let\svgwidth\undefined%
  \global\let\svgscale\undefined%
  \makeatother%
  \begin{picture}(1,0.65685954)%
    \put(0,0){\includegraphics[width=\unitlength,page=1]{forcesgravity.pdf}}%
    \put(0.52764904,0.48069366){\color[rgb]{0,0,0}\makebox(0,0)[lb]{\smash{$\theta$}}}%
    \put(0.50557298,0.03718064){\color[rgb]{0,0,0}\makebox(0,0)[lb]{\smash{$m_t \vec{g}$}}}%
    \put(0.61087275,0.07741828){\color[rgb]{0,0,0}\makebox(0,0)[lb]{\smash{}}}%
    \put(0.26099303,0.61537983){\color[rgb]{0,0,0}\makebox(0,0)[lb]{\smash{$Z_w$}}}%
    \put(0.34172956,0.45440392){\color[rgb]{0,0,0}\makebox(0,0)[lb]{\smash{}}}%
    \put(0.8498912,0.41595701){\color[rgb]{0,0,0}\makebox(0,0)[lb]{\smash{$\vec{F}^*_g$}}}%
    \put(0,0){\includegraphics[width=\unitlength,page=2]{forcesgravity.pdf}}%
    \put(0.08116085,0.31623847){\color[rgb]{0,0,0}\makebox(0,0)[lb]{\smash{$\vec{g}$}}}%
    \put(0,0){\includegraphics[width=\unitlength,page=3]{forcesgravity.pdf}}%
    \put(0.74298512,0.32758298){\color[rgb]{0,0,0}\makebox(0,0)[lb]{\smash{$\vec{K}$}}}%
    \put(0.61527338,-0.04468638){\color[rgb]{0,0,0}\makebox(0,0)[lb]{\smash{}}}%
    \put(0.52103563,0.22471809){\color[rgb]{0,0,0}\makebox(0,0)[lb]{\smash{$r$}}}%
    \put(0,0){\includegraphics[width=\unitlength,page=4]{forcesgravity.pdf}}%
    \put(0.86685646,0.60352726){\color[rgb]{0,0,0}\makebox(0,0)[lb]{\smash{$\vec{F}_a$}}}%
    \put(0.87155432,0.51664321){\color[rgb]{0,0,0}\makebox(0,0)[lb]{\smash{$m\vec{g}$}}}%
    \put(0.58445595,0.29239104){\color[rgb]{0,0,0}\makebox(0,0)[lb]{\smash{$\vec{F}^*_t$}}}%
    \put(0,0){\includegraphics[width=\unitlength,page=5]{forcesgravity.pdf}}%
    \put(0.45996548,0.15275368){\color[rgb]{0,0,0}\makebox(0,0)[rb]{\smash{$-\cos\theta\, m_t g \vec{e}_r$}}}%
    \put(0.20619752,0.02175305){\color[rgb]{0,0,0}\makebox(0,0)[lb]{\smash{$\vec{O}$}}}%
    \put(0.55015697,0.08900948){\color[rgb]{0,0,0}\makebox(0,0)[lb]{\smash{$\sin\theta\, m_t g \vec{e}_{\theta}$}}}%
    \put(0.59850001,0.1699988){\color[rgb]{0,0,0}\makebox(0,0)[lb]{\smash{$\vec{F}_t$}}}%
  \end{picture}%
\endgroup%

%% file: kinematics_d.pdf_tex
%% Creator: Inkscape inkscape 0.91, www.inkscape.org
%% Accompanies image file 'kinematics_d.pdf' (pdf, eps, ps)
\begingroup%
  \makeatletter%
  \providecommand\color[2][]{%
    \errmessage{(Inkscape) Color is used for the text in Inkscape, but the package 'color.sty' is not loaded}%
    \renewcommand\color[2][]{}%
  }%
  \providecommand\transparent[1]{%
    \errmessage{(Inkscape) Transparency is used (non-zero) for the text in Inkscape, but the package 'transparent.sty' is not loaded}%
    \renewcommand\transparent[1]{}%
  }%
  \providecommand\rotatebox[2]{#2}%
  \ifx\svgwidth\undefined%
    \setlength{\unitlength}{267.18120117bp}%
    \ifx\svgscale\undefined%
      \relax%
    \else%
      \setlength{\unitlength}{\unitlength * \real{\svgscale}}%
    \fi%
  \else%
    \setlength{\unitlength}{\svgwidth}%
  \fi%
  \global\let\svgwidth\undefined%
  \global\let\svgscale\undefined%
  \makeatother%
  \begin{picture}(1,0.74855566)%
    \put(0,0){\includegraphics[width=\unitlength,page=1]{kinematics_d.pdf}}%
    \put(0.01988711,0.00615992){\color[rgb]{0,0,0}\makebox(0,0)[lb]{\smash{$X_w$}}}%
    \put(0.91848373,0.05313809){\color[rgb]{0,0,0}\makebox(0,0)[lb]{\smash{$Y_w$}}}%
    \put(0.47287072,0.60381347){\color[rgb]{0,0,0}\makebox(0,0)[lb]{\smash{$Z_w$}}}%
    \put(0.01166967,0.34611352){\color[rgb]{0,0,0}\makebox(0,0)[lb]{\smash{$\vec{v}_a$}}}%
    \put(0.04454833,0.25914861){\color[rgb]{0,0,0}\makebox(0,0)[lb]{\smash{$\vec{v}_{a,\tau}$}}}%
    \put(0.30479408,0.38432997){\color[rgb]{0,0,0}\makebox(0,0)[lb]{\smash{$\vec{v}_{a,r}$}}}%
    \put(0.21651225,0.70989869){\color[rgb]{0,0,0}\makebox(0,0)[lb]{\smash{$\vec{F}_a$}}}%
    \put(0.36957649,0.70728158){\color[rgb]{0,0,0}\makebox(0,0)[lb]{\smash{$\vec{L}$}}}%
    \put(0.27844288,0.28954114){\color[rgb]{0,0,0}\makebox(0,0)[lb]{\smash{$\vec{D}$}}}%
    \put(0.3995636,0.21274959){\color[rgb]{0,0,0}\makebox(0,0)[lb]{\smash{$r$}}}%
    \put(0.22795231,0.05305759){\color[rgb]{0,0,0}\makebox(0,0)[lb]{\smash{$\phi$}}}%
    \put(0.22272772,0.59876883){\color[rgb]{0,0,0}\makebox(0,0)[lb]{\smash{$\vec{F}^*_g$}}}%
    \put(0.4106599,0.40855373){\color[rgb]{0,0,0}\makebox(0,0)[lb]{\smash{$\theta$}}}%
    \put(0.07345217,0.53329984){\color[rgb]{0,0,0}\makebox(0,0)[lb]{\smash{$\vec{g}$}}}%
    \put(0.48983679,0.0617913){\color[rgb]{0,0,0}\makebox(0,0)[lb]{\smash{$\vec{F}^*_t$}}}%
    \put(0.39330277,0.34219288){\color[rgb]{0,0,0}\makebox(0,0)[lb]{\smash{$\vec{K}$}}}%
    \put(0.46387789,0.1604576){\color[rgb]{0,0,0}\makebox(0,0)[lb]{\smash{$\vec{O}$}}}%
  \end{picture}%
\endgroup%

%% file: system_components.pdf_tex
%% Creator: Inkscape inkscape 0.91, www.inkscape.org
%% Accompanies image file 'system_components.pdf' (pdf, eps, ps)
\begingroup%
  \makeatletter%
  \providecommand\color[2][]{%
    \errmessage{(Inkscape) Color is used for the text in Inkscape, but the package 'color.sty' is not loaded}%
    \renewcommand\color[2][]{}%
  }%
  \providecommand\transparent[1]{%
    \errmessage{(Inkscape) Transparency is used (non-zero) for the text in Inkscape, but the package 'transparent.sty' is not loaded}%
    \renewcommand\transparent[1]{}%
  }%
  \providecommand\rotatebox[2]{#2}%
  \ifx\svgwidth\undefined%
    \setlength{\unitlength}{452bp}%
    \ifx\svgscale\undefined%
      \relax%
    \else%
      \setlength{\unitlength}{\unitlength * \real{\svgscale}}%
    \fi%
  \else%
    \setlength{\unitlength}{\svgwidth}%
  \fi%
  \global\let\svgwidth\undefined%
  \global\let\svgscale\undefined%
  \makeatother%
  \begin{picture}(1,0.54867257)%
    \put(0,0){\includegraphics[width=\unitlength,page=1]{system_components.pdf}}%
    \put(0.89036725,0.52984974){\color[rgb]{0,0,0}\makebox(0,0)[lb]{\smash{Kite}}}%
    \put(0.6494745,0.47811138){\color[rgb]{0,0,0}\makebox(0,0)[lt]{\begin{minipage}{0.14352536\unitlength}\raggedright Bridle Line\\ System\end{minipage}}}%
    \put(0.51496542,0.45020397){\color[rgb]{0,0,0}\makebox(0,0)[lt]{\begin{minipage}{0.09618946\unitlength}\raggedright Kite\\ Control\\ Unit\end{minipage}}}%
    \put(0.36687067,0.40717535){\color[rgb]{0,0,0}\makebox(0,0)[lt]{\begin{minipage}{0.10504145\unitlength}\raggedright Traction\\ Tether\end{minipage}}}%
    \put(0.18013072,0.34889818){\color[rgb]{0,0,0}\makebox(0,0)[lt]{\begin{minipage}{0.18937216\unitlength}\raggedright Drum/Generator\\ Module\\ \end{minipage}}}%
    \put(0.0864035,0.32865137){\color[rgb]{0,0,0}\makebox(0,0)[lt]{\begin{minipage}{0.11328611\unitlength}\raggedright Control\\ Center\end{minipage}}}%
    \put(0.00075926,0.29931836){\color[rgb]{0,0,0}\makebox(0,0)[lt]{\begin{minipage}{0.10691431\unitlength}\raggedright Battery\\ Module\end{minipage}}}%
    \put(0.00043241,0.18018322){\color[rgb]{0,0,0}\makebox(0,0)[lt]{\begin{minipage}{0.12787054\unitlength}\raggedright Power\\ Electronics\end{minipage}}}%
    \put(0.14245878,0.08107126){\color[rgb]{0,0,0}\makebox(0,0)[lt]{\begin{minipage}{0.12911426\unitlength}\raggedright \bf Ground \\ Station\end{minipage}}}%
    \put(0.80462708,0.25521625){\color[rgb]{0,0,0}\makebox(0,0)[lt]{\begin{minipage}{0.17730621\unitlength}\raggedright \bf Airborne\\ Components\end{minipage}}}%
    \put(0.16475954,0.47797988){\color[rgb]{0,0,0}\makebox(0,0)[lt]{\begin{minipage}{0.197019\unitlength}\raggedright Wind Meter\end{minipage}}}%
    \put(0.79990396,0.53454699){\color[rgb]{0,0,0}\makebox(0,0)[lt]{\begin{minipage}{0.11849457\unitlength}\raggedright Sensor Unit\end{minipage}}}%
    \put(0.58039229,0.18471928){\color[rgb]{0,0,0}\makebox(0,0)[lt]{\begin{minipage}{0.11781346\unitlength}\raggedright Launch Mast\end{minipage}}}%
  \end{picture}%
\endgroup%

%% file: cr_HYDRA.pdf_tex
%% Creator: Inkscape inkscape 0.91, www.inkscape.org
%% Accompanies image file 'cr_HYDRA.pdf' (pdf, eps, ps)
\begingroup%
  \makeatletter%
  \providecommand\color[2][]{%
    \errmessage{(Inkscape) Color is used for the text in Inkscape, but the package 'color.sty' is not loaded}%
    \renewcommand\color[2][]{}%
  }%
  \providecommand\transparent[1]{%
    \errmessage{(Inkscape) Transparency is used (non-zero) for the text in Inkscape, but the package 'transparent.sty' is not loaded}%
    \renewcommand\transparent[1]{}%
  }%
  \providecommand\rotatebox[2]{#2}%
  \ifx\svgwidth\undefined%
    \setlength{\unitlength}{576bp}%
    \ifx\svgscale\undefined%
      \relax%
    \else%
      \setlength{\unitlength}{\unitlength * \real{\svgscale}}%
    \fi%
  \else%
    \setlength{\unitlength}{\svgwidth}%
  \fi%
  \global\let\svgwidth\undefined%
  \global\let\svgscale\undefined%
  \makeatother%
  \begin{picture}(1,0.75)%
    \put(0,0){\includegraphics[width=\unitlength,page=1]{cr_HYDRA.pdf}}%
    \put(0.07131436,0.37419825){\color[rgb]{0,0,0}\rotatebox{90}{\makebox(0,0)[b]{\smash{Resultant force coefficient $C_R$ [-]}}}}%
    \put(0.15113299,0.6365579){\color[rgb]{0,0,0}\makebox(0,0)[lb]{\smash{\textbf{Strong wind}}}}%
  \end{picture}%
\endgroup%

%% file: cr_V3.pdf_tex
%% Creator: Inkscape inkscape 0.91, www.inkscape.org
%% Accompanies image file 'cr_HYDRA.pdf' (pdf, eps, ps)
\begingroup%
  \makeatletter%
  \providecommand\color[2][]{%
    \errmessage{(Inkscape) Color is used for the text in Inkscape, but the package 'color.sty' is not loaded}%
    \renewcommand\color[2][]{}%
  }%
  \providecommand\transparent[1]{%
    \errmessage{(Inkscape) Transparency is used (non-zero) for the text in Inkscape, but the package 'transparent.sty' is not loaded}%
    \renewcommand\transparent[1]{}%
  }%
  \providecommand\rotatebox[2]{#2}%
  \ifx\svgwidth\undefined%
    \setlength{\unitlength}{576bp}%
    \ifx\svgscale\undefined%
      \relax%
    \else%
      \setlength{\unitlength}{\unitlength * \real{\svgscale}}%
    \fi%
  \else%
    \setlength{\unitlength}{\svgwidth}%
  \fi%
  \global\let\svgwidth\undefined%
  \global\let\svgscale\undefined%
  \makeatother%
  \begin{picture}(1,0.75)%
    \put(0,0){\includegraphics[width=\unitlength,page=1]{cr_V3.pdf}}%
    \put(0.51250956,0.01601106){\color[rgb]{0,0,0}\makebox(0,0)[b]{\smash{Time [s]}}}%
    \put(0.07131436,0.37419825){\color[rgb]{0,0,0}\rotatebox{90}{\makebox(0,0)[b]{\smash{Resultant force coefficient $C_R$ [-]}}}}%
    \put(0.15113299,0.6365579){\color[rgb]{0,0,0}\makebox(0,0)[lb]{\smash{\textbf{Moderate wind}}}}%
  \end{picture}%
\endgroup%

%% file: tangentialequilibrium.pdf_tex
%% Creator: Inkscape inkscape 0.91, www.inkscape.org
%% Accompanies image file 'tangentialequilibrium.pdf' (pdf, eps, ps)
\begingroup%
  \makeatletter%
  \providecommand\color[2][]{%
    \errmessage{(Inkscape) Color is used for the text in Inkscape, but the package 'color.sty' is not loaded}%
    \renewcommand\color[2][]{}%
  }%
  \providecommand\transparent[1]{%
    \errmessage{(Inkscape) Transparency is used (non-zero) for the text in Inkscape, but the package 'transparent.sty' is not loaded}%
    \renewcommand\transparent[1]{}%
  }%
  \providecommand\rotatebox[2]{#2}%
  \ifx\svgwidth\undefined%
    \setlength{\unitlength}{232bp}%
    \ifx\svgscale\undefined%
      \relax%
    \else%
      \setlength{\unitlength}{\unitlength * \real{\svgscale}}%
    \fi%
  \else%
    \setlength{\unitlength}{\svgwidth}%
  \fi%
  \global\let\svgwidth\undefined%
  \global\let\svgscale\undefined%
  \makeatother%
  \begin{picture}(1,0.68965517)%
    \put(0,0){\includegraphics[width=\unitlength,page=1]{tangentialequilibrium.pdf}}%
    \put(0.58356308,0.26341517){\color[rgb]{0,0,0}\makebox(0,0)[lb]{\smash{$\chi$}}}%
    \put(0.7887175,0.28350636){\color[rgb]{0,0,0}\makebox(0,0)[lb]{\smash{$\vec{e}_{\phi}$}}}%
    \put(0.55824996,0.07824787){\color[rgb]{0,0,0}\makebox(0,0)[lb]{\smash{$\vec{e}_{\theta}$}}}%
    \put(0.00131423,0.2149662){\color[rgb]{0,0,0}\makebox(0,0)[lb]{\smash{$\vec{v}_{a,\tau}$}}}%
    \put(0.49932309,0.33080798){\color[rgb]{0,0,0}\makebox(0,0)[lb]{\smash{$\vec{K}$}}}%
    \put(0.9628186,0.26069816){\color[rgb]{0,0,0}\makebox(0,0)[lb]{\smash{$\vec{v}_{k,\tau}$}}}%
    \put(0.18346563,0.23108){\color[rgb]{0,0,0}\makebox(0,0)[lb]{\smash{$\vec{D}_{\tau}$}}}%
    \put(0.39509394,0.44755166){\color[rgb]{0,0,0}\makebox(0,0)[lb]{\smash{$\tau$}}}%
    \put(0.45335789,0.14691221){\color[rgb]{0,0,0}\makebox(0,0)[lb]{\smash{$\vec{v}_{w,\tau}$}}}%
    \put(0.55770401,0.59096178){\color[rgb]{0,0,0}\makebox(0,0)[lb]{\smash{$\vec{F}_{a,\tau}$}}}%
    \put(0.89203729,0.68438703){\color[rgb]{0,0,0}\makebox(0,0)[lb]{\smash{$\vec{L}_{\tau}$}}}%
    \put(0.55868326,0.01298629){\color[rgb]{0,0,0}\makebox(0,0)[lb]{\smash{$\vec{F}^*_{g,\tau}$}}}%
    \put(0.10474612,0.34571602){\color[rgb]{0,0,0}\makebox(0,0)[lb]{\smash{$\delta$}}}%
    \put(0.94074271,0.63015887){\color[rgb]{0,0,0}\makebox(0,0)[lb]{\smash{$\vec{L}_{\tau2}$}}}%
    \put(0.92655456,0.32041443){\color[rgb]{0,0,0}\makebox(0,0)[lb]{\smash{$\vec{L}_{\tau1}$}}}%
  \end{picture}%
\endgroup%

%% file: loverd_HYDRA.pdf_tex
%% Creator: Inkscape inkscape 0.91, www.inkscape.org
%% Accompanies image file 'loverd_HYDRA.pdf' (pdf, eps, ps)
\begingroup%
  \makeatletter%
  \providecommand\color[2][]{%
    \errmessage{(Inkscape) Color is used for the text in Inkscape, but the package 'color.sty' is not loaded}%
    \renewcommand\color[2][]{}%
  }%
  \providecommand\transparent[1]{%
    \errmessage{(Inkscape) Transparency is used (non-zero) for the text in Inkscape, but the package 'transparent.sty' is not loaded}%
    \renewcommand\transparent[1]{}%
  }%
  \providecommand\rotatebox[2]{#2}%
  \ifx\svgwidth\undefined%
    \setlength{\unitlength}{576bp}%
    \ifx\svgscale\undefined%
      \relax%
    \else%
      \setlength{\unitlength}{\unitlength * \real{\svgscale}}%
    \fi%
  \else%
    \setlength{\unitlength}{\svgwidth}%
  \fi%
  \global\let\svgwidth\undefined%
  \global\let\svgscale\undefined%
  \makeatother%
  \begin{picture}(1,0.75)%
    \put(0,0){\includegraphics[width=\unitlength,page=1]{loverd_HYDRA.pdf}}%
    \put(0.20900235,0.605785){\color[rgb]{0,0,0}\makebox(0,0)[lb]{\smash{$\kappa$, initial estimate}}}%
    \put(0.20900235,0.57339085){\color[rgb]{0,0,0}\makebox(0,0)[lb]{\smash{$L/D$, incl. tether drag}}}%
    \put(0.20900814,0.54099242){\color[rgb]{0,0,0}\makebox(0,0)[lb]{\smash{$L/D_k$, excl. tether drag}}}%
    \put(0.07908264,0.37575192){\color[rgb]{0,0,0}\rotatebox{90}{\makebox(0,0)[b]{\smash{Lift-to-Drag Ratio [-]}}}}%
    \put(0.15106969,0.63801379){\color[rgb]{0,0,0}\makebox(0,0)[lb]{\smash{\textbf{Strong wind}}}}%
  \end{picture}%
\endgroup%

%% file: loverd_V3.pdf_tex
%% Creator: Inkscape inkscape 0.91, www.inkscape.org
%% Accompanies image file 'loverd_V3.pdf' (pdf, eps, ps)
\begingroup%
  \makeatletter%
  \providecommand\color[2][]{%
    \errmessage{(Inkscape) Color is used for the text in Inkscape, but the package 'color.sty' is not loaded}%
    \renewcommand\color[2][]{}%
  }%
  \providecommand\transparent[1]{%
    \errmessage{(Inkscape) Transparency is used (non-zero) for the text in Inkscape, but the package 'transparent.sty' is not loaded}%
    \renewcommand\transparent[1]{}%
  }%
  \providecommand\rotatebox[2]{#2}%
  \ifx\svgwidth\undefined%
    \setlength{\unitlength}{576bp}%
    \ifx\svgscale\undefined%
      \relax%
    \else%
      \setlength{\unitlength}{\unitlength * \real{\svgscale}}%
    \fi%
  \else%
    \setlength{\unitlength}{\svgwidth}%
  \fi%
  \global\let\svgwidth\undefined%
  \global\let\svgscale\undefined%
  \makeatother%
  \begin{picture}(1,0.75)%
    \put(0,0){\includegraphics[width=\unitlength,page=1]{loverd_V3.pdf}}%
    \put(0.51200644,0.60589976){\color[rgb]{0,0,0}\makebox(0,0)[lb]{\smash{$\kappa$, initial estimate}}}%
    \put(0.51200644,0.57350561){\color[rgb]{0,0,0}\makebox(0,0)[lb]{\smash{$L/D$, incl. tether drag}}}%
    \put(0.51201222,0.54110718){\color[rgb]{0,0,0}\makebox(0,0)[lb]{\smash{$L/D_k$, excl. tether drag}}}%
    \put(0.07908264,0.37575195){\color[rgb]{0,0,0}\rotatebox{90}{\makebox(0,0)[b]{\smash{Lift-to-Drag Ratio [-]}}}}%
    \put(0.51250956,0.01601105){\color[rgb]{0,0,0}\makebox(0,0)[b]{\smash{Time [s]}}}%
    \put(0.45007144,0.63823181){\color[rgb]{0,0,0}\makebox(0,0)[lb]{\smash{\textbf{Moderate wind}}}}%
  \end{picture}%
\endgroup%

%% file: kitepathgrnddistheight_HYDRA.pdf_tex
%% Creator: Inkscape inkscape 0.91, www.inkscape.org
%% Accompanies image file 'kitepathgrnddistheight_HYDRA_weg.pdf' (pdf, eps, ps)
\begingroup%
  \makeatletter%
  \providecommand\color[2][]{%
    \errmessage{(Inkscape) Color is used for the text in Inkscape, but the package 'color.sty' is not loaded}%
    \renewcommand\color[2][]{}%
  }%
  \providecommand\transparent[1]{%
    \errmessage{(Inkscape) Transparency is used (non-zero) for the text in Inkscape, but the package 'transparent.sty' is not loaded}%
    \renewcommand\transparent[1]{}%
  }%
  \providecommand\rotatebox[2]{#2}%
  \ifx\svgwidth\undefined%
    \setlength{\unitlength}{576bp}%
    \ifx\svgscale\undefined%
      \relax%
    \else%
      \setlength{\unitlength}{\unitlength * \real{\svgscale}}%
    \fi%
  \else%
    \setlength{\unitlength}{\svgwidth}%
  \fi%
  \global\let\svgwidth\undefined%
  \global\let\svgscale\undefined%
  \makeatother%
  \begin{picture}(1,0.75)%
    \put(0,0){\includegraphics[width=\unitlength,page=1]{kitepathgrnddistheight_HYDRA.pdf}}%
    \put(0.06634026,0.28346946){\color[rgb]{0,0,0}\rotatebox{90}{\makebox(0,0)[lb]{\smash{Kite height [m]}}}}%
    \put(0.20850694,0.16406583){\color[rgb]{0,0,0}\makebox(0,0)[lb]{\smash{Gravity incl.}}}%
    \put(0.20850694,0.13182521){\color[rgb]{0,0,0}\makebox(0,0)[lb]{\smash{Gravity excl.}}}%
    \put(0.20850694,0.0995846){\color[rgb]{0,0,0}\makebox(0,0)[lb]{\smash{Experiment}}}%
    \put(0.15190992,0.19666939){\color[rgb]{0,0,0}\makebox(0,0)[lb]{\smash{\textbf{Strong wind}}}}%
  \end{picture}%
\endgroup%

%% file: kitepathgrnddistheight_V3.pdf_tex
%% Creator: Inkscape inkscape 0.91, www.inkscape.org
%% Accompanies image file 'kitepathgrnddistheight_V3_weg.pdf' (pdf, eps, ps)
\begingroup%
  \makeatletter%
  \providecommand\color[2][]{%
    \errmessage{(Inkscape) Color is used for the text in Inkscape, but the package 'color.sty' is not loaded}%
    \renewcommand\color[2][]{}%
  }%
  \providecommand\transparent[1]{%
    \errmessage{(Inkscape) Transparency is used (non-zero) for the text in Inkscape, but the package 'transparent.sty' is not loaded}%
    \renewcommand\transparent[1]{}%
  }%
  \providecommand\rotatebox[2]{#2}%
  \ifx\svgwidth\undefined%
    \setlength{\unitlength}{576bp}%
    \ifx\svgscale\undefined%
      \relax%
    \else%
      \setlength{\unitlength}{\unitlength * \real{\svgscale}}%
    \fi%
  \else%
    \setlength{\unitlength}{\svgwidth}%
  \fi%
  \global\let\svgwidth\undefined%
  \global\let\svgscale\undefined%
  \makeatother%
  \begin{picture}(1,0.75)%
    \put(0,0){\includegraphics[width=\unitlength,page=1]{kitepathgrnddistheight_V3.pdf}}%
    \put(0.34409741,0.01594462){\color[rgb]{0,0,0}\makebox(0,0)[lb]{\smash{Horizontal kite distance [m]}}}%
    \put(0.06634026,0.28346946){\color[rgb]{0,0,0}\rotatebox{90}{\makebox(0,0)[lb]{\smash{Kite height [m]}}}}%
    \put(0.20850694,0.16406583){\color[rgb]{0,0,0}\makebox(0,0)[lb]{\smash{Gravity incl.}}}%
    \put(0.20850694,0.13182521){\color[rgb]{0,0,0}\makebox(0,0)[lb]{\smash{Gravity excl.}}}%
    \put(0.20850694,0.0995846){\color[rgb]{0,0,0}\makebox(0,0)[lb]{\smash{Experiment}}}%
    \put(0.15190992,0.19666939){\color[rgb]{0,0,0}\makebox(0,0)[lb]{\smash{\textbf{Moderate wind}}}}%
  \end{picture}%
\endgroup%

%% file: kitepathelevazi_HYDRA.pdf_tex
%% Creator: Inkscape inkscape 0.91, www.inkscape.org
%% Accompanies image file 'kitepathelevazi_HYDRA_weg.pdf' (pdf, eps, ps)
\begingroup%
  \makeatletter%
  \providecommand\color[2][]{%
    \errmessage{(Inkscape) Color is used for the text in Inkscape, but the package 'color.sty' is not loaded}%
    \renewcommand\color[2][]{}%
  }%
  \providecommand\transparent[1]{%
    \errmessage{(Inkscape) Transparency is used (non-zero) for the text in Inkscape, but the package 'transparent.sty' is not loaded}%
    \renewcommand\transparent[1]{}%
  }%
  \providecommand\rotatebox[2]{#2}%
  \ifx\svgwidth\undefined%
    \setlength{\unitlength}{576bp}%
    \ifx\svgscale\undefined%
      \relax%
    \else%
      \setlength{\unitlength}{\unitlength * \real{\svgscale}}%
    \fi%
  \else%
    \setlength{\unitlength}{\svgwidth}%
  \fi%
  \global\let\svgwidth\undefined%
  \global\let\svgscale\undefined%
  \makeatother%
  \begin{picture}(1,0.75)%
    \put(0,0){\includegraphics[width=\unitlength,page=1]{kitepathelevazi_HYDRA.pdf}}%
    \put(0.06633342,0.37481517){\color[rgb]{0,0,0}\rotatebox{90}{\makebox(0,0)[b]{\smash{Elevation angle [$^\circ$]}}}}%
    \put(0.20930285,0.60590241){\color[rgb]{0,0,0}\makebox(0,0)[lb]{\smash{Gravity incl.}}}%
    \put(0.20930285,0.5736618){\color[rgb]{0,0,0}\makebox(0,0)[lb]{\smash{Gravity excl.}}}%
    \put(0.20930285,0.54142115){\color[rgb]{0,0,0}\makebox(0,0)[lb]{\smash{Experiment}}}%
    \put(0.1517516,0.63857714){\color[rgb]{0,0,0}\makebox(0,0)[lb]{\smash{\textbf{Strong wind}}}}%
  \end{picture}%
\endgroup%

%% file: kitepathelevazi_V3.pdf_tex
%% Creator: Inkscape inkscape 0.91, www.inkscape.org
%% Accompanies image file 'kitepathelevazi_V3_weg.pdf' (pdf, eps, ps)
\begingroup%
  \makeatletter%
  \providecommand\color[2][]{%
    \errmessage{(Inkscape) Color is used for the text in Inkscape, but the package 'color.sty' is not loaded}%
    \renewcommand\color[2][]{}%
  }%
  \providecommand\transparent[1]{%
    \errmessage{(Inkscape) Transparency is used (non-zero) for the text in Inkscape, but the package 'transparent.sty' is not loaded}%
    \renewcommand\transparent[1]{}%
  }%
  \providecommand\rotatebox[2]{#2}%
  \ifx\svgwidth\undefined%
    \setlength{\unitlength}{576bp}%
    \ifx\svgscale\undefined%
      \relax%
    \else%
      \setlength{\unitlength}{\unitlength * \real{\svgscale}}%
    \fi%
  \else%
    \setlength{\unitlength}{\svgwidth}%
  \fi%
  \global\let\svgwidth\undefined%
  \global\let\svgscale\undefined%
  \makeatother%
  \begin{picture}(1,0.75)%
    \put(0,0){\includegraphics[width=\unitlength,page=1]{kitepathelevazi_V3.pdf}}%
    \put(0.20930285,0.60590241){\color[rgb]{0,0,0}\makebox(0,0)[lb]{\smash{Gravity incl.}}}%
    \put(0.20930285,0.5736618){\color[rgb]{0,0,0}\makebox(0,0)[lb]{\smash{Gravity excl.}}}%
    \put(0.20930285,0.54142115){\color[rgb]{0,0,0}\makebox(0,0)[lb]{\smash{Experiment}}}%
    \put(0.51264381,0.01572388){\color[rgb]{0,0,0}\makebox(0,0)[b]{\smash{Azimuth angle [$^\circ$]}}}%
    \put(0.06633342,0.37481517){\color[rgb]{0,0,0}\rotatebox{90}{\makebox(0,0)[b]{\smash{Elevation angle [$^\circ$]}}}}%
    \put(0.1517516,0.63857714){\color[rgb]{0,0,0}\makebox(0,0)[lb]{\smash{\textbf{Moderate wind}}}}%
  \end{picture}%
\endgroup%

%% file: tetherlength_HYDRA.pdf_tex
%% Creator: Inkscape inkscape 0.91, www.inkscape.org
%% Accompanies image file 'tetherlength_HYDRA_weg.pdf' (pdf, eps, ps)
\begingroup%
  \makeatletter%
  \providecommand\color[2][]{%
    \errmessage{(Inkscape) Color is used for the text in Inkscape, but the package 'color.sty' is not loaded}%
    \renewcommand\color[2][]{}%
  }%
  \providecommand\transparent[1]{%
    \errmessage{(Inkscape) Transparency is used (non-zero) for the text in Inkscape, but the package 'transparent.sty' is not loaded}%
    \renewcommand\transparent[1]{}%
  }%
  \providecommand\rotatebox[2]{#2}%
  \ifx\svgwidth\undefined%
    \setlength{\unitlength}{576bp}%
    \ifx\svgscale\undefined%
      \relax%
    \else%
      \setlength{\unitlength}{\unitlength * \real{\svgscale}}%
    \fi%
  \else%
    \setlength{\unitlength}{\svgwidth}%
  \fi%
  \global\let\svgwidth\undefined%
  \global\let\svgscale\undefined%
  \makeatother%
  \begin{picture}(1,0.75)%
    \put(0,0){\includegraphics[width=\unitlength,page=1]{tetherlength_HYDRA.pdf}}%
    \put(0.20850694,0.16406583){\color[rgb]{0,0,0}\makebox(0,0)[lb]{\smash{Gravity incl.}}}%
    \put(0.20850694,0.13182521){\color[rgb]{0,0,0}\makebox(0,0)[lb]{\smash{Gravity excl.}}}%
    \put(0.20850694,0.0995846){\color[rgb]{0,0,0}\makebox(0,0)[lb]{\smash{Experiment}}}%
    \put(0.06600876,0.37539244){\color[rgb]{0,0,0}\rotatebox{90}{\makebox(0,0)[b]{\smash{Tether length [m]}}}}%
    \put(0.15190992,0.19666939){\color[rgb]{0,0,0}\makebox(0,0)[lb]{\smash{\textbf{Strong wind}}}}%
  \end{picture}%
\endgroup%

%% file: tetherlength_V3.pdf_tex
%% Creator: Inkscape inkscape 0.91, www.inkscape.org
%% Accompanies image file 'tetherlength_V3_weg.pdf' (pdf, eps, ps)
\begingroup%
  \makeatletter%
  \providecommand\color[2][]{%
    \errmessage{(Inkscape) Color is used for the text in Inkscape, but the package 'color.sty' is not loaded}%
    \renewcommand\color[2][]{}%
  }%
  \providecommand\transparent[1]{%
    \errmessage{(Inkscape) Transparency is used (non-zero) for the text in Inkscape, but the package 'transparent.sty' is not loaded}%
    \renewcommand\transparent[1]{}%
  }%
  \providecommand\rotatebox[2]{#2}%
  \ifx\svgwidth\undefined%
    \setlength{\unitlength}{576bp}%
    \ifx\svgscale\undefined%
      \relax%
    \else%
      \setlength{\unitlength}{\unitlength * \real{\svgscale}}%
    \fi%
  \else%
    \setlength{\unitlength}{\svgwidth}%
  \fi%
  \global\let\svgwidth\undefined%
  \global\let\svgscale\undefined%
  \makeatother%
  \begin{picture}(1,0.75)%
    \put(0,0){\includegraphics[width=\unitlength,page=1]{tetherlength_V3.pdf}}%
    \put(0.6616816,0.16406582){\color[rgb]{0,0,0}\makebox(0,0)[lb]{\smash{Gravity incl.}}}%
    \put(0.6616816,0.13182521){\color[rgb]{0,0,0}\makebox(0,0)[lb]{\smash{Gravity excl.}}}%
    \put(0.6616816,0.0995846){\color[rgb]{0,0,0}\makebox(0,0)[lb]{\smash{Experiment}}}%
    \put(0.06600876,0.37539244){\color[rgb]{0,0,0}\rotatebox{90}{\makebox(0,0)[b]{\smash{Tether length [m]}}}}%
    \put(0.51268991,0.01614773){\color[rgb]{0,0,0}\makebox(0,0)[b]{\smash{Time [s]}}}%
    \put(0.60508458,0.19666939){\color[rgb]{0,0,0}\makebox(0,0)[lb]{\smash{\textbf{Moderate wind}}}}%
  \end{picture}%
\endgroup%

%% file: reeloutvel_HYDRA.pdf_tex
%% Creator: Inkscape inkscape 0.91, www.inkscape.org
%% Accompanies image file 'reeloutvel_HYDRA_weg.pdf' (pdf, eps, ps)
\begingroup%
  \makeatletter%
  \providecommand\color[2][]{%
    \errmessage{(Inkscape) Color is used for the text in Inkscape, but the package 'color.sty' is not loaded}%
    \renewcommand\color[2][]{}%
  }%
  \providecommand\transparent[1]{%
    \errmessage{(Inkscape) Transparency is used (non-zero) for the text in Inkscape, but the package 'transparent.sty' is not loaded}%
    \renewcommand\transparent[1]{}%
  }%
  \providecommand\rotatebox[2]{#2}%
  \ifx\svgwidth\undefined%
    \setlength{\unitlength}{576bp}%
    \ifx\svgscale\undefined%
      \relax%
    \else%
      \setlength{\unitlength}{\unitlength * \real{\svgscale}}%
    \fi%
  \else%
    \setlength{\unitlength}{\svgwidth}%
  \fi%
  \global\let\svgwidth\undefined%
  \global\let\svgscale\undefined%
  \makeatother%
  \begin{picture}(1,0.75)%
    \put(0,0){\includegraphics[width=\unitlength,page=1]{reeloutvel_HYDRA.pdf}}%
    \put(0.20850694,0.60569572){\color[rgb]{0,0,0}\makebox(0,0)[lb]{\smash{Gravity incl.}}}%
    \put(0.20850694,0.57345512){\color[rgb]{0,0,0}\makebox(0,0)[lb]{\smash{Gravity excl.}}}%
    \put(0.20850694,0.54121447){\color[rgb]{0,0,0}\makebox(0,0)[lb]{\smash{Experiment}}}%
    \put(0.1518897,0.63808277){\color[rgb]{0,0,0}\makebox(0,0)[lb]{\smash{\textbf{Strong wind}}}}%
    \put(0.05894881,0.3815545){\color[rgb]{0,0,0}\rotatebox{90}{\makebox(0,0)[b]{\smash{Tether reeling velocity [m/s]}}}}%
  \end{picture}%
\endgroup%

%% file: reeloutvel_V3.pdf_tex
%% Creator: Inkscape inkscape 0.91, www.inkscape.org
%% Accompanies image file 'reeloutvel_V3_weg.pdf' (pdf, eps, ps)
\begingroup%
  \makeatletter%
  \providecommand\color[2][]{%
    \errmessage{(Inkscape) Color is used for the text in Inkscape, but the package 'color.sty' is not loaded}%
    \renewcommand\color[2][]{}%
  }%
  \providecommand\transparent[1]{%
    \errmessage{(Inkscape) Transparency is used (non-zero) for the text in Inkscape, but the package 'transparent.sty' is not loaded}%
    \renewcommand\transparent[1]{}%
  }%
  \providecommand\rotatebox[2]{#2}%
  \ifx\svgwidth\undefined%
    \setlength{\unitlength}{576bp}%
    \ifx\svgscale\undefined%
      \relax%
    \else%
      \setlength{\unitlength}{\unitlength * \real{\svgscale}}%
    \fi%
  \else%
    \setlength{\unitlength}{\svgwidth}%
  \fi%
  \global\let\svgwidth\undefined%
  \global\let\svgscale\undefined%
  \makeatother%
  \begin{picture}(1,0.75)%
    \put(0,0){\includegraphics[width=\unitlength,page=1]{reeloutvel_V3.pdf}}%
    \put(0.20850694,0.6059201){\color[rgb]{0,0,0}\makebox(0,0)[lb]{\smash{Gravity incl.}}}%
    \put(0.20850694,0.5736795){\color[rgb]{0,0,0}\makebox(0,0)[lb]{\smash{Gravity excl.}}}%
    \put(0.20850694,0.5414389){\color[rgb]{0,0,0}\makebox(0,0)[lb]{\smash{Experiment}}}%
    \put(0.51251464,0.01608075){\color[rgb]{0,0,0}\makebox(0,0)[b]{\smash{Time [s]}}}%
    \put(0.05894881,0.3815545){\color[rgb]{0,0,0}\rotatebox{90}{\makebox(0,0)[b]{\smash{Tether reeling velocity [m/s]}}}}%
    \put(0.15184111,0.63822052){\color[rgb]{0,0,0}\makebox(0,0)[lb]{\smash{\textbf{Moderate wind}}}}%
  \end{picture}%
\endgroup%

%% file: kitevel_HYDRA.pdf_tex
%% Creator: Inkscape inkscape 0.91, www.inkscape.org
%% Accompanies image file 'kitevel_HYDRA_weg.pdf' (pdf, eps, ps)
\begingroup%
  \makeatletter%
  \providecommand\color[2][]{%
    \errmessage{(Inkscape) Color is used for the text in Inkscape, but the package 'color.sty' is not loaded}%
    \renewcommand\color[2][]{}%
  }%
  \providecommand\transparent[1]{%
    \errmessage{(Inkscape) Transparency is used (non-zero) for the text in Inkscape, but the package 'transparent.sty' is not loaded}%
    \renewcommand\transparent[1]{}%
  }%
  \providecommand\rotatebox[2]{#2}%
  \ifx\svgwidth\undefined%
    \setlength{\unitlength}{576bp}%
    \ifx\svgscale\undefined%
      \relax%
    \else%
      \setlength{\unitlength}{\unitlength * \real{\svgscale}}%
    \fi%
  \else%
    \setlength{\unitlength}{\svgwidth}%
  \fi%
  \global\let\svgwidth\undefined%
  \global\let\svgscale\undefined%
  \makeatother%
  \begin{picture}(1,0.75)%
    \put(0,0){\includegraphics[width=\unitlength,page=1]{kitevel_HYDRA.pdf}}%
    \put(0.20850694,0.60587502){\color[rgb]{0,0,0}\makebox(0,0)[lb]{\smash{Gravity incl.}}}%
    \put(0.20850694,0.57363436){\color[rgb]{0,0,0}\makebox(0,0)[lb]{\smash{Gravity excl.}}}%
    \put(0.20850694,0.54139376){\color[rgb]{0,0,0}\makebox(0,0)[lb]{\smash{Experiment}}}%
    \put(0.07878411,0.37567027){\color[rgb]{0,0,0}\rotatebox{90}{\makebox(0,0)[b]{\smash{Kite velocity [m/s]}}}}%
    \put(0.15173798,0.63785738){\color[rgb]{0,0,0}\makebox(0,0)[lb]{\smash{\textbf{Strong wind}}}}%
  \end{picture}%
\endgroup%

%% file: kitevel_V3.pdf_tex
%% Creator: Inkscape inkscape 0.91, www.inkscape.org
%% Accompanies image file 'kitevel_V3_weg.pdf' (pdf, eps, ps)
\begingroup%
  \makeatletter%
  \providecommand\color[2][]{%
    \errmessage{(Inkscape) Color is used for the text in Inkscape, but the package 'color.sty' is not loaded}%
    \renewcommand\color[2][]{}%
  }%
  \providecommand\transparent[1]{%
    \errmessage{(Inkscape) Transparency is used (non-zero) for the text in Inkscape, but the package 'transparent.sty' is not loaded}%
    \renewcommand\transparent[1]{}%
  }%
  \providecommand\rotatebox[2]{#2}%
  \ifx\svgwidth\undefined%
    \setlength{\unitlength}{576bp}%
    \ifx\svgscale\undefined%
      \relax%
    \else%
      \setlength{\unitlength}{\unitlength * \real{\svgscale}}%
    \fi%
  \else%
    \setlength{\unitlength}{\svgwidth}%
  \fi%
  \global\let\svgwidth\undefined%
  \global\let\svgscale\undefined%
  \makeatother%
  \begin{picture}(1,0.75)%
    \put(0,0){\includegraphics[width=\unitlength,page=1]{kitevel_V3.pdf}}%
    \put(0.07878411,0.37567027){\color[rgb]{0,0,0}\rotatebox{90}{\makebox(0,0)[b]{\smash{Kite velocity [m/s]}}}}%
    \put(0.51248895,0.015959){\color[rgb]{0,0,0}\makebox(0,0)[b]{\smash{Time [s]}}}%
    \put(0.20850694,0.60587502){\color[rgb]{0,0,0}\makebox(0,0)[lb]{\smash{Gravity incl.}}}%
    \put(0.20850694,0.57363436){\color[rgb]{0,0,0}\makebox(0,0)[lb]{\smash{Gravity excl.}}}%
    \put(0.20850694,0.54139376){\color[rgb]{0,0,0}\makebox(0,0)[lb]{\smash{Experiment}}}%
    \put(0.15173798,0.63785738){\color[rgb]{0,0,0}\makebox(0,0)[lb]{\smash{\textbf{Moderate wind}}}}%
  \end{picture}%
\endgroup%

%% file: apparentvel_HYDRA.pdf_tex
%% Creator: Inkscape inkscape 0.91, www.inkscape.org
%% Accompanies image file 'apparentvel_HYDRA_weg.pdf' (pdf, eps, ps)
\begingroup%
  \makeatletter%
  \providecommand\color[2][]{%
    \errmessage{(Inkscape) Color is used for the text in Inkscape, but the package 'color.sty' is not loaded}%
    \renewcommand\color[2][]{}%
  }%
  \providecommand\transparent[1]{%
    \errmessage{(Inkscape) Transparency is used (non-zero) for the text in Inkscape, but the package 'transparent.sty' is not loaded}%
    \renewcommand\transparent[1]{}%
  }%
  \providecommand\rotatebox[2]{#2}%
  \ifx\svgwidth\undefined%
    \setlength{\unitlength}{576bp}%
    \ifx\svgscale\undefined%
      \relax%
    \else%
      \setlength{\unitlength}{\unitlength * \real{\svgscale}}%
    \fi%
  \else%
    \setlength{\unitlength}{\svgwidth}%
  \fi%
  \global\let\svgwidth\undefined%
  \global\let\svgscale\undefined%
  \makeatother%
  \begin{picture}(1,0.75)%
    \put(0,0){\includegraphics[width=\unitlength,page=1]{apparentvel_HYDRA.pdf}}%
    \put(0.20828234,0.60566447){\color[rgb]{0,0,0}\makebox(0,0)[lb]{\smash{Gravity incl.}}}%
    \put(0.20828234,0.57342386){\color[rgb]{0,0,0}\makebox(0,0)[lb]{\smash{Gravity excl.}}}%
    \put(0.20828234,0.54118326){\color[rgb]{0,0,0}\makebox(0,0)[lb]{\smash{Experiment}}}%
    \put(0.07908146,0.37485597){\color[rgb]{0,0,0}\rotatebox{90}{\makebox(0,0)[b]{\smash{Wind velocity [m/s]}}}}%
    \put(0.15222406,0.6383107){\color[rgb]{0,0,0}\makebox(0,0)[lb]{\smash{\textbf{Strong wind}}}}%
    \put(0.69323222,0.39374537){\color[rgb]{0,0,0}\makebox(0,0)[lb]{\smash{Apparent wind}}}%
    \put(0.69323222,0.12694681){\color[rgb]{0,0,0}\makebox(0,0)[lb]{\smash{True wind}}}%
  \end{picture}%
\endgroup%

%% file: apparentvel_V3.pdf_tex
%% Creator: Inkscape inkscape 0.91, www.inkscape.org
%% Accompanies image file 'apparentvel_V3_weg.pdf' (pdf, eps, ps)
\begingroup%
  \makeatletter%
  \providecommand\color[2][]{%
    \errmessage{(Inkscape) Color is used for the text in Inkscape, but the package 'color.sty' is not loaded}%
    \renewcommand\color[2][]{}%
  }%
  \providecommand\transparent[1]{%
    \errmessage{(Inkscape) Transparency is used (non-zero) for the text in Inkscape, but the package 'transparent.sty' is not loaded}%
    \renewcommand\transparent[1]{}%
  }%
  \providecommand\rotatebox[2]{#2}%
  \ifx\svgwidth\undefined%
    \setlength{\unitlength}{576bp}%
    \ifx\svgscale\undefined%
      \relax%
    \else%
      \setlength{\unitlength}{\unitlength * \real{\svgscale}}%
    \fi%
  \else%
    \setlength{\unitlength}{\svgwidth}%
  \fi%
  \global\let\svgwidth\undefined%
  \global\let\svgscale\undefined%
  \makeatother%
  \begin{picture}(1,0.75)%
    \put(0,0){\includegraphics[width=\unitlength,page=1]{apparentvel_V3.pdf}}%
    \put(0.20828234,0.60566447){\color[rgb]{0,0,0}\makebox(0,0)[lb]{\smash{Gravity incl.}}}%
    \put(0.20828234,0.57342386){\color[rgb]{0,0,0}\makebox(0,0)[lb]{\smash{Gravity excl.}}}%
    \put(0.20828234,0.54118326){\color[rgb]{0,0,0}\makebox(0,0)[lb]{\smash{Experiment}}}%
    \put(0.15222406,0.6383107){\color[rgb]{0,0,0}\makebox(0,0)[lb]{\smash{\textbf{Moderate wind}}}}%
    \put(0.51268991,0.01614772){\color[rgb]{0,0,0}\makebox(0,0)[b]{\smash{Time [s]}}}%
    \put(0.07908146,0.37485597){\color[rgb]{0,0,0}\rotatebox{90}{\makebox(0,0)[b]{\smash{Wind velocity [m/s]}}}}%
    \put(0.69378138,0.42335338){\color[rgb]{0,0,0}\makebox(0,0)[lb]{\smash{Apparent wind}}}%
    \put(0.69378138,0.27384456){\color[rgb]{0,0,0}\makebox(0,0)[lb]{\smash{True wind}}}%
  \end{picture}%
\endgroup%

%% file: cabletension_HYDRA.pdf_tex
%% Creator: Inkscape inkscape 0.91, www.inkscape.org
%% Accompanies image file 'cabletension_HYDRA_weg.pdf' (pdf, eps, ps)
\begingroup%
  \makeatletter%
  \providecommand\color[2][]{%
    \errmessage{(Inkscape) Color is used for the text in Inkscape, but the package 'color.sty' is not loaded}%
    \renewcommand\color[2][]{}%
  }%
  \providecommand\transparent[1]{%
    \errmessage{(Inkscape) Transparency is used (non-zero) for the text in Inkscape, but the package 'transparent.sty' is not loaded}%
    \renewcommand\transparent[1]{}%
  }%
  \providecommand\rotatebox[2]{#2}%
  \ifx\svgwidth\undefined%
    \setlength{\unitlength}{576bp}%
    \ifx\svgscale\undefined%
      \relax%
    \else%
      \setlength{\unitlength}{\unitlength * \real{\svgscale}}%
    \fi%
  \else%
    \setlength{\unitlength}{\svgwidth}%
  \fi%
  \global\let\svgwidth\undefined%
  \global\let\svgscale\undefined%
  \makeatother%
  \begin{picture}(1,0.75)%
    \put(0,0){\includegraphics[width=\unitlength,page=1]{cabletension_HYDRA.pdf}}%
    \put(0.20850694,0.60561524){\color[rgb]{0,0,0}\makebox(0,0)[lb]{\smash{Gravity incl.}}}%
    \put(0.20850694,0.57337464){\color[rgb]{0,0,0}\makebox(0,0)[lb]{\smash{Gravity excl.}}}%
    \put(0.20850694,0.54113404){\color[rgb]{0,0,0}\makebox(0,0)[lb]{\smash{Experiment}}}%
    \put(0.05350164,0.37513402){\color[rgb]{0,0,0}\rotatebox{90}{\makebox(0,0)[b]{\smash{Tether force [N]}}}}%
    \put(0.15190652,0.6383107){\color[rgb]{0,0,0}\makebox(0,0)[lb]{\smash{\textbf{Strong wind}}}}%
  \end{picture}%
\endgroup%

%% file: cabletension_V3.pdf_tex
%% Creator: Inkscape inkscape 0.91, www.inkscape.org
%% Accompanies image file 'cabletension_V3_weg.pdf' (pdf, eps, ps)
\begingroup%
  \makeatletter%
  \providecommand\color[2][]{%
    \errmessage{(Inkscape) Color is used for the text in Inkscape, but the package 'color.sty' is not loaded}%
    \renewcommand\color[2][]{}%
  }%
  \providecommand\transparent[1]{%
    \errmessage{(Inkscape) Transparency is used (non-zero) for the text in Inkscape, but the package 'transparent.sty' is not loaded}%
    \renewcommand\transparent[1]{}%
  }%
  \providecommand\rotatebox[2]{#2}%
  \ifx\svgwidth\undefined%
    \setlength{\unitlength}{576bp}%
    \ifx\svgscale\undefined%
      \relax%
    \else%
      \setlength{\unitlength}{\unitlength * \real{\svgscale}}%
    \fi%
  \else%
    \setlength{\unitlength}{\svgwidth}%
  \fi%
  \global\let\svgwidth\undefined%
  \global\let\svgscale\undefined%
  \makeatother%
  \begin{picture}(1,0.75)%
    \put(0,0){\includegraphics[width=\unitlength,page=1]{cabletension_V3.pdf}}%
    \put(0.51243565,0.01589349){\color[rgb]{0,0,0}\makebox(0,0)[b]{\smash{Time [s]}}}%
    \put(0.05350164,0.37513402){\color[rgb]{0,0,0}\rotatebox{90}{\makebox(0,0)[b]{\smash{Tether force [N]}}}}%
    \put(0.20850694,0.60561524){\color[rgb]{0,0,0}\makebox(0,0)[lb]{\smash{Gravity incl.}}}%
    \put(0.20850694,0.57337464){\color[rgb]{0,0,0}\makebox(0,0)[lb]{\smash{Gravity excl.}}}%
    \put(0.20850694,0.54113404){\color[rgb]{0,0,0}\makebox(0,0)[lb]{\smash{Experiment}}}%
    \put(0.15190652,0.6383107){\color[rgb]{0,0,0}\makebox(0,0)[lb]{\smash{\textbf{Moderate wind}}}}%
  \end{picture}%
\endgroup%

%% file: power_HYDRA.pdf_tex
%% Creator: Inkscape inkscape 0.91, www.inkscape.org
%% Accompanies image file 'power_HYDRA_weg.pdf' (pdf, eps, ps)
\begingroup%
  \makeatletter%
  \providecommand\color[2][]{%
    \errmessage{(Inkscape) Color is used for the text in Inkscape, but the package 'color.sty' is not loaded}%
    \renewcommand\color[2][]{}%
  }%
  \providecommand\transparent[1]{%
    \errmessage{(Inkscape) Transparency is used (non-zero) for the text in Inkscape, but the package 'transparent.sty' is not loaded}%
    \renewcommand\transparent[1]{}%
  }%
  \providecommand\rotatebox[2]{#2}%
  \ifx\svgwidth\undefined%
    \setlength{\unitlength}{576bp}%
    \ifx\svgscale\undefined%
      \relax%
    \else%
      \setlength{\unitlength}{\unitlength * \real{\svgscale}}%
    \fi%
  \else%
    \setlength{\unitlength}{\svgwidth}%
  \fi%
  \global\let\svgwidth\undefined%
  \global\let\svgscale\undefined%
  \makeatother%
  \begin{picture}(1,0.75)%
    \put(0,0){\includegraphics[width=\unitlength,page=1]{power_HYDRA.pdf}}%
    \put(0.20827748,0.60588037){\color[rgb]{0,0,0}\makebox(0,0)[lb]{\smash{Gravity incl.}}}%
    \put(0.20827748,0.57363976){\color[rgb]{0,0,0}\makebox(0,0)[lb]{\smash{Gravity excl.}}}%
    \put(0.20827748,0.54139911){\color[rgb]{0,0,0}\makebox(0,0)[lb]{\smash{Experiment}}}%
    \put(0.15222406,0.6383107){\color[rgb]{0,0,0}\makebox(0,0)[lb]{\smash{\textbf{Strong wind}}}}%
    \put(0.05895426,0.37571245){\color[rgb]{0,0,0}\rotatebox{90}{\makebox(0,0)[b]{\smash{Mechanical power [kW]}}}}%
  \end{picture}%
\endgroup%

%% file: power_V3.pdf_tex
%% Creator: Inkscape inkscape 0.91, www.inkscape.org
%% Accompanies image file 'power_V3_weg.pdf' (pdf, eps, ps)
\begingroup%
  \makeatletter%
  \providecommand\color[2][]{%
    \errmessage{(Inkscape) Color is used for the text in Inkscape, but the package 'color.sty' is not loaded}%
    \renewcommand\color[2][]{}%
  }%
  \providecommand\transparent[1]{%
    \errmessage{(Inkscape) Transparency is used (non-zero) for the text in Inkscape, but the package 'transparent.sty' is not loaded}%
    \renewcommand\transparent[1]{}%
  }%
  \providecommand\rotatebox[2]{#2}%
  \ifx\svgwidth\undefined%
    \setlength{\unitlength}{576bp}%
    \ifx\svgscale\undefined%
      \relax%
    \else%
      \setlength{\unitlength}{\unitlength * \real{\svgscale}}%
    \fi%
  \else%
    \setlength{\unitlength}{\svgwidth}%
  \fi%
  \global\let\svgwidth\undefined%
  \global\let\svgscale\undefined%
  \makeatother%
  \begin{picture}(1,0.75)%
    \put(0,0){\includegraphics[width=\unitlength,page=1]{power_V3.pdf}}%
    \put(0.20827748,0.60588037){\color[rgb]{0,0,0}\makebox(0,0)[lb]{\smash{Gravity incl.}}}%
    \put(0.20827748,0.57363976){\color[rgb]{0,0,0}\makebox(0,0)[lb]{\smash{Gravity excl.}}}%
    \put(0.20827748,0.54139911){\color[rgb]{0,0,0}\makebox(0,0)[lb]{\smash{Experiment}}}%
    \put(0.51268991,0.01614775){\color[rgb]{0,0,0}\makebox(0,0)[b]{\smash{Time [s]}}}%
    \put(0.05895426,0.37571245){\color[rgb]{0,0,0}\rotatebox{90}{\makebox(0,0)[b]{\smash{Mechanical power [kW]}}}}%
    \put(0.15222406,0.6383107){\color[rgb]{0,0,0}\makebox(0,0)[lb]{\smash{\textbf{Moderate wind}}}}%
  \end{picture}%
\endgroup%